\documentclass[preprint2]{aastex6}
\usepackage{cancel}	
\usepackage{color}
\usepackage{apjfonts}
\usepackage{affils}

\newcommand{\kms}{km\,s$^{-1}$}

\def\M{M$_{\odot}$}

\def\Ha{H{$\alpha$}}
\def\Hb{H{$\beta$}}

\def\Ni{$^{56}$Ni}
 \def\Co{$^{56}$Co}
 \def\Fe{$^{56}$Fe}
 
 \def\Mej{$M_{\rm ej}$}
\def\Mcsm{$M_{\rm CSM}$}
\def\Mni{$M_{\rm Ni}$}

\def\E{$E_{\rm k}$}

\def\ergs{erg\,s$^{-1}$}

\def\a{\textit{a}}
\def\b{\textit{b}}
\def\c{\textit{c}}
\def\d{\textit{d}}
\def\e{\textit{e}}

\begin{document}


\title{SN 2015bn: a detailed multi-wavelength view of a nearby superluminous supernova}

\shortauthors{Nicholl et al.}

\shorttitle{SN 2015bn: a nearby SLSN}


\DeclareAffil{cfa}{Harvard-Smithsonian Center for Astrophysics, 60 Garden Street, Cambridge,
  Massachusetts 02138, USA; \href{mailto:matt.nicholl@cfa.harvard.edu}{matt.nicholl@cfa.harvard.edu}}
\DeclareAffil{queens}{Astrophysics Research Centre, School of Mathematics and Physics, Queens University Belfast, Belfast BT7 1NN, UK}
\DeclareAffil{ccpp}{Center for Cosmology and Particle Physics, New York University, 4 Washington Place, New York, NY 10003, USA}
\DeclareAffil{mpifep}{Max-Planck-Institut f{\"u}r Extraterrestrische Physik, Giessenbachstra\ss e 1, 85748, Garching, Germany}
\DeclareAffil{lcogt}{Las Cumbres Observatory Global Telescope, 6740 Cortona Dr, Suite 102, Goleta, CA 93111, USA}
\DeclareAffil{kitp}{Kavli Institute for Theoretical Physics, University of California, Santa Barbara, CA 93106, USA}
\DeclareAffil{soton}{School of Physics and Astronomy, University of Southampton, Southampton, SO17 1BJ, UK}
\DeclareAffil{hawaii}{Institute for Astronomy, University of Hawaii at Manoa, Honolulu, HI 96822, USA}
\DeclareAffil{ohio}{Astrophysical Institute, Department of Physics and Astronomy, 251B Clippinger Lab, Ohio University, Athens, OH 45701, USA}
\DeclareAffil{cambridge}{Institute of Astronomy, University of Cambridge, Madingley Road, Cambridge, CB3 0HA}
\DeclareAffil{benoziyo}{Benoziyo Center for Astrophysics, Weizmann Institute of Science, Rehovot 76100, Israel}
\DeclareAffil{millennium}{Millennium Institute of Astrophysics, Vicu\~{n}a Mackenna 4860, 7820436 Macul, Santiago, Chile}
\DeclareAffil{chile}{Departamento de Astronom\'ia, Universidad de Chile, Camino El Observatorio 1515, Las Condes, Santiago, Chile}
\DeclareAffil{tuorla}{Tuorla Observatory, Department of Physics and Astronomy, University of Turku, V\"ais\"al\"antie 20, FI-21500 Piikkiö, Finland}
\DeclareAffil{osu}{Department of Astronomy, The Ohio State University, 140 West 18th Avenue, Columbus, OH 43210, USA}
\DeclareAffil{ccapp}{Center for Cosmology and AstroParticle Physics (CCAPP), The Ohio State University, 191 W. Woodruff Ave., Columbus, OH 43210, USA}
\DeclareAffil{ucsb}{Department of Physics, University of California, Santa Barbara, Broida Hall, Mail Code 9530, Santa Barbara, CA 93106-9530, USA}
\DeclareAffil{taiwan}{Institute of Astronomy, National Central University, Chung-Li 32054, Taiwan}
\DeclareAffil{columbia}{Columbia Astrophysics Laboratory, Columbia University, New York, NY 10027, USA}
\DeclareAffil{sorbonne}{Sorbonne Universit\'es, UPMC, Paris VI,UMR 7585, LPNHE, F-75005, Paris, France}
\DeclareAffil{cnrs}{CNRS, UMR 7585, Laboratoire de Physique Nucleaire et des Hautes Energies, 4 place Jussieu, 75005 Paris, France}
\DeclareAffil{not}{Nordic Optical Telescope, Apartado 474, E-38700 Santa Cruz de La Palma, Spain}
\DeclareAffil{carnegie}{Carnegie Observatories, 813 Santa Barbara Street, Pasadena, CA 91101, USA}
\DeclareAffil{hcpf}{Hubble, Carnegie-Princeton Fellow}
\DeclareAffil{davis}{Department of Physics, University of California, Davis, CA 95616, USA}
\DeclareAffil{mit}{Kavli Institute for Astrophysics and Space Research, Massachusetts Institute of Technology, 77 Massachusetts Avenue, Cambridge, MA 02139}
\DeclareAffil{caltech}{Astronomy Department, California Institute of Technology, Pasadena, California 91125, USA}

\affilauthorlist{M.~Nicholl\affils{cfa},
E.~Berger\affils{cfa},
S.~J.~Smartt\affils{queens},
R.~Margutti\affils{ccpp},
A.~Kamble\affils{cfa},
K.~D.~Alexander\affils{cfa},
T.-W.~Chen\affils{mpifep},
C.~Inserra\affils{queens},
I.~Arcavi\affils{lcogt,kitp},
P.~K.~Blanchard\affils{cfa},
R.~Cartier\affils{soton},
K.~C.~Chambers\affils{hawaii},
M.~J.~Childress\affils{soton},
R.~Chornock\affils{ohio},
P.~S.~Cowperthwaite\affils{cfa},
M.~Drout\affils{cfa},
H.~A.~Flewelling\affils{hawaii},
M.~Fraser\affils{cambridge},
A.~Gal-Yam\affils{benoziyo},
L.~Galbany\affils{millennium,chile},
J.~Harmanen\affils{tuorla},
T.~W.-S.~Holoien\affils{osu,ccapp},
G.~Hosseinzadeh\affils{lcogt,ucsb},
D.~A.~Howell\affils{lcogt,ucsb},
M.~E.~Huber\affils{hawaii},
A.~Jerkstrand\affils{queens},
E.~Kankare\affils{queens},
C.~S.~Kochanek\affils{osu,ccapp},
Z.-Y.~Lin\affils{taiwan},
R.~Lunnan\affils{caltech},
E.~A.~Magnier\affils{hawaii},
K.~Maguire\affils{queens},
C.~McCully\affils{lcogt,ucsb},
M.~McDonald\affils{mit},
B.~D.~Metzger\affils{columbia},
D.~Milisavljevic\affils{cfa},
A.~Mitra\affils{sorbonne,cnrs},
T.~Reynolds\affils{tuorla,not},
J.~Saario\affils{tuorla,not},
B.~J.~Shappee\affils{carnegie,hcpf},
K.~W.~Smith\affils{queens},
S.~Valenti\affils{lcogt,davis},
V.~A.~Villar\affils{cfa},
C.~Waters\affils{hawaii},
D.~R.~Young\affils{queens}
}

\begin{abstract}

We present observations of SN 2015bn (= PS15ae = CSS141223-113342+004332 = MLS150211-113342+004333), a Type I superluminous supernova (SLSN) at redshift $z=0.1136$. As well as being one of the closest SLSNe I yet discovered, it is intrinsically brighter ($M_U\approx-23.1$) and in a fainter  galaxy ($M_B\approx-16.0$) than other SLSNe at $z\sim0.1$. We used this opportunity to collect the most extensive dataset for any SLSN I to date, including densely-sampled spectroscopy and photometry, from the UV to the NIR, spanning $-$50 to +250 days from optical maximum. SN 2015bn fades slowly, but exhibits surprising undulations in the light curve on a timescale of 30-50 days, especially in the UV. The spectrum shows extraordinarily slow evolution except for a rapid transformation between +7 and +20-30 days. No narrow emission lines from slow-moving material are observed at any phase. We derive physical properties including the bolometric luminosity, and find slow velocity evolution and non-monotonic temperature and radial evolution. A deep radio limit rules out a healthy off-axis gamma-ray burst, and places constraints on the pre-explosion mass loss. The data can be consistently explained by a $\ga10$\,\M~stripped progenitor exploding with $\sim 10^{51}$\,erg kinetic energy, forming a magnetar with a spin-down timescale of $\sim20$\,d (thus avoiding a gamma-ray burst) that reheats the ejecta and drives ionization fronts. The most likely alternative scenario -- interaction with $\sim20$\,\M~of dense, inhomogeneous circumstellar material -- can be tested with continuing radio follow-up.

 \end{abstract}


\section{Introduction}\label{sec:intro}

The parameter space of observed supernovae (SNe), in terms of luminosity, duration and energy, has expanded dramatically since the turn of the century. Much of this progress has occurred thanks to a new generation of transient surveys. One of the most surprising results has been the discovery of `superluminous supernovae' \citep[SLSNe;][]{qui2011,gal2012}, which reach peak absolute magnitudes $-21\ga M\ga-23$\,mag and thus are intrinsically brighter than normal Type I and Type II SNe by at least 2 magnitudes. These were initially discovered by the ROTSE-III robotic telescope \citep{ake2003}, and are now detected by untargeted sky surveys such as Pan-STARRS1 \citep[PS1;][]{kai2010}, the Palomar Transient Factory \citep[PTF;][]{rau2009}, the Catalina Real-Time Transient Survey \citep[CRTS;][]{dra2009}, the La Silla QUEST survey \citep[LSQ;][]{balt2013}, and the All-Sky Automated Survey for Supernovae \citep[ASAS-SN;][]{sha2014}. Unveiling the progenitors and explosion mechanisms of SLSNe is crucial for our understanding of massive star evolution. Due to their luminosity, they have also generated interest as potential standard candles for high-redshift cosmology \citep{qui2013,ins2014}.

Traditionally, bright SNe indicate that a large mass of radioactive \Ni~was synthesized in the explosion; its subsequent decay to \Co~and then \Fe~provides the sustained power input to keep the ejecta hot despite rapid expansion. The peak luminosity in SLSNe ($>10^{44}$\,erg\,s$^{-1}$) would require several solar masses of \Ni, yet late-time observations suggest \Mni\,$<1$\,\M~\citep{pas2010,qui2011,ins2013,chen2013}. Furthermore, estimates of the ejected mass in SLSNe suggest $\langle$\Mej$\rangle \sim 10$\,\M~\citep{nic2015b}, whereas a helium core mass of $\ga40$\,\M~is required to synthesize a sufficiently high \Ni~mass \citep{mor2010}.

SLSNe are divided observationally into hydrogen-poor (Type I) and hydrogen-rich (Type II) subtypes \citep{gal2012}. When hydrogen is present in the spectrum, it usually appears in the form of strong, multi-component emission lines \citep[e.g.][]{smi2007b}, almost certainly indicating interaction with the circumstellar medium (CSM). Such objects are classified as `SLSNe IIn', by analogy with the fainter SNe IIn, which show narrow Balmer lines from shocked CSM. In SLSNe IIn, the luminosity is generated in the conversion of kinetic energy to radiative energy by shocks from the ejecta-CSM collision. However, a small fraction of SLSNe II are dominated by broader Balmer lines \citep{mil2009,gez2009,ben2014}, and in such cases the power source is less clear-cut.

Hydrogen-poor SLSNe show spectral similarities to normal Type Ic core-collapse SNe, and hence have been termed SLSNe I or Ic. Interaction models can also match the light curves here \citep[e.g.][]{che2011,gin2012,cha2013}, although presumably the CSM must be deficient in hydrogen to hide its spectral signature (in fact, no narrow lines of any kind are detected). The other model commonly invoked for SLSNe I is one in which the ejecta are heated internally by a central engine, in many ways similar to long-duration gamma-ray bursts \citep[LGRBs;][]{dunc1992,woo1993,macf1999}. The essential difference is that the engine timescale in SLSNe is comparable to the ejecta diffusion timescale, whereas in LGRBs, most of the energy is input very early \citep{met2015}. In this way, more of the energy in SLSNe goes into late-time observable radiation rather than driving a jet. The leading candidate for this engine is the spin-down of a nascent millisecond magnetar \citep{kas2010,woo2010}, though fallback accretion on to a black hole has also been proposed \citep{dex2013}.

\begin{figure*}
\centering
\includegraphics[width=15cm]{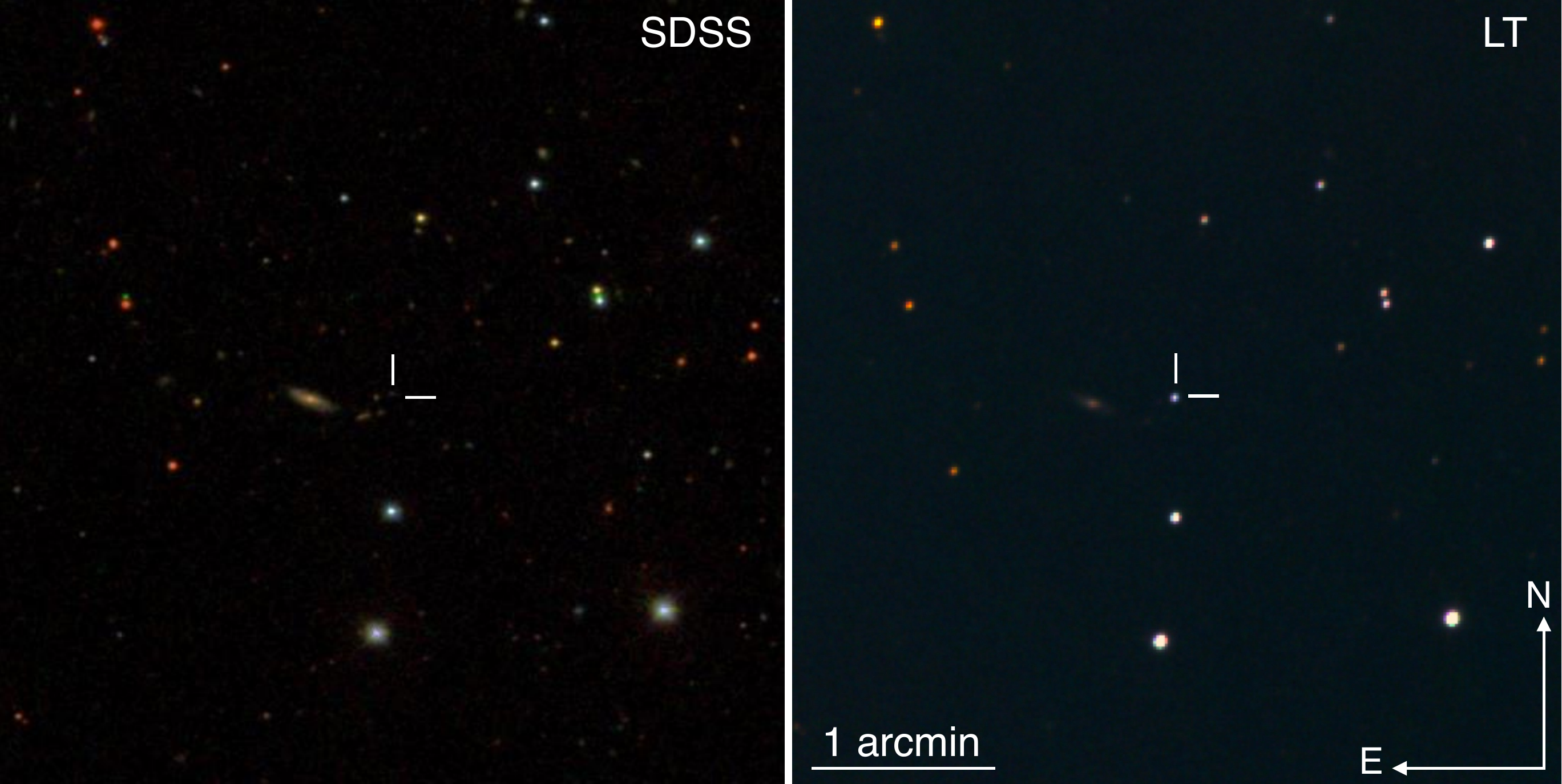}
\figcaption{Left: SDSS DR12 \citep{alam2015} pre-explosion colour image from 1999-03-22. The host galaxy of SN 2015bn is just visible as a faint smudge. Right: LT follow-up image of SN 2015bn. The colour image was made by combining $gri$ data obtained on 2015-03-01.\label{fig:pic}}
\end{figure*}

A few SLSNe I do exhibit slow light curve decline rates that appear consistent with \Co~decay, and have been suggested to arise from pair-instability SNe (PISNe) in progenitors with initial mass $M_{\rm ZAMS}>130$\,\M~\citep{gal2009}. However, other properties, such as their blue colours \citep{des2012} and relatively fast rise times \citep{nic2013} argue against this interpretation. Moreover, their spectroscopic evolution in the photospheric phase closely resembles other, faster SLSNe I, suggesting that the slow-decliners may be high-mass analogues of the more typical events \citep{nic2013,nic2015b}. Recently, \citet{jer2016} computed nebular phase spectra for PISN explosion models, finding them to be highly discrepant with observed spectra of these objects at $\sim400$-500\,d after explosion.

In this paper, we present and analyse SN 2015bn, which is one of the closest SLSNe I yet discovered. Its redshift of $z=0.1136$ puts it behind only PTF10hgi \citep[$z=0.098$;][]{ins2013} and PTF12dam \citep[$z=0.107$;][]{nic2013} in terms of published SLSNe I. We have used this rare opportunity to collect one of the largest and most densely sampled datasets for any SLSN. The paper is structured as follows. We discuss the discovery of the SN in section \ref{sec:class}, photometry in \ref{sec:phot}, and spectroscopic observations in section \ref{sec:spec}. In section \ref{sec:bol}, we construct the bolometric light curve and use it to derive physical properties, which we model in section \ref{sec:mod}. Our optical analysis is supplemented with the first late-time radio observation of a SLSN, presented in section \ref{sec:radio}.  We describe the host galaxy of SN 2015bn in section \ref{sec:host}, before concluding with a discussion of our findings in section \ref{sec:conc}.

\section{Discovery and classification}\label{sec:class}

The transient was first discovered by the Catalina Sky Survey on 2014 December 23, by the Mount Lemmon Survey (a division of CRTS) on 2015 February 11, and the Pan-STARRS Survey for Transients \citep[PSST;][]{hub2015} on 2015 February 15. It therefore has three different survey designations: CSS141223-113342+004332, MLS150211-113342+004333, and PS15ae. The IAU name from the Transient Name Server\footnote{http://wis-tns.weizmann.ac.il/} is SN 2015bn. We will use this name throughout the paper. The sky coordinates are $\alpha = 11^{\rm h}33^{\rm m}41^{\rm s}.55$, $\delta = +00^{\rm o}43'33''.4$ (J2000.0). Images of the field before and after explosion are shown in Figure \ref{fig:pic}.

In PESSTO \citep[Public ESO Spectroscopic Survey of Transient Objects;][]{sma2015} we ingest all publicly available transients into a database and cross-match and combine photometry from multiple surveys. Within the PESSTO marshall, SN 2015bn was identified as an unusual transient by human scanners: it exhibited a long rise time and bright apparent magnitude, and was coincident with a faint host galaxy. It was therefore prioritized for spectroscopic classification, and was observed by PESSTO on 2015 February 17 using EFOSC2 on the ESO New Technology Telescope (NTT) at La Silla Observatory. The classification spectrum is available on WISeREP\footnote{http://www.weizmann.ac.il/astrophysics/wiserep/} \citep{yar2012}. It showed good matches with young SLSNe I such as PTF12dam and PTF09cnd \citep{gui2015}, and indicated a redshift $z\sim0.11$. Subsequent detection of narrow lines from the host galaxy confirmed $z=0.1136$ -- making it one of the most nearby SLSNe to date. On 2015 February 27, \citet{dra2015} also reported a spectrum, agreeing with our classification.

\section{Photometry}\label{sec:phot}

\subsection{Imaging data reduction}\label{sec:phot_re}

Follow-up imaging observations of SN 2015bn were triggered with a number of instruments. Optical data in \textit{ugriz} were obtained with the 2.0-m Liverpool Telescope (LT) using the IO:O imager, and the 1.0-m Las Cumbres Observatory Global Telescope network (LCOGT). After $\sim200$\,d, deep \textit{ugriz} images were obtained with EFOSC2 on the NTT. These images were reduced by respective facility pipelines \citep[the PESSTO pipeline is described by][]{sma2015}, including de-biasing and flat-fielding. PS1 data obtained in the $w_{\rm P1}$-band were converted to the more standard $r$-band by a shift of $+0.1$ magnitudes. The same shift was applied to CRTS $R$-band data to match the better-sampled $r$-band. These shifts were calculated from the observed spectra. \textit{UBVRI} data were obtained from the 16" Ritchey-Chretien telescope (SLT) at Lulin Observatory, Taiwan. These were reduced using standard techniques in \textsc{iraf}, correcting for bias, flat-field and dark current. Near-infrared (NIR) images were taken using SOFI on the NTT and NOTCam on the Nordic Optical Telescope (NOT). These were reduced using the PESSTO pipeline, and the external \textsc{iraf} package \textsc{notcam} (version 2.5)\footnote{http://www.not.iac.es/instruments/notcam/guide/observe.html
}, applying flat-fielding and sky-subtraction on dithered frames.

\begin{figure}
\includegraphics[width=8.5cm]{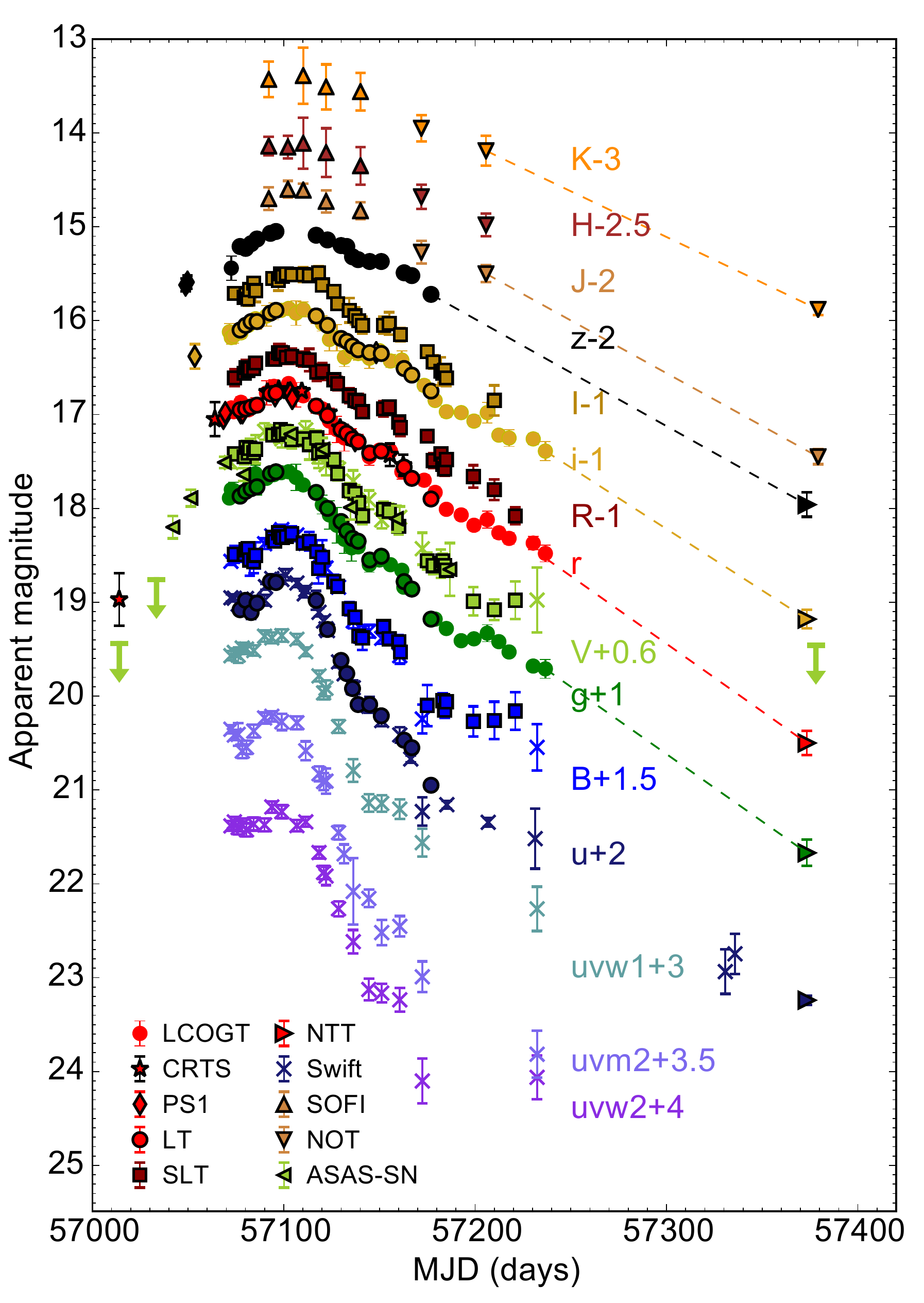}
\figcaption{The observed multicolour light curves of SN 2015bn from the UV to NIR. Magnitudes are given in their `natural' systems: AB for $ugriz$; Vega for the others. Constant offsets (labeled) have been added for clarity.
\label{fig:obs_lcs}}
\end{figure}

Photometric measurements were made on the reduced frames using a custom \textsc{pyraf} photometry pipeline, calling standard \textsc{daophot} routines to fit the point-spread function (PSF) and capture the SN flux. The nightly zero-points, in AB magnitudes, were calculated using a sequence of local field stars in the Sloan Digital Sky Survey (SDSS) Data Release 12 \citep{alam2015}. For the SLT photometry, the SDSS \textit{ugriz} magnitudes of the sequence stars were transformed into \textit{UBVRI} using the equations from \citet{jor2006}. These are reported in the Vega magnitude system. Errors in the SN magnitudes include both the scatter in the zero-points and the uncertainty in the PSF fit returned by \textsc{daophot}. ASAS-SN photometry was computed using a dedicated pipeline. Because SN 2015bn was close to the limiting magnitude for ASAS-SN ($\approx17$\,mag), we increased the signal-to-noise ratio of the data using a noise-weighted stacking of neighbouring epochs.

For the latest epochs of $ugrizJHK$ imaging at $\sim250$\,d, the host galaxy contributes $\sim20\%$ of the observed flux. Thus we remove this flux as follows: for $griz$, we align the images with SDSS templates and apply image subtraction using \textsc{hotpants}\footnote{http://www.astro.washington.edu/users/becker/v2.0/hotpants.html} \citep[based on the algorithms of][]{ala1998}. For $uJHK$, where template images without the SN are either not available or are too shallow to effectively remove the host flux, we do not use 2D image subtraction, but instead subtract fluxes based on a model galaxy spectral energy distribution (see section \ref{sec:host}) from our measured magnitudes. This method works reasonably well since the host of SN 2015bn is faint and compact, and the SN does not show a large offset from the centroid.

We activated our approved \textit{Swift}-GI program dedicated to the study of nearby SLSNe (PI: Margutti), as well as a number of other Target-of-Opportunity programs (PI: Inserra). We also include public data obtained under a separate program (PI: Brown). Imaging was taken with the Ultraviolet and Optical Telescope (UVOT), in the filters \textit{uvw2}, \textit{uvm2}, \textit{uvw1}, \textit{u}, \textit{b} and \textit{v}. The SN flux was extracted following standard methods \citep{bro2009} and using a 5'' aperture, and calibrated in the \textit{Swift} photometric system \citep{poo2008}. A shift of +1.02\,mag\footnote{http://swift.gsfc.nasa.gov/analysis/uvot\_digest/zeropts.html} was used to convert the \textit{Swift} $u$ zeropoint from Vega to AB system. While the transmission functions of the \textit{Swift} and ground-based $u$-filters are not identical, the photometry obtained matches that from LT perfectly, hence we feel justified in combining these data into a single, well-sampled light curve.
All photometric measurements, including epochs of observation, instruments, and reference star magnitudes, are given in the Appendix.

\subsection{Observed light curves}

The observed light curves of SN 2015bn, from UV to NIR, are shown in Figure \ref{fig:obs_lcs}. SN 2015bn reached $r$-band maximum light on MJD 57102. At $r=16.7$\,mag, it exhibits the brightest apparent magnitude of any SLSN I yet discovered. The first detection from CSS is 88\,d earlier, which translates to 79\,d in the rest-frame at $z=0.1136$. This makes SN 2015bn a slowly evolving SLSN with one of the longest rise times to date, similar to the estimate for iPTF13ehe \citep{yan2015}. To fill in the gap from this first detection until our classification and follow-up campaign beginning at $-28$\,d, we searched archival images from PS1 and ASAS-SN. We recovered detections in $V$- (ASAS-SN), $i$-, and $z$-bands (PS1), confirming the long rise. While the wavelength and temporal sampling is extremely dense during our follow-up campaign, the pre-classification photometry is relatively sparse. This makes it difficult to constrain the shape of the light curve at early times. We will address this issue in the next section.

\begin{figure}
\centering
\includegraphics[width=8.5cm,angle=0]{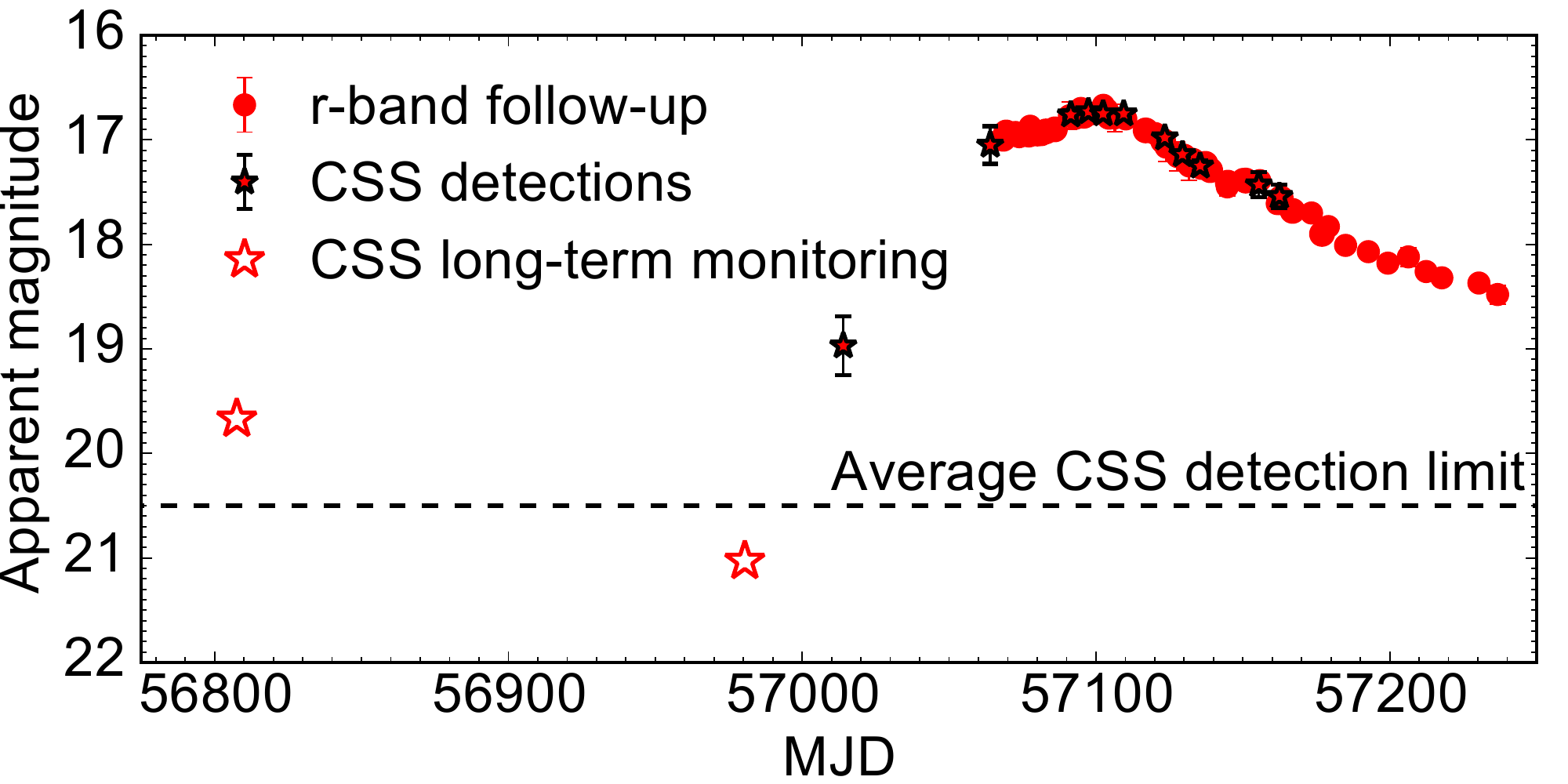}
\figcaption{Long-term CRTS/CSS \citep{dra2009} monitoring of the location of SN 2015bn, compared to our observed $r$-band photometry.\label{fig:css}}
\end{figure}

In Figure \ref{fig:css}, we show the historical magnitudes measured at the location of SN 2015bn by CRTS/CSS \citep{dra2009}. For clarity, we plot only the average detection limit, although there are many historic non-detections (the SN 2015bn host, at $R\sim22$\,mag, is well below the survey limit -- see section \ref{sec:host}). The CSS photometry shows two possible source detections (marked as unfilled stars on Figure \ref{fig:css}) prior to the unambiguous SN photometry (filled stars and circles). 
%The later point appears to be consistent with a detection of the SN itself, and could fit with a detection soon after explosion for both a smooth rise or a non-monotonic bump. Unfortunately, the image itself is not available for analysis, and the detection may not be real. As can be seen in Figure \ref{fig:pic}, there is a star with $r_{\rm SDSS}=21.53$\,mag and a probable galaxy with similar or slightly fainter brightness (though not listed in the SDSS photometric catalogue) within $\approx6$'' of the SN 2015bn host, as well as a galaxy with $r_{\rm SDSS}=20.83$\,mag located 12'' away (also visible in the CSS reference stack). It is possible that poor seeing conspired to blend some of these nearby sources together, giving the appearance of a single source close to the survey limit (A.~Drake, private communication). There is a similar detection at $R=20.83$\,mag on MJD 53005 (not shown on this scale). The other point, at MJD 56807 (2014 May 30) is well above the detection limit. However, other CSS images taken on the same night show no source brighter than $R=20.6$\,mag, suggesting that this detection is most likely not real. 
Without access to the CSS data, we cannot check whether these points are related to SN 2015bn. If real, this would be the first detection of historic variability at the location of a SLSN I (either a pre-explosion outburst or another SN in the same galaxy). We will not speculate further on the nature these early `detections', and exclude them from the rest of our analysis.

\subsection{Rest-frame light curves and colour evolution}\label{sec:rf}

To convert our photometry to absolute magnitudes in the rest-frame of the SN, we correct for distance modulus, extinction, and differences in rest-frame filter wavelengths (the $K$-correction)\footnote{We assume a flat $\Lambda$CDM Cosmology with $H_0=70\,{\rm km\,s}^{-1}\,{\rm Mpc}^{-1},\,\Omega_M=0.27,\,\Omega_\Lambda = 0.73$ throughout this work.}. We use the dust maps of \citet{schlaf2011} to correct for Milky Way extinction with $E(B-V)=0.022$, and assume negligible extinction in the SN host. This assumption is necessary, as the host emission lines are too weak to use reliably for an estimate of internal reddening (see section \ref{sec:host}), but also reasonable since we do not see any strong Na\,\textsc{i} D absorption from the host, which is generally thought to correlate strongly with dust extinction \citep{poz2012}. $K$-corrections were calculated in the optical and NIR by comparing synthetic photometry on our spectra (section \ref{sec:spec}) in the observer and rest frames. This was facilitated by the $K$-correction code \textsc{snake} \citep{ins2016}. At this redshift, the size of the correction was typically $\la0.2$\,mag. In the NUV, where we have no spectroscopic coverage, the correction was estimated as $K_{\rm UV}=-2.5\log_{10}(1+z)=-0.12$\,mag.

\begin{figure}
\centering
\includegraphics[width=8.5cm,angle=0]{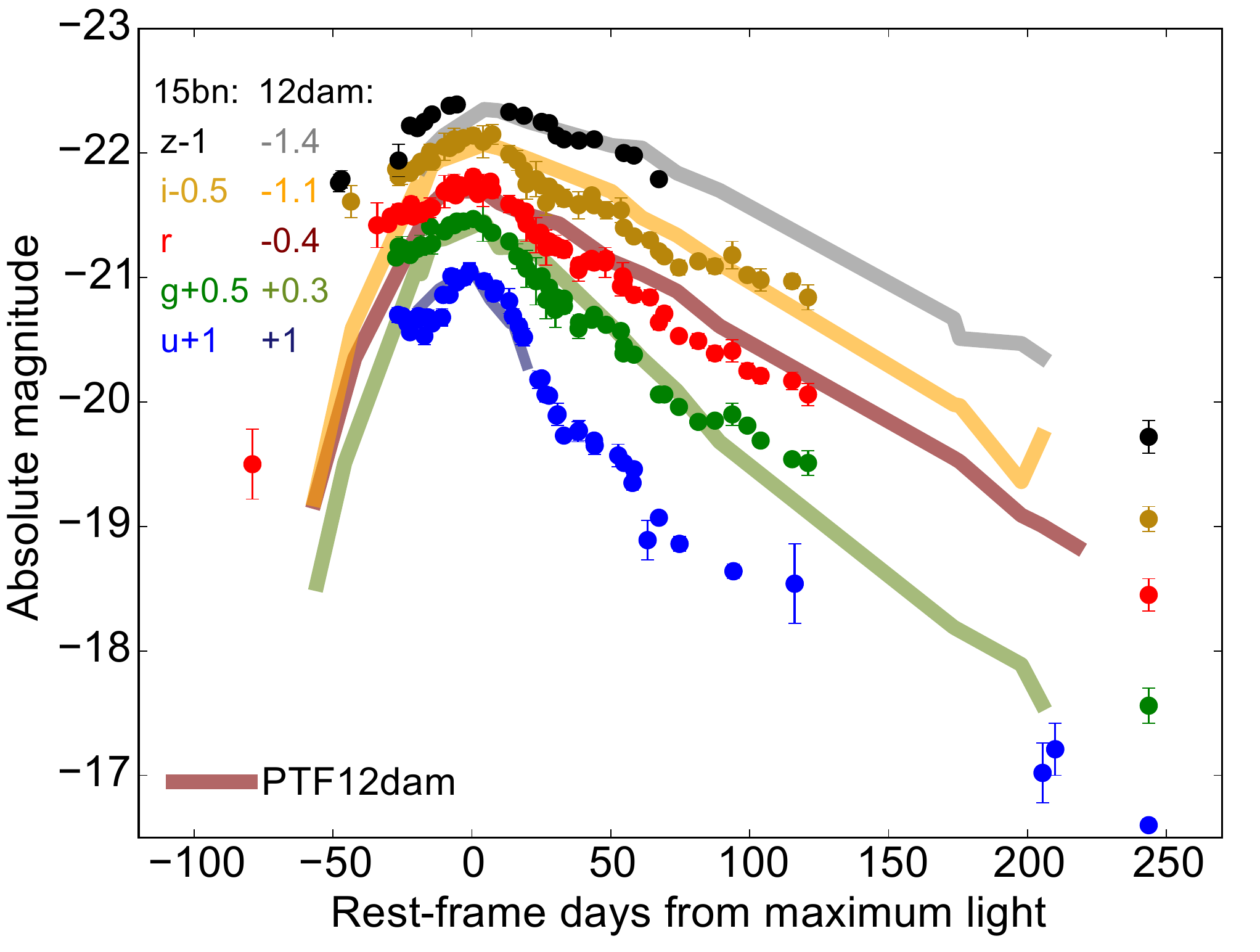}
\figcaption{Absolute rest-frame light curves in \textit{ugriz}, including extinction corrections based on the infrared dust maps of \citet{schlaf2011}, and $K$-corrections from our spectra. Also shown for comparison are the \textit{ugriz} light curves of PTF12dam. The two SLSNe show a similar decline rate beyond $\approx50$\,d, but SN 2015bn has a shallower rise, and an initially steeper decline around maximum.\label{fig:rf_lcs}}
\end{figure}

SN 2015bn has $M_g=-22.0\pm0.08$\,mag (AB) and $M_U=-23.07\pm0.09$\,mag (Vega) at maximum light (where the uncertainty includes a systematic error dominated by the $K$-correction as well as the statistical error), making it comparable to some of the most luminous SLSNe I \citep{vre2014,nic2015b}. However, this is about 1.5\,mag fainter than ASASSN-15lh, the brightest known SN \citep{dong2016}. The rest-frame optical light curves are shown in Figure \ref{fig:rf_lcs}. Overlaid for comparison is another well-observed, low-redshift SLSN I: PTF12dam \citep{nic2013}. The decline rate after $\approx50$\,d is similar between these two objects, but the light curves do have a number of differences. Most obvious is the slower rise exhibited by SN 2015bn, where the broad peak looks more symmetrical around maximum. SN 2015bn is also brighter than PTF12dam by $\approx0.5$\,mag in the red filters, but the peak luminosities match in $u$ band. Thus SN 2015bn is brighter but redder than PTF12dam. However, after $\sim100$\,d from peak, SN 2015bn is brighter in the $g$-band (PTF12dam was not observed in $u$-band at this epoch), demonstrating a very slow colour evolution.

\begin{figure}
\centering
\includegraphics[width=8.5cm,angle=0]{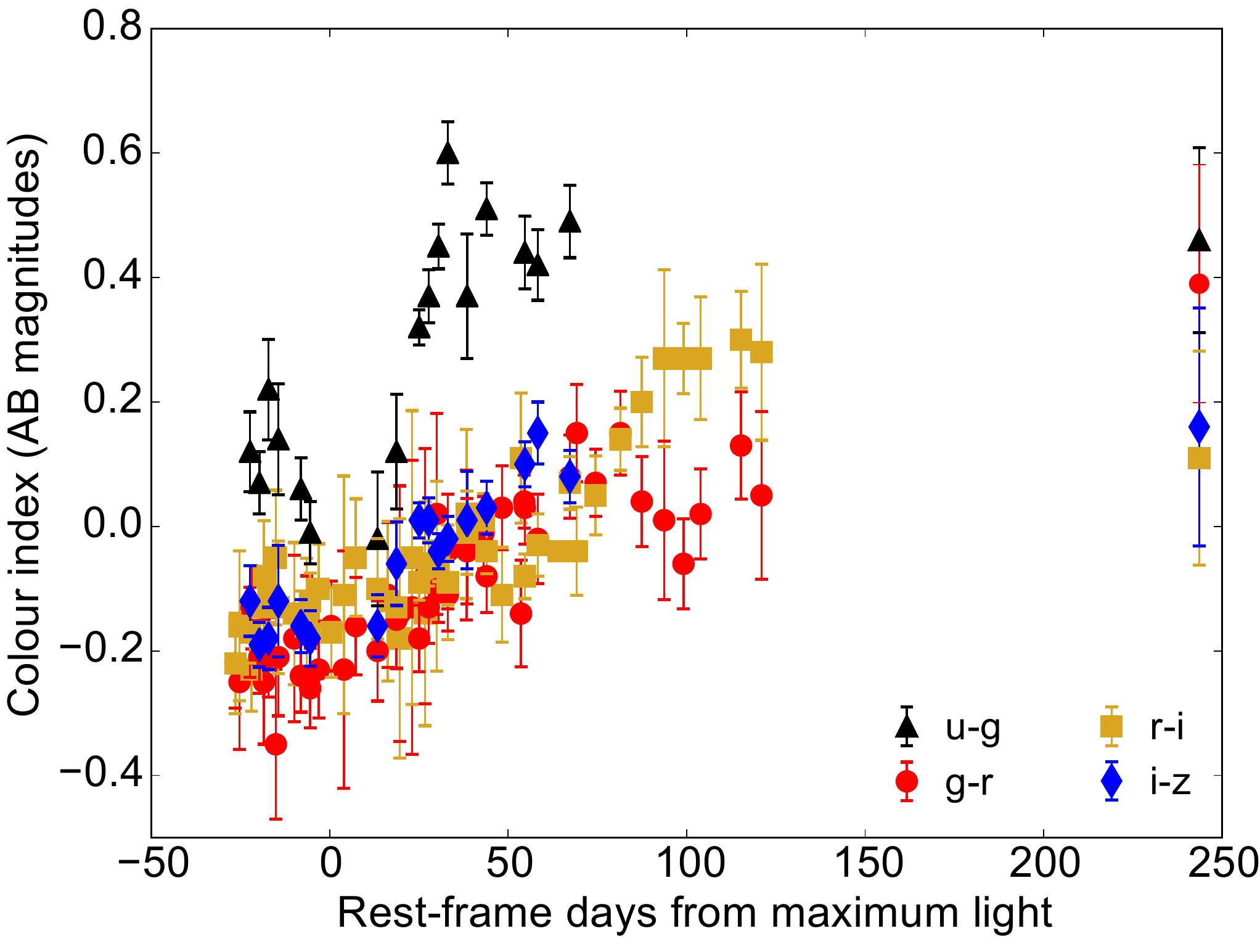}
\includegraphics[width=8.5cm,angle=0]{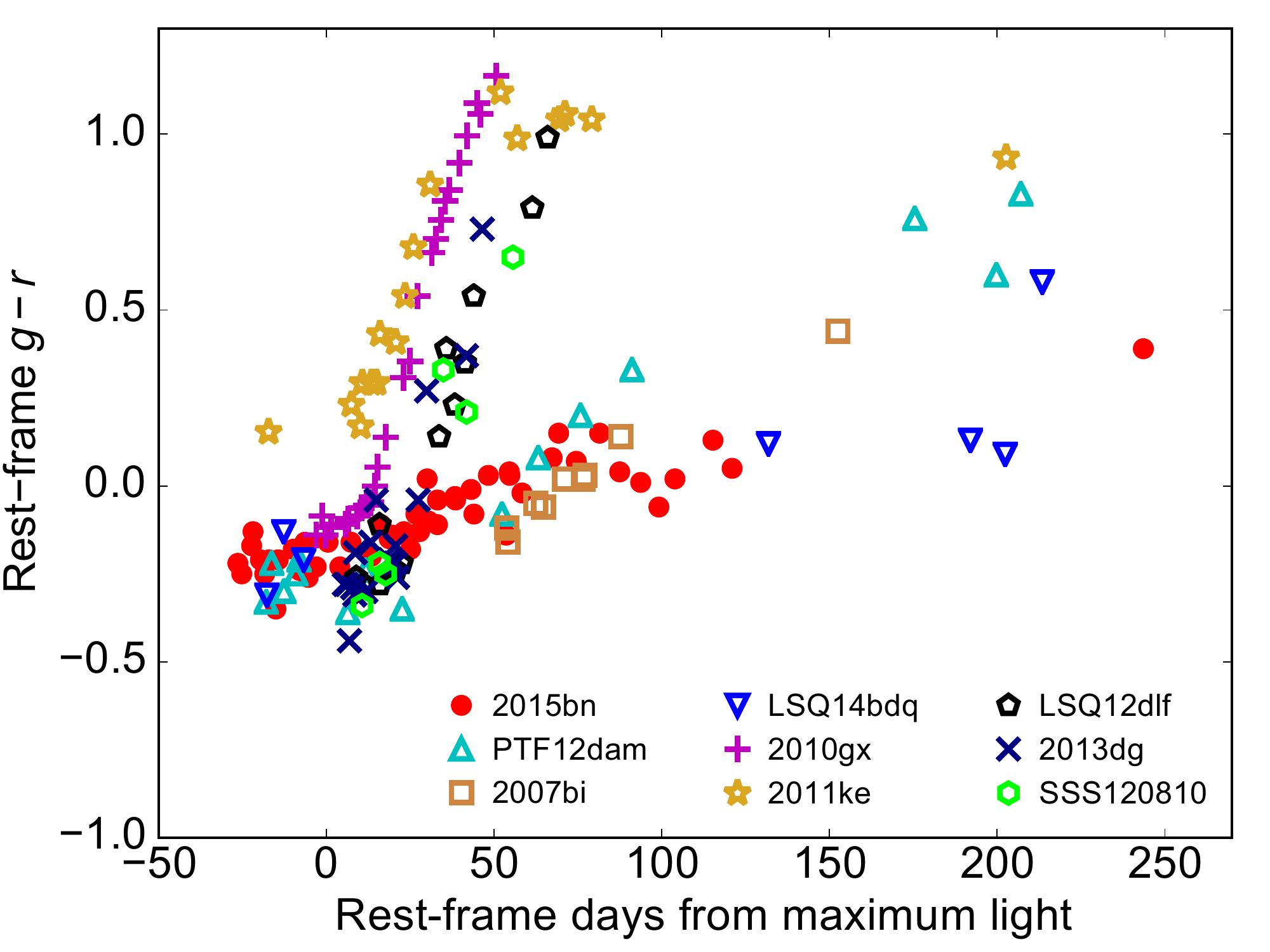}
\figcaption{Rest-frame colour evolution of SN 2015bn. Top: colour evolution in optical filters with respect to $g$-band. Bottom: Comparison of rest-frame ($K$-corrected) $g-r$ colour with other SLSNe. SN 2015bn shows a very shallow gradient. Data sources: \citet{you2010,pas2010,ins2013,nic2013,nic2014,nic2015a}.\label{fig:col}}
\end{figure}

The slow colour evolution is shown explicitly in Figure \ref{fig:col}, and compared to other SLSNe (all objects have been $K$-corrected to rest-frame using their observed spectra). The colour at peak is similar for most objects, with $-0.1\ga g-r \ga-0.4$\,mag (though SN 2011ke seems to have $g-r\ga0$\,mag). However all the objects with slowly declining light curves remain bluer for much longer than their fast-declining counterparts. SN 2011ke, which settled onto a slow light curve tail after around 50\,d from maximum light, has a comparable $g-r$ colour to the slowly declining objects at 200\,d. The $u-g$ colour of SN 2015bn initially evolves to the blue, reaching a minimum of $\sim-0.1$\,mag at maximum light, before declining relatively steeply compared to the optical-only colours. The faster decline in the UV is expected as the SN cools. Optical colours are roughly constant from our first observations at $-25$\,d until shortly after peak, with $g-r\approx r-i\approx i-z\approx-0.2$\,mag. The optical colours then slowly evolve to the red. SN 2015bn reaches $g-r\approx0.4$\,mag by +250\,d. This evolution is slower and bluer than any literature SLSN with such late observations, as is seen in the bottom panel of Figure \ref{fig:col}, and from the $g$-band excess compared to PTF12dam in Figure \ref{fig:rf_lcs}.

\begin{figure}
\centering
\includegraphics[width=8.5cm,angle=0]{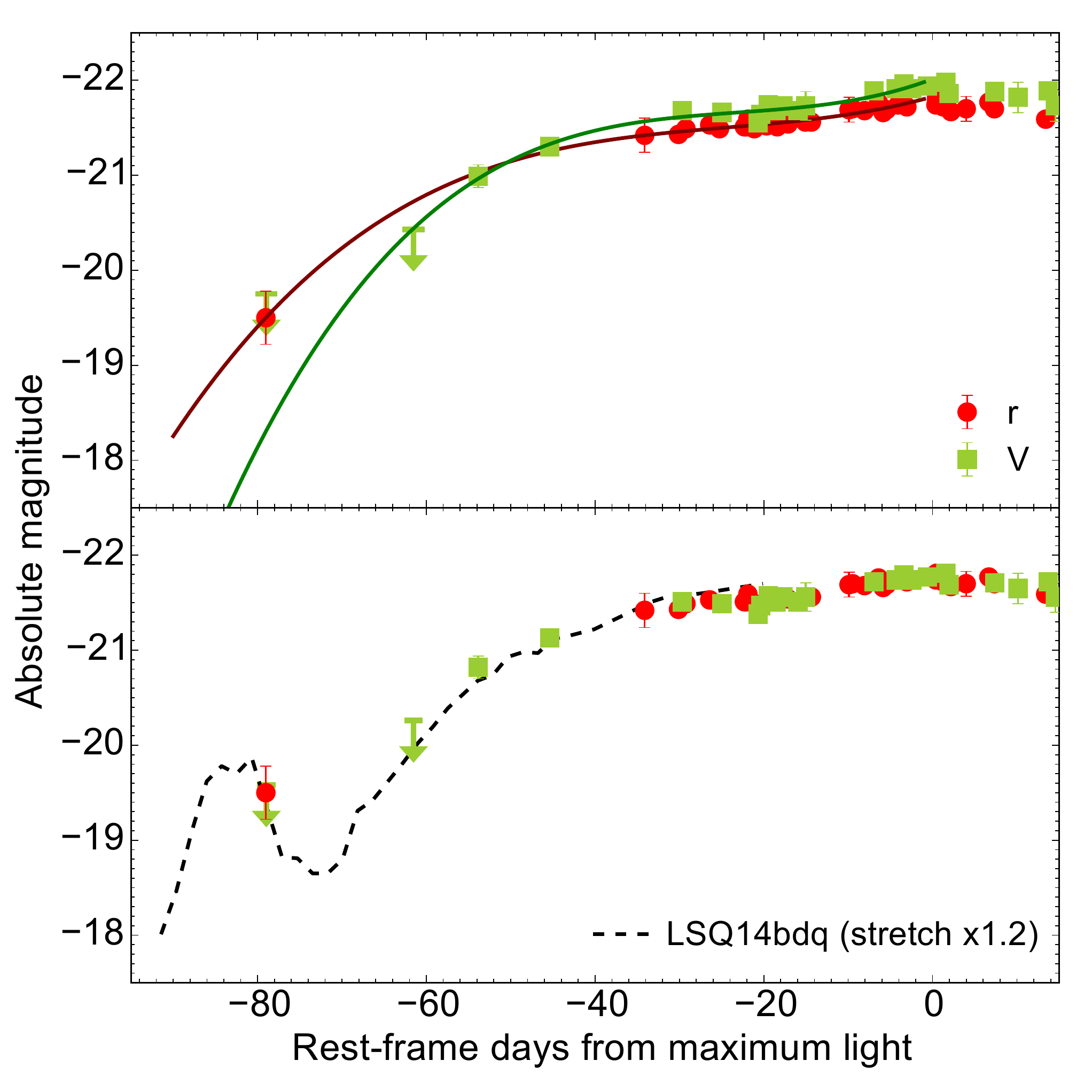}
\figcaption{The early absolute $V$- and $r$-band light curves of SN 2015bn. Top: Fits to the rising light curves with third order polynomials. The $V$-band points from ASAS-SN suggest a steeper rise than the early $R$-band detection by CSS. The implied colour at the earliest epoch -- $V-r\approx1$\,mag -- would be much redder, and the evolution much steeper, than other SLSNe at early times. Bottom: assuming a constant colour in $V-r$, the pre-maximum light curve is well matched by LSQ14bdq, with a stretch in time by a factor of 1.2. The early point therefore seems to be more consistent with an initial bump.\label{fig:rise}}
\end{figure}

Many SLSNe I exhibit a non-monotonic rise to maximum brightness, with a fainter initial peak, or `bump'. These bumps have been clearly detected for a number of individual objects \citep{lel2012,nic2015a,smi2016}. Recently, \citet{nic2016a} have shown that a number of literature objects are also consistent with an undersampled bump, and the relative faintness of this first peak makes it difficult to exclude in most cases. In Figure \ref{fig:rise}, we plot the rising part of the light curve in the two filters with the best early sampling: $r$- and $V$-bands. We fitted a third-order polynomial (top panel) to each light curve. While either band alone would be insufficient to test for the presence of a bump, the two bands together show that if we were to assume a smooth rise, we would derive a very surprising colour evolution. The $V$-band data clearly show a steeper decline than that implied by a smooth fit to $r$-band. The $V-r$ colour between $-30$\,d and peak is constant at $-0.15$\,mag, but the fits suggest the colour at early times would be more like $V-r \approx 1$\,mag. Such a dramatic change in colour over the rising phase (and such a red colour at early times) is highly inconsistent with other SLSNe I that have multicolour data available at this phase \citep{lel2012,how2013,ins2013,smi2016}. Such objects generally show blue colours, similar to what we see around the light curve peak here, and a flat colour evolution before maximum. By contrast, if we assume that the colour evolution is relatively flat, as in other SLSNe I, and combine our $V$- and $r$-band data into a single light curve using the observed colour at peak, we find that the rising light curve of LSQ14bdq \citep{nic2015a} gives an excellent match to the data, after applying a temporal stretch factor of 1.2 (bottom panel). Thus the multicolour data seem to be more consistent with a bump similar to LSQ14bdq and other SLSNe I observed at very early times. However, we note that this is a larger stretch than was required for any of the objects in \citet{nic2016a}.

\begin{figure}
\centering
\includegraphics[width=8.5cm,angle=0]{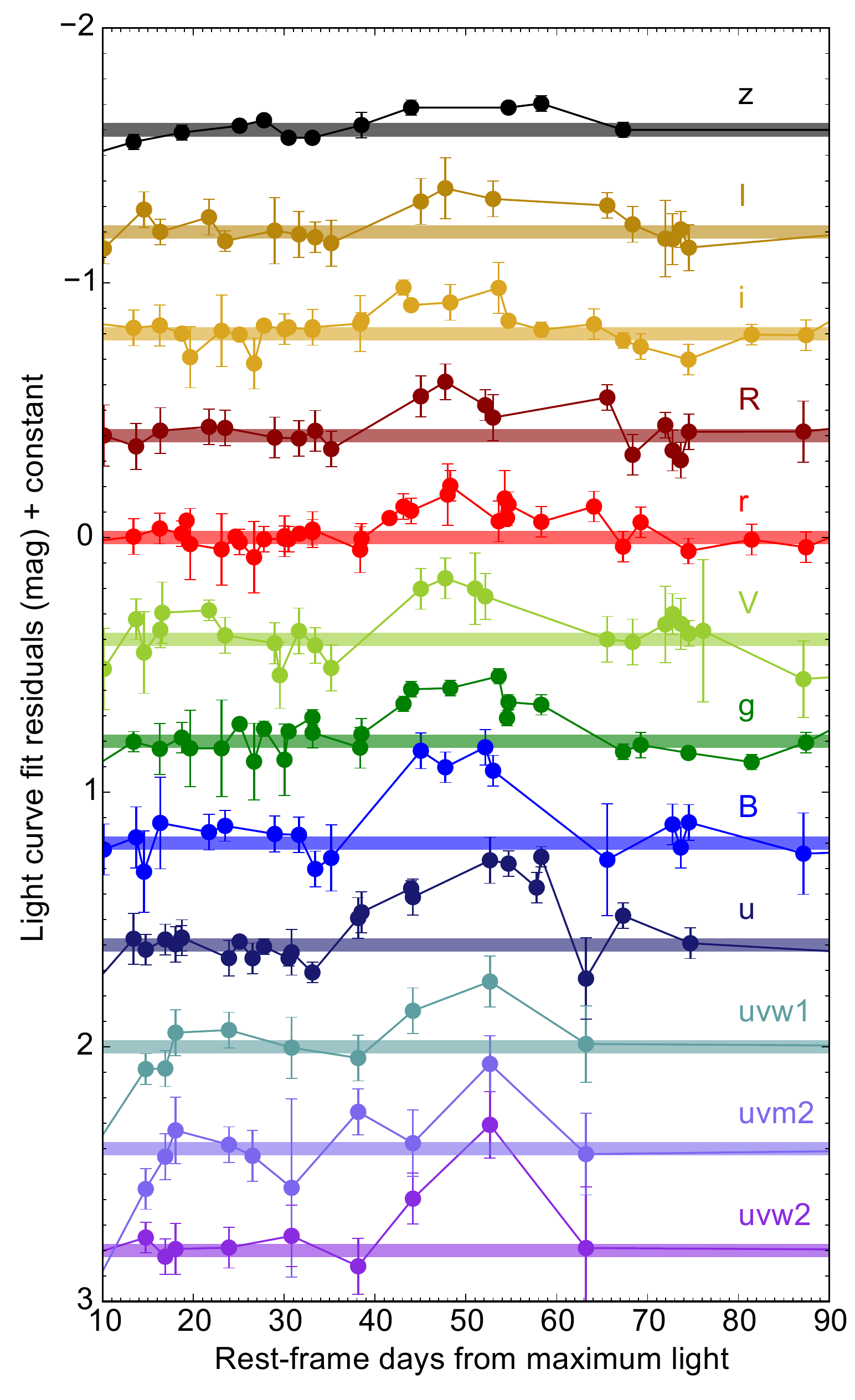}
\figcaption{Residuals after subtracting fits to the declining rest-frame UV and optical light curves, showing the `knee' at +50\,d. The feature is more pronounced in the bluer and UV bands.\label{fig:bump}}
\end{figure}

Our high-cadence photometry reveals the presence of several more distinctive bumps/plateaus in the light curve, most prominent in the UV bands. First the $u$-band shows a plateau lasting 14\,d in the rest-frame, until 10\,d before maximum light (Figures \ref{fig:obs_lcs} and \ref{fig:rf_lcs}). This shows up clearly as a dip in the UVOT \textit{uvm2} light curve, which has an effective wavelength $\sim2230$\,\AA~in the observer frame, or $\sim2000$\,\AA~at the redshift of SN 2015bn. We designate this feature `the shoulder' for all subsequent discussion. A second undulation is present in all optical and UV filters at 50\,d after maximum light, or MJD 57158 -- we call this `the knee'. 
It is known that UVOT observations can occasionally be affected by artificial drops in flux \citep{ede2015}. However the variability exhibited by SN 2015bn is reliable for a number of reasons. These flux `drop-outs' are not expected to show correlated deviations spanning numerous successive epochs. Furthermore, the biggest deviation is seen in \textit{uvm2}, which was found by \citet{ede2015} to have a lower drop-out rate than \textit{uvw2} or \textit{uvw1}. Most importantly, these light curve fluctuations are seen at the same phases not only across all of the UVOT filters, but also match perfectly (in time and brightness) with the fluctuations in our ground-based data.

To see the undulations more clearly, we fitted third-order polynomials to the rest-frame light curve decline in each UV and optical filter (after masking epochs between +40 and +60\,d), and subtracted the fit from the full declining light curve. Unfortunately, the NIR cadence is too sparse to show if the knee exists there too. The residuals are shown in Figure \ref{fig:bump}. The excess is more pronounced in the blue and UV bands. This temporary evolution to the blue can also be seen in Figure \ref{fig:col}, where the $u-g$ colour evolves from $\sim 0.6$\,mag at +30\,d, to $\sim 0.4$\,mag at +50\,d. A third undulation may occur at +100\,d. This is quite prominent in $g$-band, but the UV light curves are not well sampled at this epoch. This type of behaviour -- a temporary late-time re-brightening, more pronounced in the blue -- has been seen before in one other SLSN, SSS120810 \citep{nic2014}. SN 2007bi also seemed to exhibit a bump at $\approx100$\,d after maximum, but no colour information is available at this phase \citep{gal2009}. The implications of these fluctuations will be discussed in sections \ref{sec:bol} and \ref{sec:mod}.

\begin{figure*}
\centering
\includegraphics[width=12.5cm,angle=0]{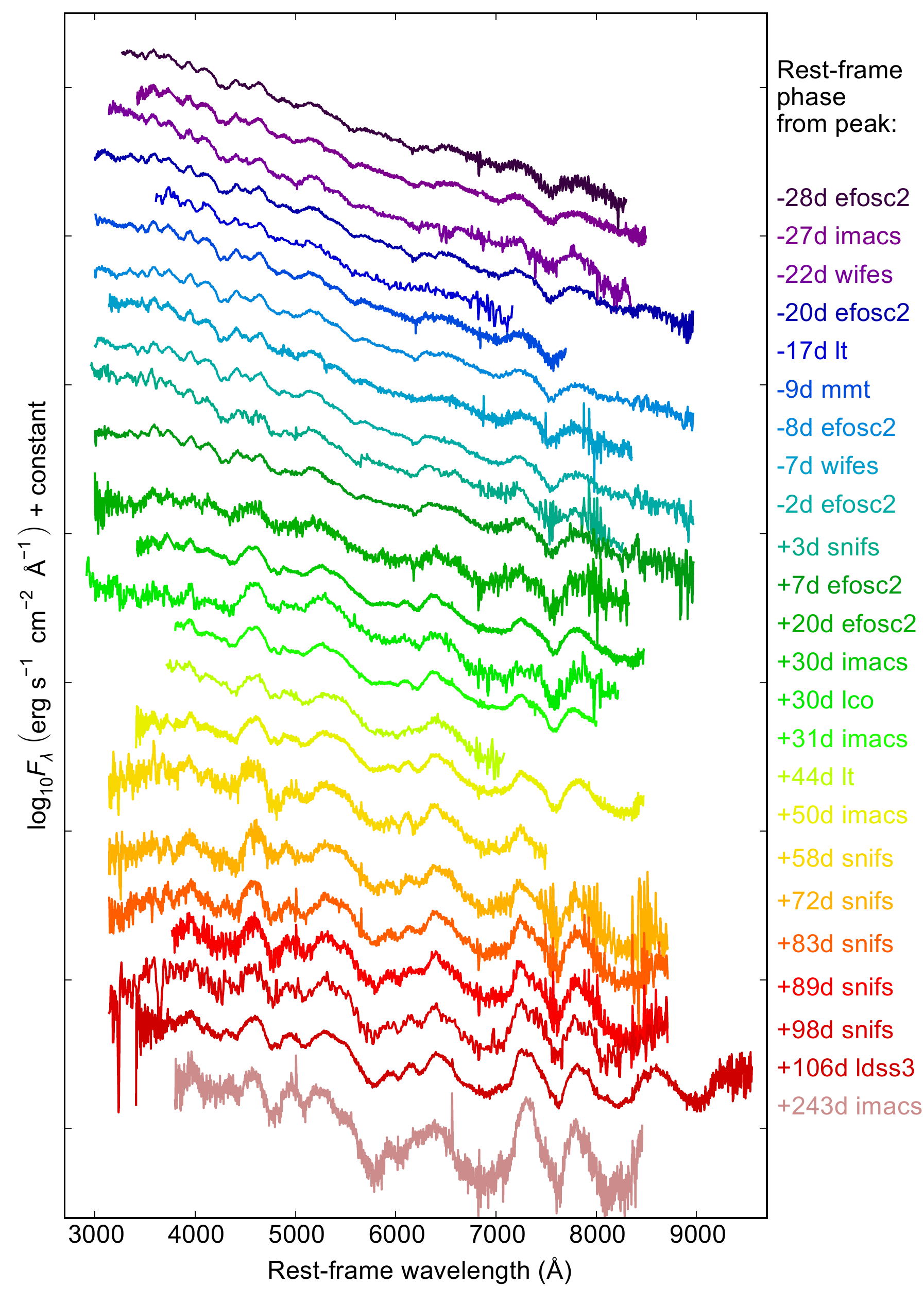}
\figcaption{Complete rest-frame spectroscopic evolution of SN 2015bn. Data have been corrected for extinction using $E(B-V)=0.022$ \citep{schlaf2011}. The labels on the right give the phase with respect to optical maximum, as well as the instrument used in each observation.\label{fig:sp_ev}}
\end{figure*}

\begin{figure*}
\centering
\includegraphics[width=18cm,angle=0]{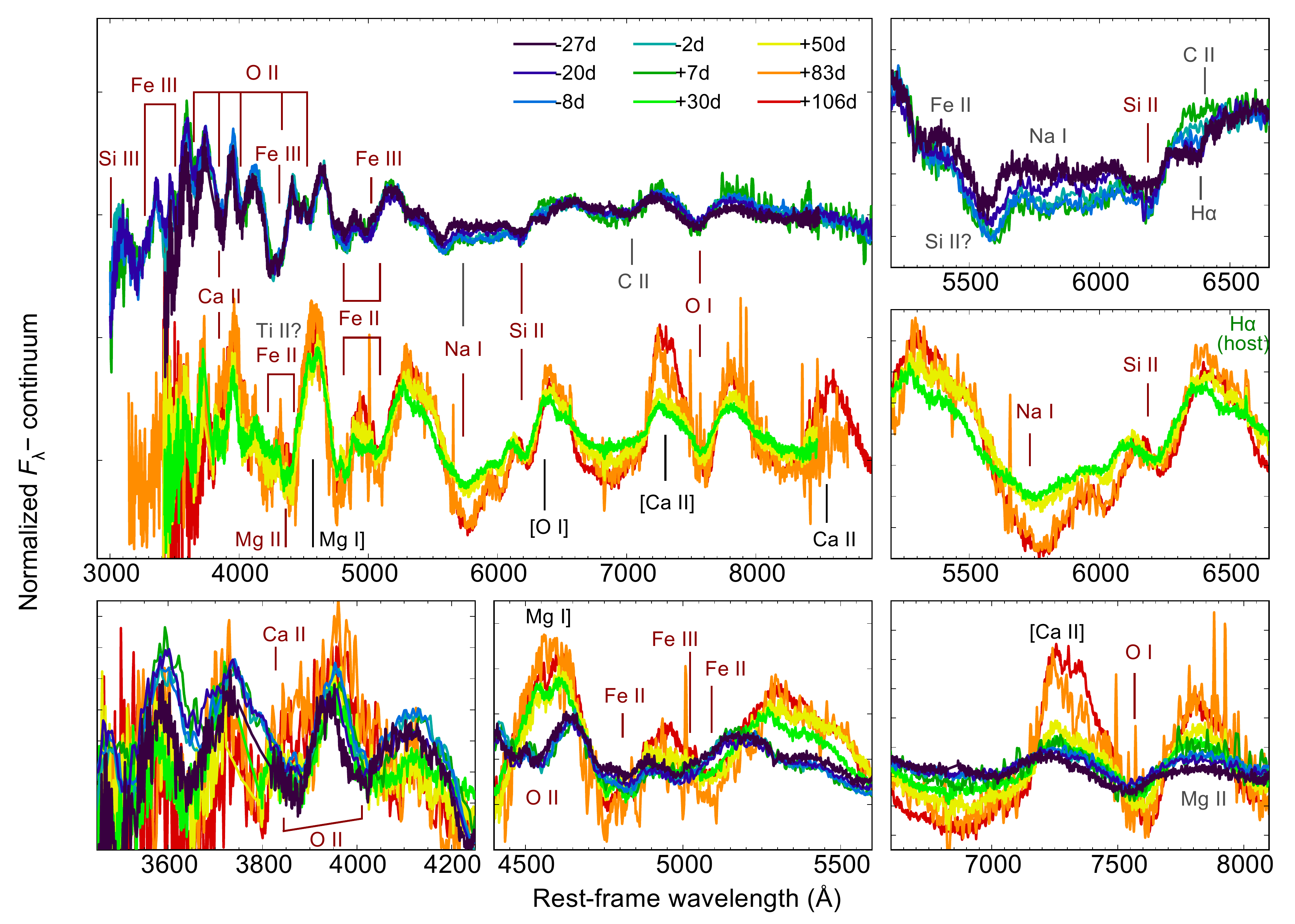}
\figcaption{Normalized and continuum-subtracted spectra of SN 2015bn. Line identifications are labeled using the following convention: absorption components are marked in dark red assuming a velocity of 8000\,\kms; emission lines are marked in black assuming they are centred on zero velocity; host emission is marked in green; uncertain identifications are marked in grey. Subplots contain zooms to various parts of the spectrum to show line evolution. Little spectroscopic evolution occurs from $-$27\,d to +7\,d; the spectrum then undergoes something of a transformation at some time between +7\,d and +20\,d; the spectrum beyond 20-30\,d evolves very slowly once again.\label{fig:lines}}
\end{figure*}

\section{Spectroscopy}\label{sec:spec}

\subsection{Spectroscopic reductions}\label{sec:spec_re}

Optical spectra of SN 2015bn were acquired using the NTT and EFOSC2, the 6.5-m Magellan telescopes with the IMACS and LDSS3 spectrographs, WiFeS on the Australian National University 2.3-m telescope \citep{dop2007}, LT with SPRAT, the 2.0-m LCOGT Faulkes Telescopes and FLOYDS, and the University of Hawaii 2.2-m telescope with SNIFS. All spectra were reduced, including de-biasing, flat-fielding, object extraction, wavelength and flux calibration, using instrument-specific pipelines or standard routines in \textsc{iraf}. The relative flux calibrations were achieved using standard star observations taken on the same nights as the SN spectra. Details of the complete list of spectroscopic observations are given in the Appendix. NIR spectra were obtained using SOFI on the NTT, and reduced using the PESSTO pipeline to flat-field, sky-subtract and co-add the frames, apply wavelength calibration to the 2D frames, as well as a flux calibration and telluric correction from standard star observations, and finally to extract the 1D spectrum. For every spectrum, the absolute flux calibration was checked against contemporaneous multicolour photometry; spectra were then scaled to match the photometry if necessary. PESSTO raw data are immediately available in the ESO archive\footnote{http://archive.eso.org/eso/eso\_archive\_main.html}, and the reduced data for the season in which SN 2015bn was observed are due for bulk public release in late 2016\footnote{See www.pessto.org}. In the meantime, all spectra are available on WISeREP.

\begin{figure}
\centering
\includegraphics[width=8.7cm,angle=0]{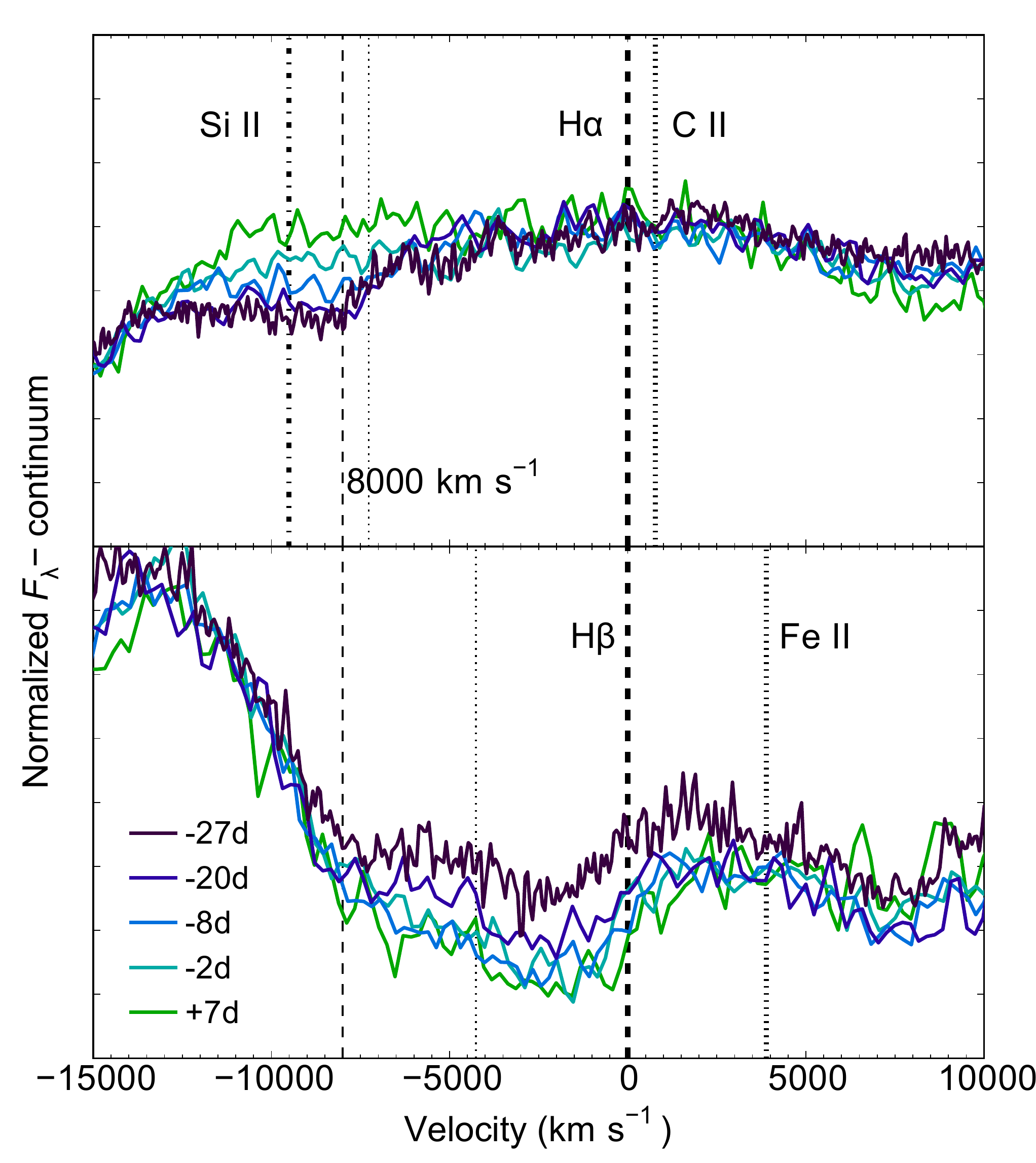}
\figcaption{Close-ups of the early spectra of SN 2015bn around the wavelengths of \Ha~and \Hb. Thick lines show the rest-frame locations of lines in velocity space, while thin lines show the same lines redshifted by 8000\,\kms~from their rest wavelengths. The earliest spectra show possible absorptions due to hydrogen, which disappear before maximum light. However, the origin of these lines is not totally certain, due to blending with blueshifted C\,\textsc{ii} and Fe\,\textsc{ii}.\label{fig:H}}
\end{figure}

\subsection{Spectroscopic evolution}

The complete set of optical spectra are shown chronologically in Figure \ref{fig:sp_ev}. SN 2015bn retains a blue continuum throughout the observing season, and this is still present as late as 240\,d after maximum light, i.e.~the ejecta are not yet fully nebular. The lines in the spectrum also evolve very slowly. To show more clearly how the individual lines evolve, and to identify the ions responsible, we normalized each spectrum by dividing by its mean flux, and fitted the continuum with a third-order polynomial, which was then subtracted off. Plotting the normalized, continuum-subtracted spectra in Figure \ref{fig:lines}, we can see directly how the line centres, velocities and strengths evolve with time.

The line identifications in this section were made through detailed comparison with identifications and SYNOW fits \citep{tho2011} given for other SLSNe I by \citet{gal2009,you2010,pas2010,qui2011,chom2011,lel2012,ins2013,lun2013,nic2013,nic2014}. Furthermore, we examined similar modelling of normal Type Ic SN spectra by \citet{mil1999,sau2006,val2008,hun2009}. Line wavelengths in the rest frame were determined by consulting the NIST Atomic Spectra Database \citep{nist}, and compared to absorption lines assuming a photospheric velocity of 8000\,\kms~(see section \ref{sec:vel}). For the most common ions, multiple lines of the same species were checked for consistency using the interactive plotting functionality of WISeREP \citep{yar2012}. A full set of self-consistent models to reproduce the spectral evolution of SN 2015bn will be the subject of future work, but is beyond the scope of this paper. In the discussion below, we use the symbol $\lambda$ when referring to lines by their rest-frame wavelength, and \AA~when referring to redshifted absorption.

It is immediately clear from Figures \ref{fig:sp_ev} and \ref{fig:lines} that the spectra divide fairly cleanly into two phases. Very little evolution is seen from our first spectrum at $-$27\,d through to +7\,d. Our highest signal-to-noise spectra taken over this period are overlaid in Figure \ref{fig:lines} (purple to dark green lines). Between +7 and +20\,d, the spectrum undergoes something of a transformation, after which it evolves very slowly for the next 200\,d. This result is independent of any model assumptions or line identifications. Sample spectra from this second phase are also overlaid in the figure (light green to red lines).

The early ($t\le7$\,d) spectra are dominated by blue continuum emission, superimposed with fairly weak, broad absorption features and P Cygni lines. In the blue, we clearly see the O\,\textsc{ii} absorption lines that have come to characterise the SLSN I class; however, in this case the absorption at 4300\,\AA~is stronger and broader than the other lines. This could indicate a contribution from Fe\,\textsc{iii}\,$\lambda$4430. Between 3000-3600\,\AA, we detect absorptions that have been attributed to Fe\,\textsc{iii} and Si\,\textsc{iii} in other SLSNe I \citep[e.g.][]{lel2012,lun2013}. Moving to redder wavelengths, we see signatures of Fe\,\textsc{ii} at $\sim5000$-6000\,\AA, along with Na\,\textsc{i}\,D and Si\,\textsc{ii}\,$\lambda$6355. We tentatively associate the feature at $\sim7000$\,\AA~with C\,\textsc{ii}\,$\lambda$7234. To the red of this we see a fairly strong O\,\textsc{i}\,$\lambda$7774, which may include some contribution from a Mg\,\textsc{ii} line of similar wavelength.

We may also see some spectroscopic indication of hydrogen at early times. \citet{yan2015} presented an \Ha~emission line in the nebular spectrum of iPTF13ehe, caused by late-time ($\ga200$\,d) interaction with hydrogen-rich material at a large radius, but no hydrogen was seen in the photospheric phase. The upper right panel of Figure \ref{fig:lines} includes the region around \Ha~(rest-wavelength 6563\,\AA). There is an absorption in the earliest spectrum at 6300\,\AA, which is consistent with \Ha~at a velocity $\approx 8000$\,\kms. However, an alternative identification for this line is C\,\textsc{ii}\,$\lambda$6580, particularly if the line at 7000\,\AA~during this phase is also carbon. On the other hand, the absorption at 6300\,\AA~vanishes by the time of maximum, light, whereas the 7000\,\AA~line persists.

In a further attempt to break the degeneracy, we look more closely at the region around \Ha~and \Hb~(in velocity space) in Figure \ref{fig:H}. The strongest line in the vicinity of \Hb~is Fe\,\textsc{ii}\,$\lambda$4924, which was identified for SN 2007bi by \citet{gal2009,you2010}, and is consistent with a SYNOW fit to LSQ12dlf by \citet{nic2014}. The initial blueshift of the P Cygni peak may indicate that the line is at first blended with \Hb; this blueshift quickly disappears along with \Ha. There is also an absorption in the earliest spectrum located at $\approx8000$\,\kms~bluewards of \Hb. However, the effects of line blending make a firm association with \Hb~difficult. Overall, we consider it plausible, though by no means certain, that these lines are from unburned hydrogen. \citet{par2015} showed that the line profiles around 6200-6500\,\AA~in superluminous and normal Type Ic SNe may be better reproduced by including hydrogen as well as Si\,\textsc{ii}. Although their analysis was based on a SLSN spectrum well after maximum light, the resolved absorption features in the early spectrum of SN 2015bn may provide independent evidence of this claim.

\begin{figure}
\centering
\includegraphics[width=8.7cm,angle=0]{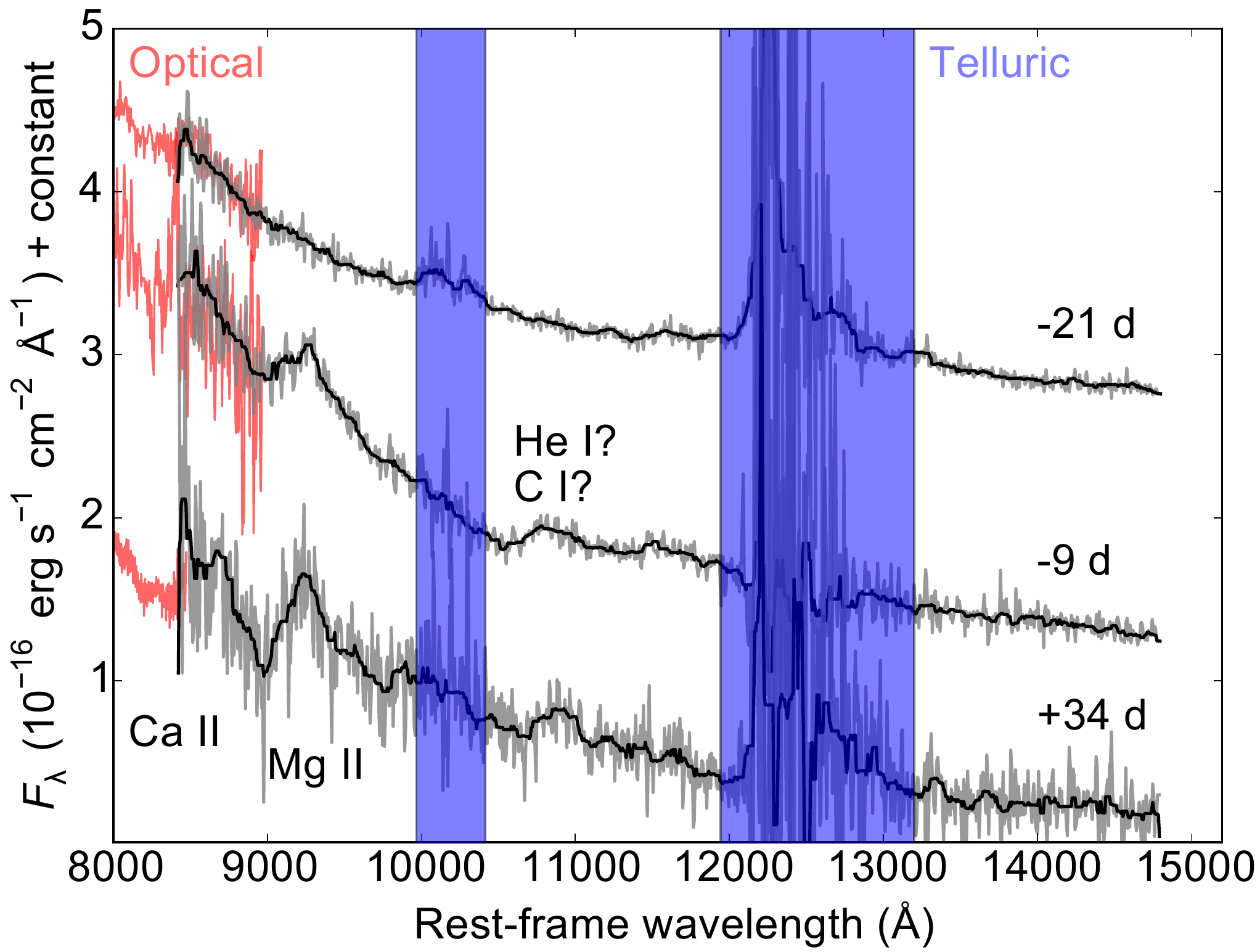}
\figcaption{NIR spectra of SN 2015bn from SOFI. Regions with strong telluric lines have been masked. The black curves are smoothed using a 15-pixel median filter for clarity. The main lines in the NIR are the Ca\,\textsc{ii} NIR triplet, Mg\,\textsc{ii}, and an unidentified line that could be either He\,\textsc{i} or C\,\textsc{i}.\label{fig:nir}}
\end{figure}

If hydrogen is present in the spectrum of SN 2015bn, indicating that some small amount of the progenitor's envelope managed to remain bound before explosion, we might also expect to see some helium (though this is much more difficult to excite). While the strongest optical He\,\textsc{i} line is generally hard to distinguish from the Na\,\textsc{i}\,D lines, our NIR spectra (Figure \ref{fig:nir}) show a clear line at $\approx10800$\,\AA~that could be He\,\textsc{i}\,$\lambda$10830. This line is expected to be 2-10 times stronger than the strongest optical line of He\,\textsc{i} \citep{ins2013}. Few NIR spectra of SLSNe I exist, but \citet{ins2013} saw a similar feature in SN 2012il. Normal SNe Ic generally exhibit P Cygni lines with a deep absorption component at this wavelength. However, the identification of this line has been hotly debated in the literature. \citep{pat2001} claimed a detection of He in the NIR spectrum of the LGRB-associated Type Ic SN 1998bw, but this has been challenged by \citet{tau2006,hac2011}. It has been argued that this line may contain a large contribution from C\,\textsc{i}, since the other strong helium line in the NIR ($\sim2.1\,\mu$m) tends to be weak or absent in SNe Ic, unlike in He-rich SNe Ib \citep{hun2009}. Unfortunately, our wavelength coverage does not extend to 2\,$\mu$m for SN 2015bn. While the line could plausibly be He\,\textsc{i}, a firm identification is not possible. The presence of carbon in the optical spectrum may suggest that C\,\textsc{i} is a more likely explanation for this feature. Other species that could contribute at this wavelength are Mg\,\textsc{ii} and Si\,\textsc{ii}. Future spectral modelling will be required to determine the nature of the 10800\,\AA~line.

Returning from our digression to the NIR, we now examine the optical evolution after maximum light. Looking again at Figure \ref{fig:lines}, the most obvious changes occurring between 7 and 20\,d after maximum are the emergence of Mg\,\textsc{i}]\,$\lambda$4571 emission, strong Fe\,\textsc{ii} lines between 4000-5500\,\AA, and Na\,\textsc{i}\,D absorption. There is also an unidentified line at around 6100\,\AA~-- this could perhaps be a detached, high-velocity component of Si\,\textsc{ii} with a velocity of $\sim16000$\,\kms, but it would be very surprising for a high-velocity line to appear only at late times. At the same time, the O\,\textsc{ii} and Fe\,\textsc{iii} lines that previously dominated the blue part of the spectrum weaken and disappear as the ejecta cool and these species recombine. The P Cygni line at 7000\,\AA~is replaced by an asymmetrical emission line due to [Ca\,\textsc{ii}]\,$\lambda\lambda$7291,7323. The [Ca\,\textsc{ii}] shows a steep blue side but a broad red wing. One explanation for this asymmetry could be blending with C\,\textsc{ii} or Fe\,\textsc{ii}. At the same time, we start to see likely emission from [O\,\textsc{i}]\,$\lambda\lambda$6300,6364 and the Ca\,\textsc{ii} NIR triplet at 8500\,\AA. The strengths of the calcium lines, as well as O\,\textsc{i}\,$\lambda$7774 and the Na\,\textsc{i}\,D absorption, increase substantially after +50\,d. The final spectrum obtained at +243\,d (and after SN 2015bn had returned from behind the sun) remains largely similar, with significant continuum persisting in the blue. The only noticeable change over this period is that some emission lines (especially calcium) increase in luminosity, and absorption lines (especially sodium) increase in depth. The spectra at $>100$-200\,d, with necessarily longer exposure times, finally reveal the presence of faint nebular emission lines originating in the host galaxy, from which we securely derive the redshift of $z=0.1136$.

\begin{figure}
\centering
\includegraphics[width=8.7cm,angle=0]{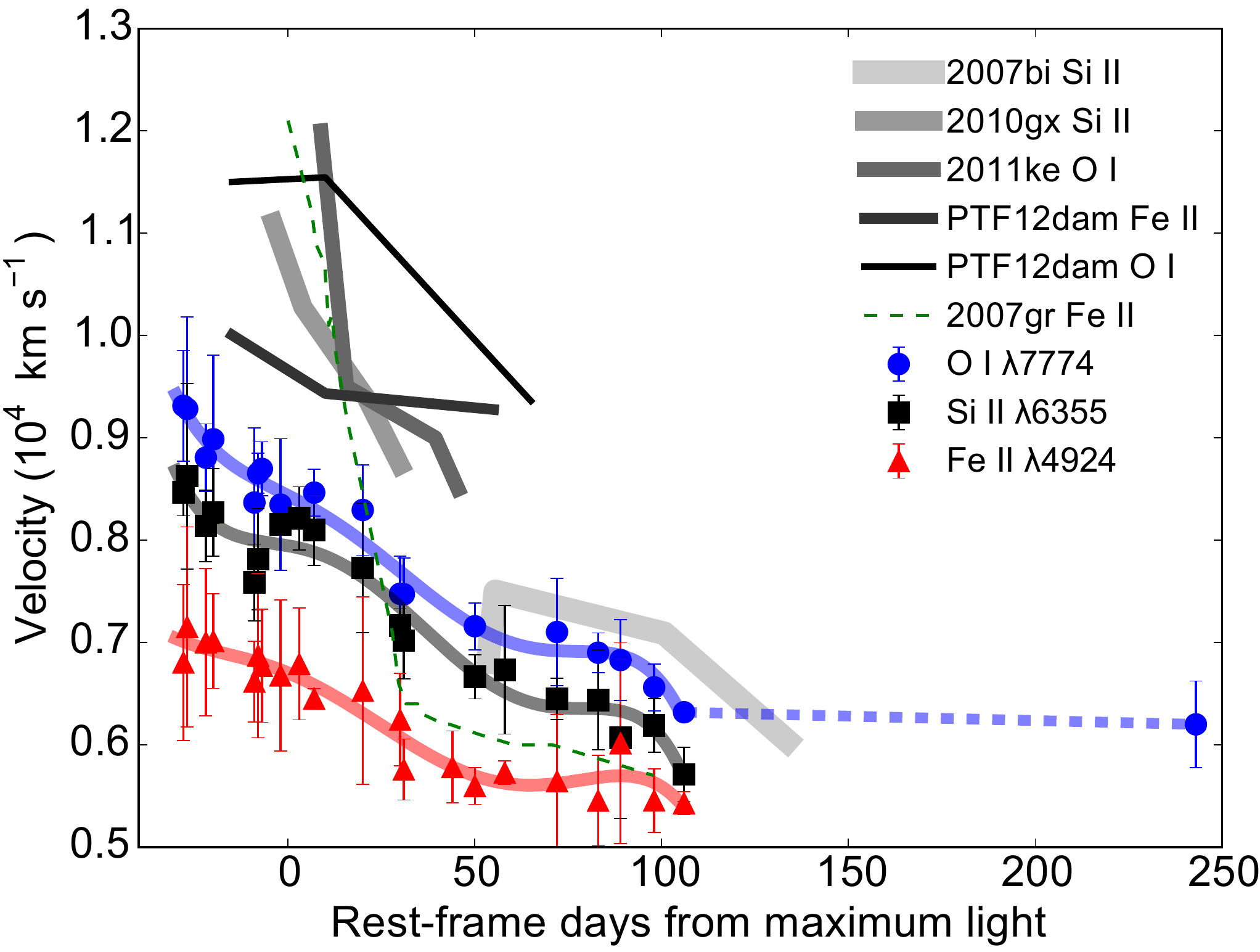}
\figcaption{Velocity evolution compared to other SLSNe. Measurements were made by fitting Gaussian curves to absorption minima, using lines that appeared to be largely unblended throughout the photospheric evolution. Comparisons with other SLSNe indicate relatively low velocities. The flat velocity curve is markedly different from the normal Type Ic SN 2007gr. Data sources: \citet{you2010,pas2010,ins2013,nic2013}.\label{fig:v}}
\end{figure}

\subsection{Line velocities}\label{sec:vel}

The various subplots in Figure \ref{fig:lines} show close-ups of different spectral lines. We can see that the minima of all absorption lines, as well as the peaks of some P Cygni profiles, move gradually to lower velocities (i.e.~redder wavelength) as the spectrum evolves. We measured the photospheric velocity as a function of time using a number of ions by fitting Gaussian profiles to each line and determining the blueshift of the absorption minima. It is common to use the Fe\,\textsc{ii}\,$\lambda$5169 line to measure photospheric velocities \citep[e.g.][]{nic2015b}. However, in SN 2015bn this line seems to be blended with Fe\,\textsc{iii} at early times. Moreover, our densely sampled spectra, spanning $\sim200$\,d in the rest-frame, shows that the peak of the P Cygni profile moves significantly to the red as the spectrum evolves, such that we cannot be confident that the same line dominates the absorption at all times. To get the most consistent velocity measurements, we choose lines that appear to be unblended for the majority of spectra. The smooth evolution towards lower velocity is particularly clear for the Si\,\textsc{ii}\,$\lambda$6355 and O\,\textsc{i}\,$\lambda$7774 lines, so we used these two lines, along with Fe\,\textsc{ii}\,$\lambda$4924, to measure the photospheric velocity of SN 2015bn.

The velocity measurements are shown in Figure \ref{fig:v}, and compared with other SLSNe (using the same lines). We performed the measurements ourselves for PTF12dam and SN 2010gx, whereas the velocities for SNe 2007bi and 2011ke were taken from \citet{you2010} and \citet{ins2013} respectively. All of the ions indicate relatively lower velocities in SN 2015bn: $\approx 8000$\,\kms~at maximum light compared to 9000-12000\,\kms~in the other objects. \citet{nic2015b} estimated a median maximum-light velocity of $10500\pm3000$\,\kms~for SLSNe I, so SN 2015bn falls at the low-end of the expected range. The O\,\textsc{i} line shows the highest velocity, and Fe\,\textsc{ii} the lowest, as we would expect for a sensible stratification of the ejecta, with heavier nuclei towards the centre. The effect is compounded as the O\,\textsc{i} line is relatively strong, such that absorption in the outermost, rarified material (which has the highest velocity) is still significant.

The temporal evolution in the measured velocities is very slow, decreasing by only $\sim3000$\,\kms~during a period of nearly 150\,d, and seemingly staying constant thereafter. For comparison, the Fe\,\textsc{ii} velocity of the normal Type Ic SN 2007gr is also shown. Although these SNe show similar post-maximum velocity, SN 2007gr underwent a steep deceleration around maximum light. The relatively flat velocity curves of SLSNe I have been interpreted as one possible signature of a central engine accelerating the inner ejecta \citep{kas2010,chom2011,nic2015b}. The unprecedented spectroscopic coverage of SN 2015bn -- together with low velocities that reduce some of the effects of line blending -- reveals the clearest indication yet that at least some SLSNe have very shallow velocity gradients in the ejecta.

\subsection{Comparisons}

\begin{figure}
\centering
\includegraphics[width=8.7cm,angle=0]{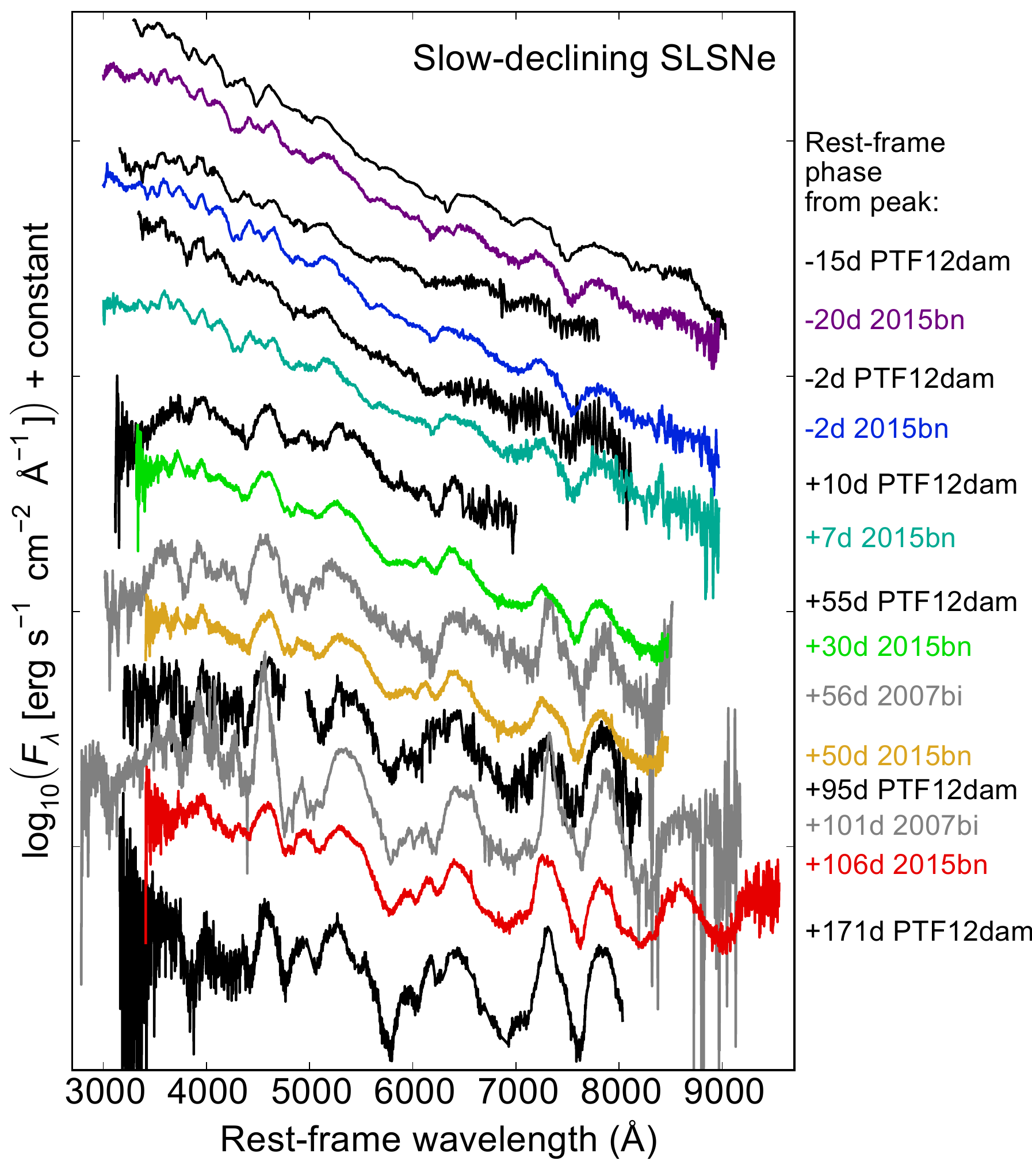}
\figcaption{Comparison of selected spectra with other slow-fading SLSNe at similar phases: PTF12dam \citep{nic2013} and SN 2007bi \citep{gal2009,you2010}. The spectroscopic evolution is nearly identical, although SN 2015bn shows Fe\,\textsc{iii} at $\sim4000$\,\AA~at earlier phases than PTF12dam, which shows only O\,\textsc{ii} at this wavelength at $-$15\,d. SN 2007bi exhibits the same lines at late times, but the features are more pronounced at $\sim100$\,d.\label{fig:slow}}
\end{figure}

\begin{figure}
\centering
\includegraphics[width=8.7cm,angle=0]{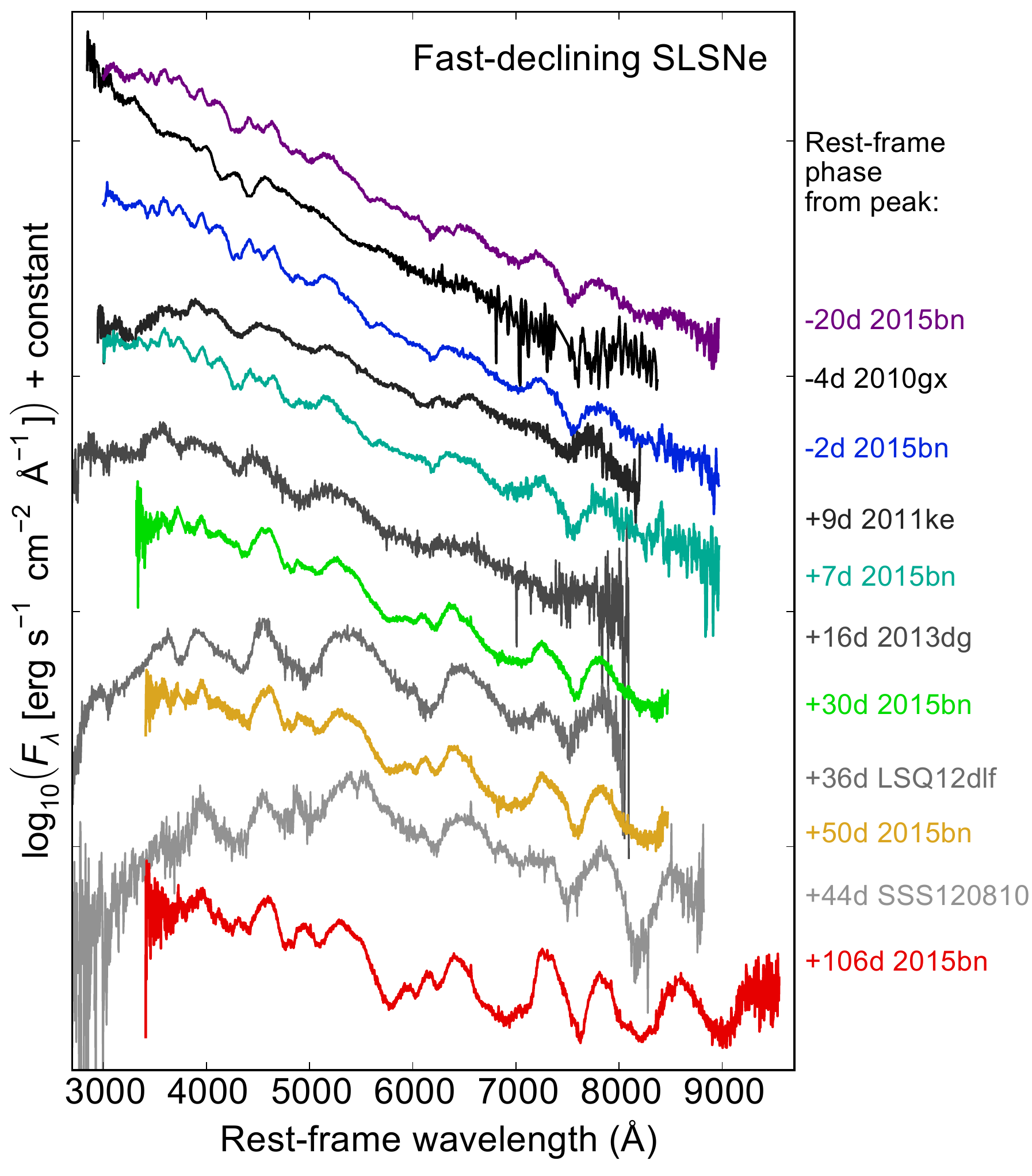}
\figcaption{Comparison of selected spectra with the more common fast-evolving SLSNe. Although the objects share all important lines, SN 2015bn shows a number of differences from these objects: it shows stronger absorption above 5000\,\AA~before maximum light, whereas 2010gx is relatively featureless in this region; lines are noticeably less broad in SN 2015bn, corresponding to lower velocities (c.f.~Figure \ref{fig:v}); SN 2015bn retains a very blue continuum for much longer (c.f.~Figure \ref{fig:col}). Data from \citet{pas2010,ins2013,nic2014}.\label{fig:fast}}
\end{figure}

In Figures \ref{fig:slow} and \ref{fig:fast}, we compare our spectra of SN 2015bn to other SLSNe from the literature. First we consider objects with broad light curves similar to SN 2015bn (Figure \ref{fig:rf_lcs}). SN 2007bi \citep{gal2009,you2010} and PTF12dam \citep{nic2013} are some of the best observed examples. As demonstrated by Figure \ref{fig:slow}, the spectroscopic evolution of SN 2015bn is an excellent match to these objects. In particular, all three objects exhibit a distinctive `trio' of lines between $\sim5700$-6500\,\AA~at late times. However, there are some differences. SN 2007bi shows stronger emission lines at $\sim100$\,d, especially in Mg\,\textsc{i}] and [Ca\,\textsc{ii}], suggesting that SN 2007bi develops a significant nebular component earlier than the other two objects.

SN 2015bn also exhibits a slight difference compared to PTF12dam at very early times. While the two objects look virtually identical at peak, the shapes and equivalent widths of the O\,\textsc{ii} lines at 4000-4600\,\AA~in PTF12dam at $-$15\,d were very similar to those in fast-declining SLSNe I, whereas in SN 2015bn these lines are weaker and may be blended with Fe\,\textsc{iii}\,$\lambda$4430 at the time of our earliest spectra $>20$\,d before maximum. Given that the continuum temperature looks virtually identical to PTF12dam, the differences may be due to velocity structure or composition. That the spectrum of SN 2015bn looks unchanged for longer around maximum light than PTF12dam is consistent with its shallower rise, and may be linked to the pre-peak shoulder in the light curve. In summary, the early detection and high-cadence spectra reveal a small degree of diversity among these objects, but overall the spectroscopic evolution of SN 2015bn confirms the similarity of this object to PTF12dam and SN 2007bi. Given the relatively low velocities and high signal-to-noise spectra of SN 2015bn, this is an ideal dataset for detailed spectroscopic modelling in the future, to confirm our line identifications and explain the slight differences between the SLSNe. However, such modelling is beyond the scope of this paper.

We now compare and contrast the same sequence of spectra with some typical fast-declining SLSNe I (the same objects used for the colour comparison in Figure \ref{fig:col}). In fact, the most obvious difference in Figure \ref{fig:fast} was already revealed by Figure \ref{fig:col}: that while most SLSNe I evolve to the red after maximum light, SN 2015bn (and other slow decliners) maintain a strong blue continuum for much longer. The other important difference is in the line widths; the fast-declining objects show significantly more Doppler-broadening and blending compared to SN 2015bn, as expected from their higher velocities (Figure \ref{fig:v}). Despite these differences, the slow- and fast- declining SLSNe I do display nearly all the same lines, and in the same order -- particularly O\,\textsc{ii} before maximum light and Ca\,\textsc{ii}, Mg\,\textsc{i}], Fe\,\textsc{ii} and Si\,\textsc{ii} afterwards -- but the lines only become prominent at $+30-50$\,d in objects like SN 2015bn, in keeping with their slower photometric evolution. This degree of similarity clearly suggests a link between the fast and slow SLSNe I. Higher ejecta mass could simultaneously explain the broader light curve, lower velocity, and later formation of lines in the slow objects like SN 2015bn. In order to account for the persistent blue continuum, more late-time heating may be required.

\begin{figure}
\centering
\includegraphics[width=8.7cm,angle=0]{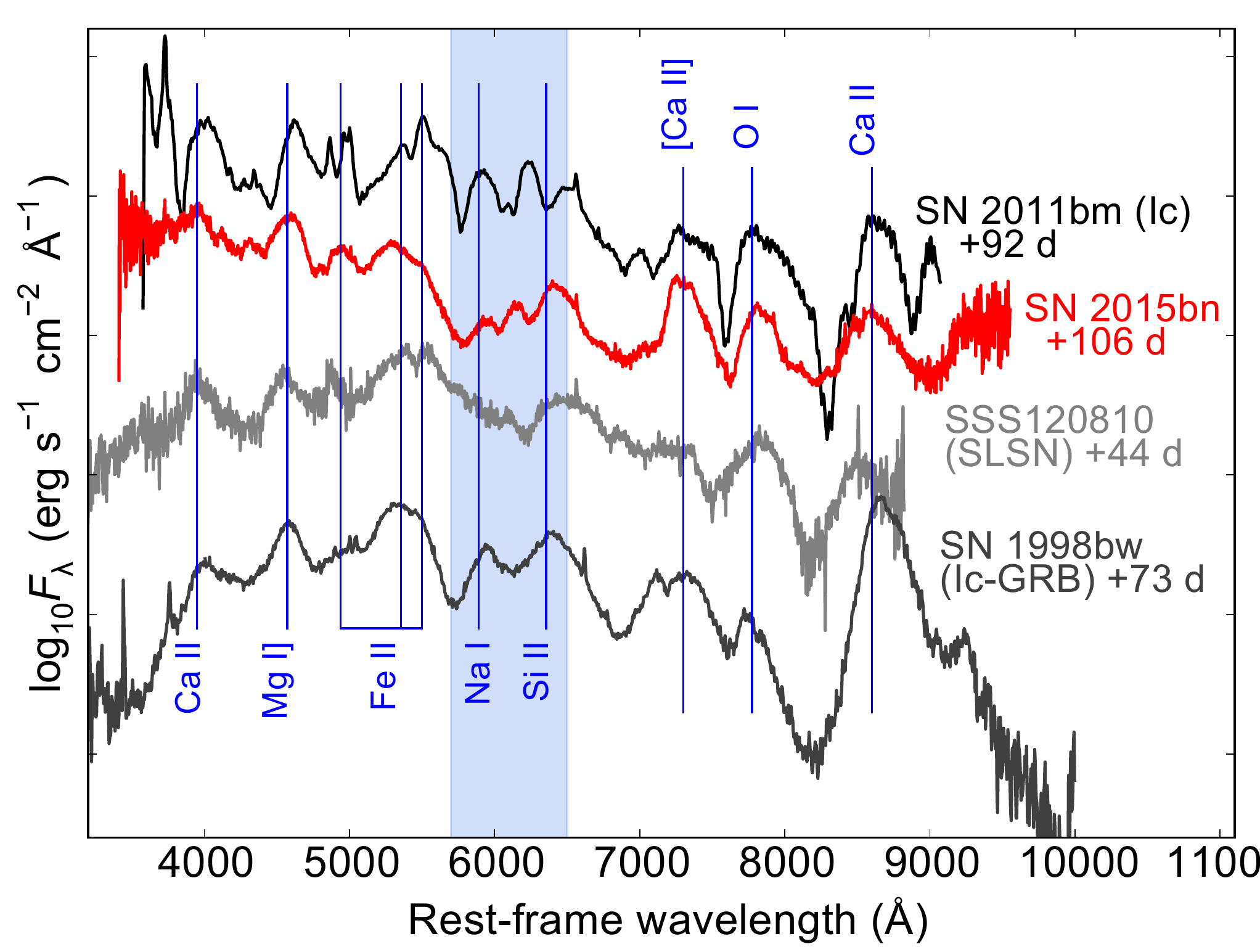}
\figcaption{Comparison of late-phase spectra of SLSNe and other SNe Ic. All spectra have been corrected for extinction. Objects are plotted in order of increasing line velocities (top to bottom). The spectrum of SN 2011bm has been de-reddened by an additional $E(B-V)=0.4$ for ease of comparison. The four objects show all the same major spectral features (marked on the plot), apart from in the shaded blue area between $\sim5700$-6500\,\AA, where SN 2015bn shows a distinctive trio of lines. The vertical lines mark the rest-frame wavelengths of the transitions labelled. The region between 3000-4000\,\AA~is likely also affected by absorption due to Mg\,\textsc{ii}, Ti\,\textsc{ii} and Fe-group elements \citep[e.g.][]{mil1999}. Data from \citet{pat2001,val2012,nic2014}.\label{fig:ic}}
\end{figure}

The literature contains numerous comparisons between the early (photospheric) spectra of SLSNe I and of normal and broad-lined SNe Ic \citep[e.g.][]{pas2010,ins2013}. Given the high signal-to-noise late-time spectra of SN 2015bn, we now have an opportunity to do a similar comparison times when the objects are evolving towards the nebular phase. A full nebular comparison is not possible yet, as SN 2015bn retains significant continuum and absorption lines even at +243\,d. In Figure \ref{fig:ic}, we plot the +106\,d spectrum of SN 2015bn along with the spectroscopically normal (but high mass) Type Ic SN 2011bm \citep{val2012} and the broad-lined, LGRB-associated Type Ic SN 1998bw \citep{pat2001} at comparable phases. We also show the spectrum of the fast-declining SLSN I SSS120810 at +44\,d. Unfortunately, no high signal-to-noise spectra of this kind of event is available beyond $\approx +50$\,d, as high redshifts make observations of fast-declining SLSNe challenging at this phase. A later spectrum of SSS120810 at +60\,d does exist \citep{nic2014} and looks virtually identical to this one, but we use the earlier spectrum as it has a much higher signal-to-noise ratio. Given the faster spectroscopic (and photometric) evolution of these events, it is not unreasonable to compare a fast event at $\approx +50$\,d with SN 2015bn at $\approx +100$\,d.

The four objects show all the same major spectral features, apart from in the region between $\sim5700$-6500\,\AA, where SN 2015bn shows the distinctive trio of lines. The fast-declining SLSN does not show Na\,\textsc{i} D or the unidentified line at this phase. We note that SN 2015bn started to exhibit the trio of lines as early as +50\,d, despite the much slower evolution in all of the lines it shows in common with SSS120810. This gives us some confidence that this trio may represent a real spectroscopic difference rather than an artefact of the phases we choose for the comparison. The GRB-SN 1998bw does not exhibit the central, unidentified line of the trio, but agrees in the other two lines (Si\,\textsc{ii} and Na\,\textsc{i}). This could be a consequence of line blending due to the higher velocities in this object. SN 2011bm does show three absorptions within the shaded region, but apart from Na\,\textsc{i}, the profiles seem to be redshifted relative to SN 2015bn, and may not actually be the same lines. The other distinctive difference is in the Ca\,\textsc{ii} lines. SN 2015bn is the only object that shows a larger luminosity in the forbidden line at 7300\,\AA~than in the allowed NIR triplet at 8600\,\AA. A detailed explanation of these line differences is beyond the scope of our study, but may hold clues to whether there are significant differences in the ejecta conditions between different stripped SNe. In particular, the strong [Ca\,\textsc{ii}] emission in all slowly-declining SLSNe has yet to be understood \citep{gal2009}.

\section{Bolometric light curve and physical properties}\label{sec:bol}

\subsection{Flux integration and blackbody fits}

The next step in our analysis was to construct the bolometric light curve of SN 2015bn. Because the photometric observations span the full wavelength range from the UV to the NIR, the bolometric luminosity could be estimated relatively straightforwardly. After applying reddening corrections and $K$-corrections to the measured magnitudes, and correcting for distance modulus at $z=0.1136$, the absolute rest-frame light curve in each filter was interpolated to match the epochs with $g$- and $V$-band observations. Magnitudes were then converted to spectral luminosity ($L_\lambda$) to give a spectral energy distribution (SED) at each point on the light curve. To get the luminosity, we integrated each SED numerically, assuming fluxes go to zero at the blue edge of the \textit{uvw2} band and the red edge of $K$-band.

\begin{figure}
\centering
\includegraphics[width=8.7cm,angle=0]{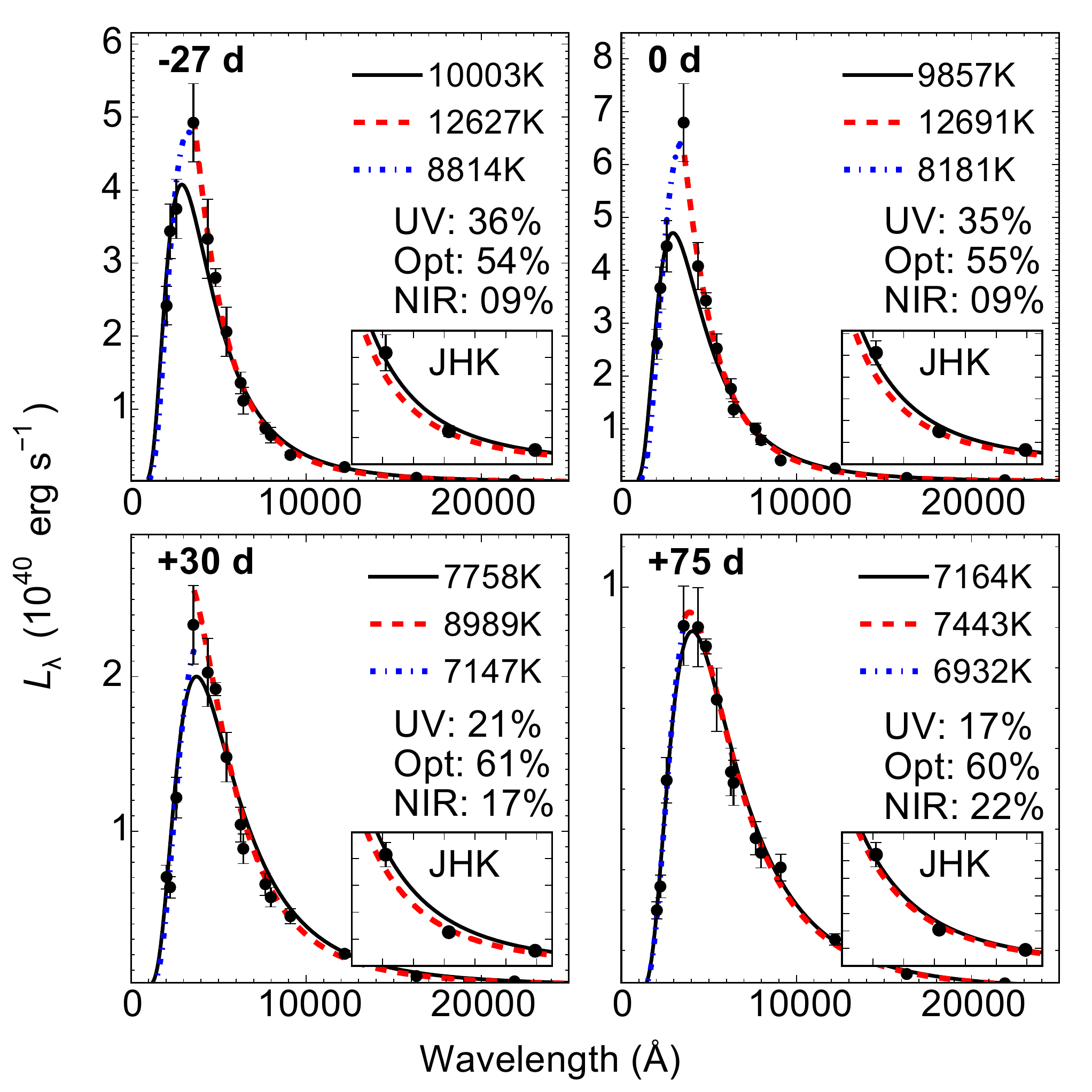}
\figcaption{Spectral energy distribution of SN 2015bn at representative points during its evolution. Fractions of flux emitted in the UV (bluewards of $u$-band), optical ($u$- to $z$-band) and NIR (redwards of $z$-band) are labelled. The SED peaks in the $u$-band at all epochs shown, but can be seen moving into $g$-band beyond $\sim75$\,d. Also shown are blackbody fits. Solid line: fit to all points; dashed line: fit $u$- to $K$-band; dash-dotted line: fit \textit{uvw2}- to $u$-band. The separate fit to the UV shows a lower colour temperature at early times due to line blocking by metals.\label{fig:bb}}
\end{figure}

We also fitted blackbody curves to each SED. As well as enabling a check that the SED looks sensible, these fits were used to derive temperatures and radii. In Figure \ref{fig:bb}, we plot the SED and blackbody fits at 4 representative epochs: our earliest epoch with multicolour photometry at $-$27\,d; maximum light; +30\,d, sampling the steeper decline prior to the knee; and +75\,d, sampling the slower post-knee decline. At early times (top row of figure), the blackbody fits are poor around the peak of the SED. This has been observed before \citep[e.g.][]{chom2011}, and is a consequence of metal line absorption in the UV, which significantly suppresses the flux bluewards of $\sim3000$\,\AA. Therefore we also fitted a `two-temperature' blackbody model, using two components that meet at the $u$-band. These curves (shown in Figure \ref{fig:bb}) are not intended to suggest two physically distinct temperature zones, but simply to demonstrate the very different colour temperature in the UV due to the strong absorption. The red component ($u$- to $K$-band), is not subject to substantial line blocking, and should thus give a better overall representation of the colour temperature of the underlying continuum emission. The amount of line absorption increases with time across the UV and optical (Figure \ref{fig:lines}). However, as the ejecta cool and the peak of the SED moves into the optical, we find that by $\sim75$\,d the SED can be described well by a single blackbody fit from the UV to the NIR. At all epochs, integrating the two-component blackbody model from 1000-25000\,\AA~gives a luminosity estimate that is consistent (to within $<5$\%) with that obtained by directly integrating the observed fluxes.

In the UV and optical filters, our photometric sampling is sufficient that linear interpolation between epochs was generally only over periods of a few days, and therefore does not wash out the substructure in the light curve (shoulder and knee; see section \ref{sec:phot}). The NIR is not sampled quite so densely, but this is not problematic, as the NIR contribution to the overall flux is relatively modest. As demonstrated by Figure \ref{fig:bb}, at maximum light the NIR contribution is $\la10\%$, rising to $\sim20\%$ during the decline phase. Since we lack NIR observations earlier than $-$9\,d, we assumed constant colours in $g-J$, $g-H$ and $g-K$ for the early points. The blackbody fit in the first panel of Figure \ref{fig:bb} shows that this is a reasonable assumption. At the very earliest epochs, we have only a single filter -- in this case we assumed the same bolometric correction as at $-$27\,d. This assumption should be perfectly reasonable for small extrapolations, but is questionable for the earliest CRTS point -- for example, \citet{nic2016a} and \citet{smi2016} found temperatures $\approx 25000$\,K during the early bump phases of SLSNe, compared to temperatures of 10000-15000\,K at maximum light. For our latest photometric point at +243\,d, we lack UV data. The bolometric luminosity here was estimated by integrating a one-temperature blackbody fit to the optical and NIR data.

\subsection{Luminosity, temperature and radius}

\begin{figure}
\centering
\includegraphics[width=8.7cm,angle=0]{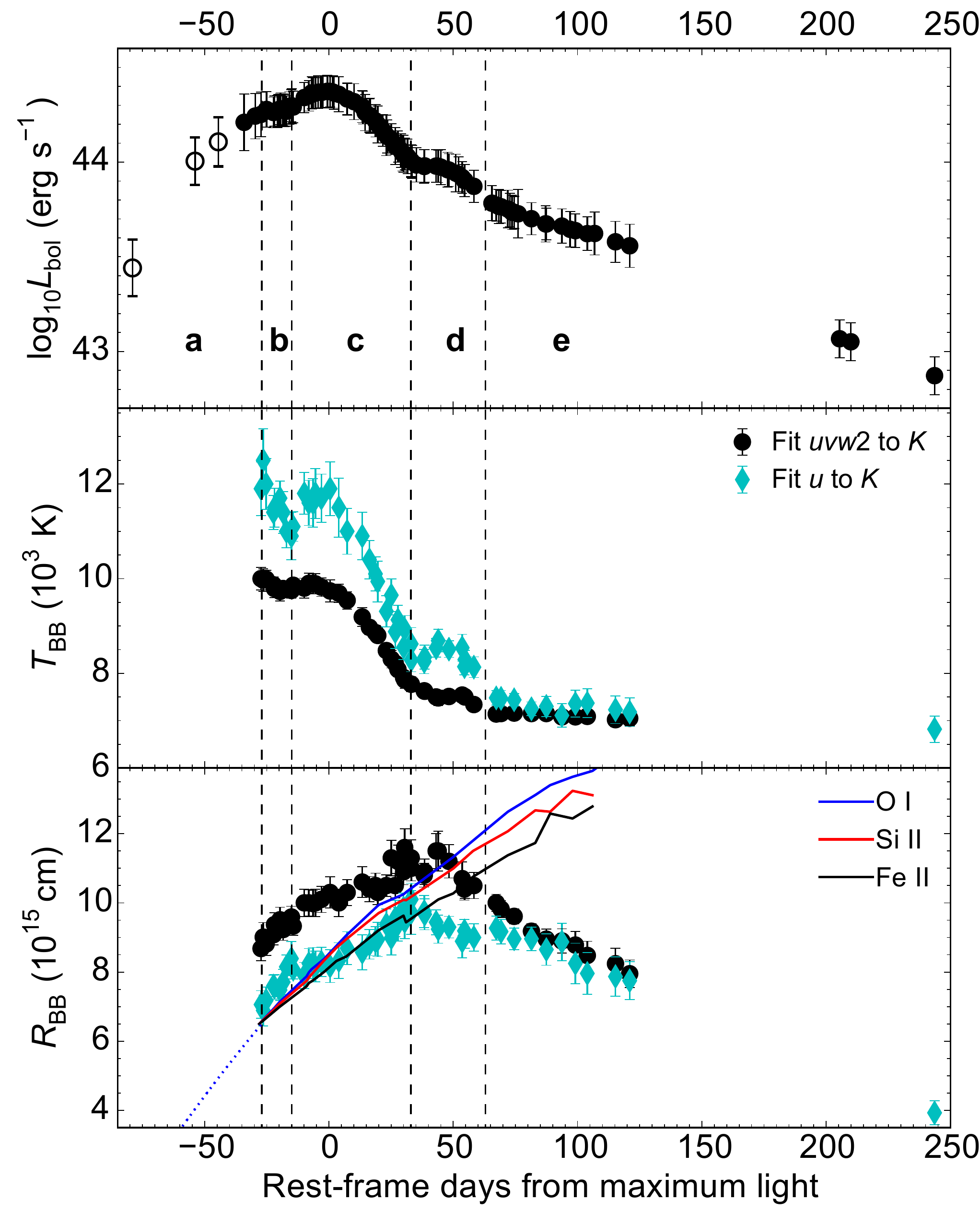}
\figcaption{Top: bolometric light curve of SN 2015bn, obtained by integrating the flux from the \textit{uvw2} to $K$-band. Empty symbols mark epochs where the luminosity has been estimated using a large extrapolation in colour. The thick grey line is a linear fit to the last 10 data points. Middle: Temperature evolution of best-fit blackbody models -- black circles correspond to the solid lines in Figure \ref{fig:bb}; cyan diamonds correspond to the dashed lines in the same figure. Bottom: Evolution of blackbody radius. Symbols have same meanings as in temperature plot. We overplot lines calculated from the velocity evolution in Figure \ref{fig:v}, choosing an initial radius to match the blackbody evolution between $-$27 to +30\,d.\label{fig:bol}}
\end{figure}

The bolometric light curve, as well as the temperatures and radii inferred from the blackbody fits, are plotted in Figure \ref{fig:bol}. We divide the evolution into five phases for our analysis; these are labelled \a, \b, \c, \d\ and \e\ in the figure. In region \a, the evolution is only partially observed. We have an isolated point at $-$79\,d with a large error due to the reliance on an uncertain bolometric correction. At this phase, the point could be part of a long smooth rise, but our analysis in section \ref{sec:phot} suggested that it was more likely to be connected with the initial bump phase exhibited shortly after explosion by many SLSNe I \citep{lel2012,nic2015a,nic2016a,smi2016}. When we pick up the object again with ASAS-SN, it is $-$54\,d from peak and rising. As we only have single-filter photometry at this time, we have no information about the temperature or radius during region \a.

Region \b\ encompasses the pre-maximum shoulder, first pointed out in section \ref{sec:phot}. This lasts for around 12\,d, and seems to be a flat plateau in bolometric space. The blackbody temperature decreases from $\sim 12000$\,K to $\sim 11000$\,K during this phase (based on optical-NIR blackbody fit). This drop is compensated by an increase in the radius of the emitting surface. The change in radius over time during this phase is consistent with our measured velocities in Figure \ref{fig:v} -- these curves are overplotted here. If we extrapolate the O\,\textsc{i} velocity backwards to $R_{\rm BB}=0$ (assuming a deceleration of 26\,\kms\,d$^{-1}$, based on the velocity evolution shown in Figure \ref{fig:v}), we arrive at an explosion date 92\,d before maximum light. However, as the ejecta will likely have undergone steeper deceleration at early times, the 92\,d rise should be considered an upper limit only. If the explosion occurred $<92$\,d before peak, then the detection at $-$78\,d would most likely be during the initial bump phase, assuming SN 2015bn follows a similar morphology to LSQ14bdq. This agrees with our analysis in section \ref{sec:phot}.

The next phase of the light curve, labelled \c, is the broad, smooth peak. Over 15\,d, SN 2015bn rises to a bolometric maximum with $L_{\rm peak}=2.3\pm0.4\times10^{44}$\,erg\,s$^{-1}$. The light curve is quite symmetric close to maximum, then settles onto a smooth decline until $\sim30$\,d later. The luminosity closely tracks the temperature evolution, which shows an increase back to 12000\,K at maximum and then declines smoothly afterwards. The radius largely continues to track the line velocities, indicating that photospheric recession is relatively unimportant during this phase. Therefore the changes in the spectrum occurring a few weeks after maximum light are primarily driven by the decrease in temperature. That the spectrum shows so little evolution prior to this can also be understood in terms of the temperature evolution, which varies by only around 10\% from our first measurement until $\sim15$\,d after maximum.

The most interesting behaviour occurs at $\ga30$\,d. This is the beginning of the knee in the light curve, which we designate region \d. The blue excess visible in Figure \ref{fig:bump} shows up clearly as a brief plateau in the bolometric output, during which time the temperature stays constant at around 8500\,K. The radius stops tracking the measured line velocities, and begins to decrease from its maximum value of $10^{16}$\,cm, indicating that the photosphere is receding through the ejecta faster than the ejecta are expanding. Up to this point, the measured velocities were slowly decreasing, but during region \d\ they settle onto a constant $\sim6500$\,\kms~(see also Figure \ref{fig:v}). This suggests a change in the ejecta structure, with a flatter velocity profile inside some mass coordinate. This will be discussed in more detail in section \ref{sec:bumps}.

Finally, in region \e\ the bolometric luminosity settles onto a long, slow decline, which we followed for 60\,d before SN 2015bn disappeared behind the sun. Since its return just before the end of 2015, our observations are consistent with the same slow decline rate, as demonstrated by a linear fit across this part of the light curve. The temperature is approximately constant at 7000\,K. Meanwhile the radius looks to have decreased slowly and smoothly to $\sim4\times10^{15}$\,cm. During this phase, the blackbody fits to the full UV-NIR wavelength range give much the same colour temperature as the optical-NIR fits; this is expected given Figure \ref{fig:bb}. The total radiative output integrated across regions \a-\e\ is $E_{\rm rad}=2.3\pm0.5 \times 10^{51}$\,erg.

\begin{figure}
\centering
\includegraphics[width=8.7cm,angle=0]{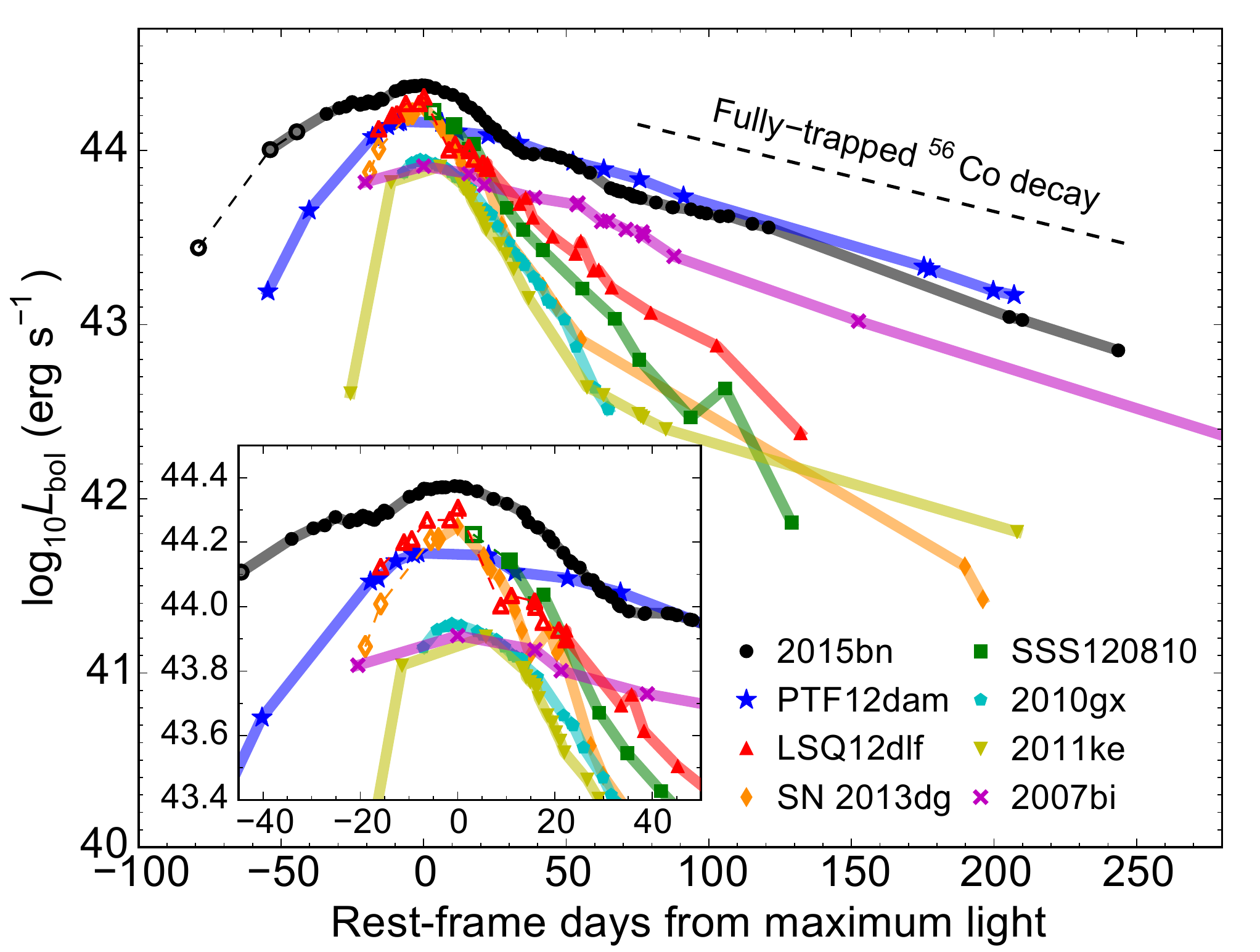}
\includegraphics[width=8.7cm,angle=0]{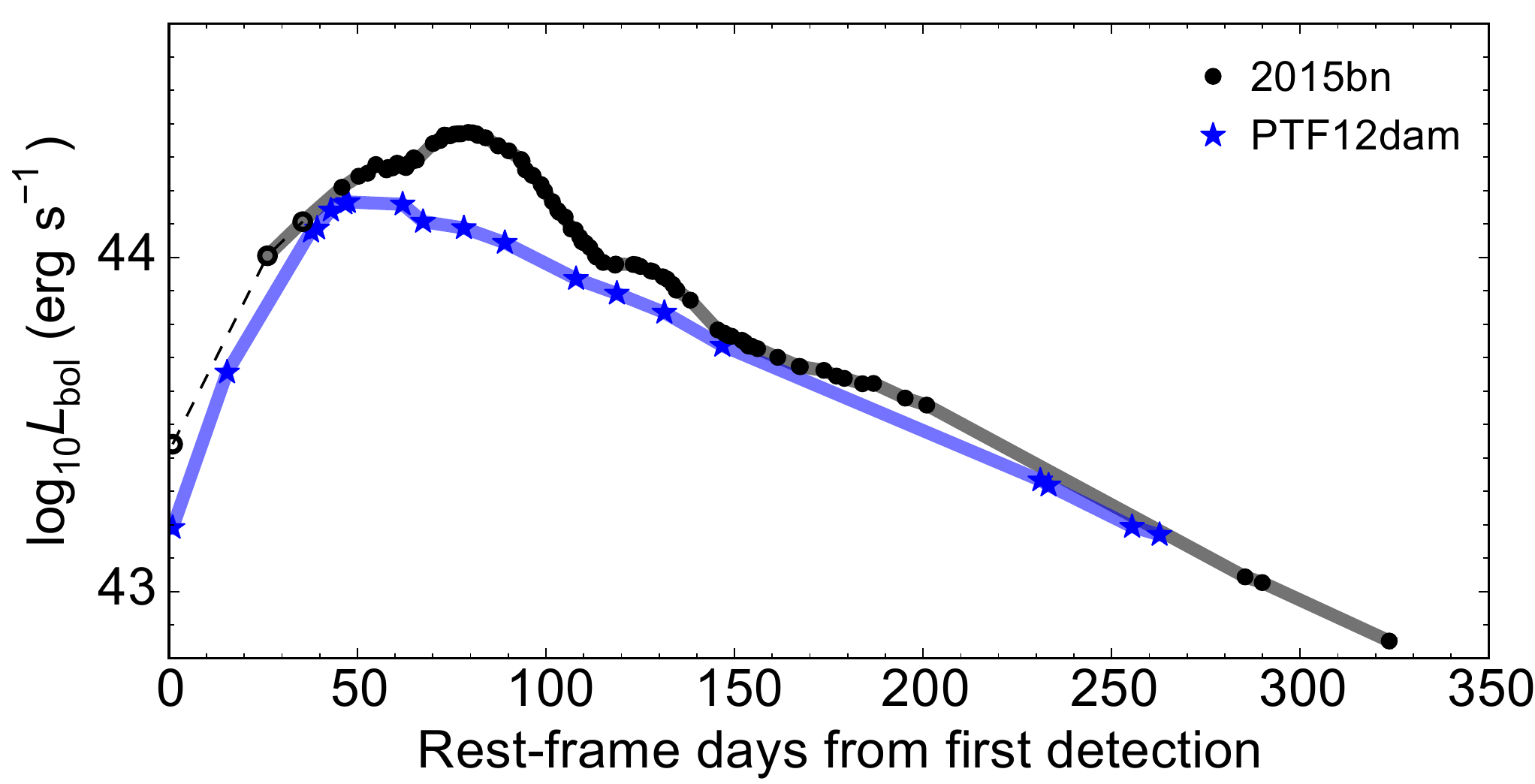}
\figcaption{Top: Comparison of SN 2015bn bolometric light curve with other SLSNe I. Empty symbols and dashed lines indicated luminosities estimated from single-filter photometry. The slow decline very closely matches that seen in PTF12dam and SN 2007bi. We note that the tail phase of SN 2011ke (which initially shows a fast decline) has a very similar slope. The slope is approximately consistent with fully-trapped cobalt decay. Inset: The drop from light curve peak is initially steeper for SN 2015bn compared to the other slow-decliners. Bottom: Assuming we can map the first detections to explosion dates, the shoulder in SN 2015bn closely matches the peak in PTF12dam. SN 2015bn then shows 3 undulations of decreasing amplitude during the `decline' phase (though the first of these is the absolute bolometric maximum). Data sources: \citet{pas2010,you2010,ins2013,nic2013,nic2014,chen2014}.\label{fig:bc}}
\end{figure}

\subsection{Bolometric comparisons}

In Figure \ref{fig:bc}, we compare the bolometric light curve of SN 2015bn to other SLSNe I. The late-time decline rate of SN 2015bn is a near-perfect match for PTF12dam and SN 2007bi. The fast-declining SLSN 2011ke transitioned to a tail phase after around 50\,d, which also matches this slope (though not all SLSNe seem to show this behaviour). The decline rate is suspiciously similar to the decay of \Co; this will be discussed in more detail in section \ref{sec:mod}. Looking more closely at the light curve shape around peak, SN 2015bn does show a marked difference from other slow-declining SLSNe I observed so far. PTF12dam and SN 2007bi appear to begin their slow-declining tail phase immediately after maximum light, whereas SN 2015bn initially shows a steeper decline more reminiscent of faster SLSNe. The measured bolometric decline rate between days +10 and +30 is 0.038\,mag\,d$^{-1}$, which is almost identical to LSQ12dlf around the same phase, whereas PTF12dam declines at a rate of only 0.011\,mag\,d$^{-1}$. The initially steeper decline rate may indicate a lower mass compared to PTF12dam; however, the rise time may be longer, and argues in the opposite direction. Mass estimates from model fits will be presented in section \ref{sec:mod}.

While on the subject of the rise time, another useful comparison can be made to PTF12dam if we assume that the earliest detection of each event is close to the time of explosion; this is shown in the bottom panel of Figure \ref{fig:bc}. In this case, the time to peak of PTF12dam matches the time to the shoulder in SN 2015bn. The discrepancies between the light curves then appear between $\sim60$-140\,d after explosion, as SN 2015bn undergoes a number of undulations of decreasing amplitude (there is potentially a third, weak undulation around 150\,d), before joining smoothly onto a tail phase that again matches PTF12dam. This shows that, without a clear picture of the physical mechanism underlying the light curve undulations, it is difficult to know which phases are most appropriate to make like-for-like comparisons between the SNe. That the spectrum of SN 2015bn during the shoulder phase looks like PTF12dam at maximum light may indicate that this alignment gives a fairer comparison than the `days-from-peak' approach used elsewhere in this work. Fortunately, this is not crucial for the analysis to follow.

\section{Physical models}\label{sec:mod}

\subsection{Light curve fits}

To better understand the nature of SN 2015bn, we fitted the bolometric light curve with diffusion models based on the various power sources proposed for SLSNe: a magnetar engine; circumstellar interaction; and radioactive decay. The necessary equations and fitting procedures have been extensively described by \citet{ins2013,cha2012,nic2014}, and are based on the original diffusion solution of \citet{arn1982} with constant opacity. There is a degree of uncertainty as to what value to use for the opacity when modelling SLSNe. In the optical, the dominant source of opacity is electron scattering. For normal SNe Ic, most authors assume $\kappa=0.1$\,cm$^{2}$\,g$^{-1}$. In the case of SLSNe, ionization may be more complete, especially before and around maximum when the temperature is higher than in normal SNe Ic; full ionization gives $\kappa=0.2$\,cm$^{2}$\,g$^{-1}$. In the present work we will assume this latter value. For fixed kinetic energy, the value of \Mej~derived from fitting the light curve scales as \Mej\,$\propto \kappa^{-2/3}$. Therefore our ejecta mass estimates can easily be compared with literature models that use $\kappa=0.1$\,cm$^{2}$\,g$^{-1}$ simply by multiplying them by a factor of 1.6.

\subsection{Magnetar models}

We first consider models in which the luminosity is powered by magnetar spin-down. This is a specific example of the more general class of `central engine' models, which also include accretion-powered SNe \citep{dex2013}. The magnetar model assumes that the core-collapse of the SLSN progenitor leaves behind a highly magnetised neutron star ($B\sim10^{14}$\,G) rotating with a period $P\sim1-10$\,ms. The nascent pulsar thus has a large reservoir of rotational energy, which is tapped via the $B$-field and heats the ejecta at the rate prescribed by the standard magnetic dipole formula \citep{ost1971,kas2010,woo2010}. The magnetar emission is expected to be in the form of energetic electron-positron pairs, which generate a hard input spectrum through a pair-cascade \citep{met2014}. We treat $\gamma$-ray leakage from the ejecta following \citet{wang2015} \citep[see also][]{chen2014}.

As discussed in sections \ref{sec:phot} and \ref{sec:bol}, because the early light curve is not well sampled, we do not know for sure whether SN 2015bn rises monotonically from discovery to maximum light, or if the first detection is during the initial bump phase that seems to be common in SLSNe I \citep{nic2016a}. The colour evolution suggested that the bump interpretation is more likely (see section \ref{sec:phot}). The basic form of the magnetar model employed here does not accommodate a non-monotonic rise (though \citealt{kas2015} have shown that including delayed shock breakout driven by the magnetar can give a good fit to the double-peaked light curve of LSQ14bdq). \citet{nic2015b} showed that, for both magnetar models and observed SLSNe, the rise and decline timescales obey a fairly tight correlation. Thus the nature of the first detection, driving the fit at early times, will have a strong influence on the fit at peak, and therefore any derived parameters. For this reason, we calculated the best-fitting model under two different constraints: including the first point as part of a smooth rise; and introducing a synthetic point around where we would expect the monotonic rise to begin if the first detection is in the bump phase -- we do this by scaling the light curve of LSQ14bdq as in Figure \ref{fig:obs_lcs}.

\begin{table}
\begin{center}
\caption{Magnetar fit parameters for Figure \ref{fig:mag}.\label{tab:mag}} 
\begin{tabular}{ccccc}
\hline
Rise time/d   & \Mej/\M$^a$    & $P$/ms  &  $B$/$10^{14}$\,G	&	$\chi^2_{\rm red}$ \\
\hline	
69$^b$	&	8.4	&	2.1	&	0.9	&	0.31	\\
85	&	15.1	&	1.7	&	1.0	&	0.48	\\
\hline
\end{tabular}
\end{center}
$^a$ Assuming $\kappa=0.2$\,cm$^{2}$\,g$^{-1}$\\
$^b$ Estimated from scaled light curve of LSQ14bdq\\

\end{table}

\begin{figure}
\centering
\includegraphics[width=8.7cm,angle=0]{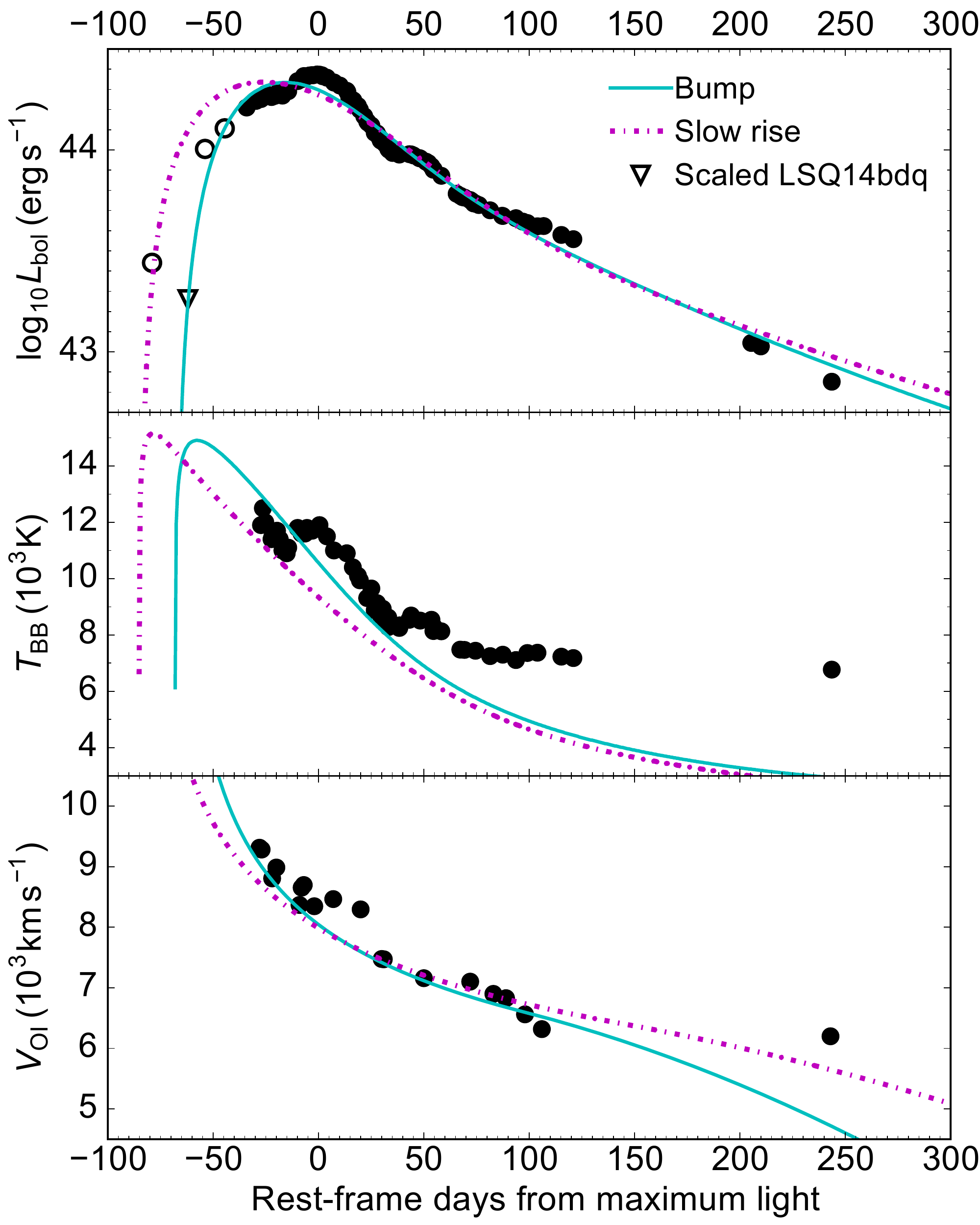}
\figcaption{Magnetar-powered model fits to SN 2015bn. We calculate two models under different assumptions about the first point: including it as part of a smooth rise; and treating it as a bump similar to LSQ14bdq \citep{nic2015a}. Top: Modelling the bolometric luminosity. Middle: Comparisons of model temperatures with the blackbody fits. Bottom: Comparisons of model velocities with those measured from the O\,\textsc{i}\,$\lambda$7774 line. The magnetar model gives a good overall fit to the properties of SN 2015bn, particularly if the first detection is during the bump phase. Parameters of all models are listed in Table \ref{tab:mag}.\label{fig:mag}}
\end{figure}

The model fits are shown in Figure \ref{fig:mag}, with the fit parameters listed in Table \ref{tab:mag}. In general, the fits are seen to be reasonable; however these simple models are not able to replicate the detailed structure in terms of the knee and shoulder. The value of the derived mass is driven by two competing factors. The relatively narrow shape around the main light curve peak can be more easily fit with a lower-mass model, while the overall broader shape and long rise favours higher mass. If we assume a smooth rise, the best-fit ejecta mass is 15\,\M. The slow tail phase is well-matched due to a long spin-down time, given by the relatively weak $B$-field \citep{kas2010,nic2013}. This model gives a fairly good match to the velocity evolution, especially the slow velocity decline at late times. On the other hand, the broad luminosity peak in this case is a rather poor fit to the data.

If we instead assume that the first point is part of the bump phase, we get a more convincing fit between $-50$\,d and maximum, with a lower ejecta mass of \Mej\,$\approx8.4$\,\M. The model still peaks a little earlier than the data, which is interesting in light of the comparison to PTF12dam in the lower panel of Figure \ref{fig:bc}. This model also gives an excellent match to the photospheric velocity. While the temperature in this model fits the data well from $-$30\,d to +50\,d, it is too low by up to a factor of $\sim2$ by 250\,d. However, the fact that the better overall fit to the light curve, temperature and velocity is the one for which we assume an initial bump phase is consistent with the analysis in section \ref{sec:rf}.

\begin{table*}
\begin{center}
\caption{Interaction fit parameters for Figure \ref{fig:csm}.\label{tab:csm}} 
\begin{tabular}{ccccccccc}
\hline
Rise time/d	&	Profile		& \Mej/\M$^a$	& \Mcsm/\M$^a$	&  \Mni/\M		&	$E_{\rm k}/10^{51}$erg	& $R_{\rm int}$/$10^{15}$cm$^b$ 	& $\log (\rho_0$/g\,cm$^{-3}$)$^{c}$	& $\chi^2_{\rm red}$\\
\hline
67			&	shell$^d$		&	49.0	&	18.5		&	--		&	1.9					&	3.0					    	&	$-$10.94		&	0.51	 \\    
87			&	shell			&	21.9	&	19.2		&	--		&	1.8					&	3.0					    	&	$-$12.00		&	0.35	 \\
82			&	shell			&	13.8	&	13.0		&	3.7		&	1.3					&	3.0					    	&	$-$12.34		&	0.15	 \\
61			&	wind$^e$		&	30.1		&	15.1			&	--		&	4.3					&	0.3					    	&	$-$10.49		&	0.37	 \\    
85			&	wind			&	33.3		&	18.0			&	--		&	4.5					&	0.3					    	&	$-$10.61		&	0.40	 \\
59			&	wind			&	29.4		&	16.1			&	2.3		&	4.4					&	3.1					    	&	$-$12.19		&	0.17	 \\
\hline
\end{tabular}
\end{center}
$^a$Assuming $\kappa=0.2$\,cm$^{2}$\,g$^{-1}$;
$^b$Interaction radius = inner radius of CSM density profile;
$^c$$\rho_0 = \rho_{\rm CSM}(r = R_{\rm int})$;
$^d$$\rho_{\rm CSM}(r) = \rho_0$;
$^e$$\rho_{\rm CSM}(r) = \rho_0 r^{-2}$.

\end{table*}

However, inspection of Table 2 shows that this model has a total rise
time of 69\,d, seemingly in contradiction to the presumed bump
detection at $-$79\,d. The widths of the bumps may be 10-15 days \citep{nic2015a,nic2016a}, which could then imply a
discrepancy between the explosion time and the magnetar fit of up to 20-25\,d. There are two possibilities. The first is that this is a real inconsistency, and the magnetar fit must be fixed with an explosion time at the first detection at the latest (i.e.~our `slow rising' model). The second is that thermalisation of the magnetar wind energy in the ejecta is not 100\% efficient from the time of explosion. Our model does naively assume that the magnetar energy input is 100\% efficient but this is by no means certain. The detailed physical processes by which the energy is emitted and thermalised are not yet well understood \citep{met2015,kas2015}. 
Therefore there could be a delay between core-collapse and the onset of efficient energy transfer between the magnetar and the expanding ejecta. Given these uncertainties, we conclude that a magnetar spin-down model with relatively high ejecta mass and low B-field is a good candidate to explain the properties of SN 2015bn.

\subsection{Interaction models}\label{sec:csm}

Next we investigate models powered by a collision between the fast SN ejecta and dense CSM. In this context, the observed luminosity comes from reprocessing and thermalisation of the kinetic energy by a forward shock wave propagating in the CSM and a reverse shock in the ejecta. For the energy conversion to be efficient, in general the mass contained in the CSM must be an appreciable fraction of that ejected by the explosion -- i.e.~at least a few solar masses \citep{cha2013,nic2014,ins2015}. The progenitors of interacting SLSNe would thus require exceptionally high mass-loss rates, perhaps as much as 0.1-1\,\M\,yr$^{-1}$ \citep[e.g.][]{ben2014}. This is difficult to achieve via stellar winds, and could instead point towards large discrete mass ejections, such as the eruptions of luminous blue variables \citep{jus2014} or hypothetical pulsational-PISNe \citep{woo2007}. Thus the expected CSM density profile around the star is quite uncertain. Following \citet{cha2012}, we calculate models with two representative density profiles: $\rho_{\rm CSM} \propto r^{-2}$, as expected for a wind; and $\rho_{\rm CSM} =$\,constant, which may be more appropriate for a CSM `shell' from a discrete mass ejection. Because the interaction model has significantly more free parameters than the magnetar models, there is no tight relationship between the rise and decline rates \citep{nic2015b}. Therefore it is not so important in this case how we treat the first detection of SN 2015bn at $-$79\,d; typically, increasing the ejecta or CSM mass can extend the rise time sufficiently to include this point as desired, without significantly altering the shape during the decline phase. For each density profile, we calculate a slow- and fast-rising model to demonstrate this (see fit parameters in Table \ref{tab:csm}).

\begin{figure}
\centering
\includegraphics[width=8.7cm,angle=0]{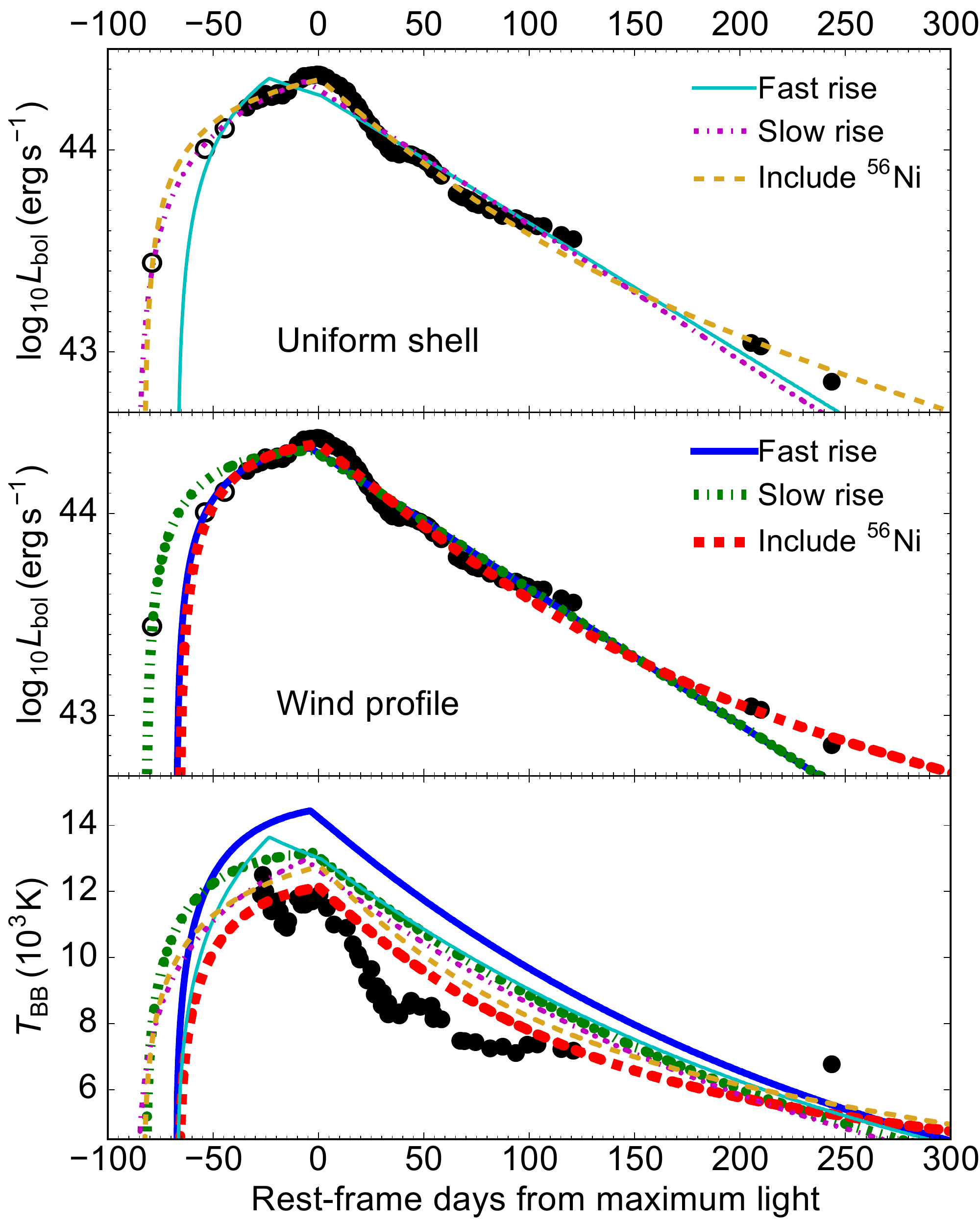}
\figcaption{Interaction-powered model fits to SN 2015bn. The rise time can be varied by increasing the mass of CSM or ejecta. Top: Modelling the bolometric luminosity assuming a constant-density shell profile for the CSM. Middle: Same as top, but assuming an inverse square wind profile. Bottom: Comparisons of model temperatures with the blackbody fits. We find that wind models give much better fits to the temperature, and that adding $\approx5$\,\M~of \Ni~gives a good match to the tail phase. Parameters of all models are listed in Table \ref{tab:csm}.\label{fig:csm}}
\end{figure}

The fits are shown in Figure \ref{fig:csm}. In the top panel we plot models with uniform shells of CSM. We fit for \Mej, \Mcsm, $\rho_{\rm CSM}$, the radius of the ejecta-CSM interface and kinetic energy (\E). We note that the radius, $R_{\rm int}$, is only weakly constrained by the light curve fitting routine, but can be checked against the observed temperature evolution\footnote{The CSM model is assumed to radiate as a blackbody at fixed radius} (bottom panel). The best-fit slow-and fast rising shell models both have \Mcsm\,$\approx19$\,\M~and \E\,$\approx2\times10^{51}$\,erg. The ratio \Mej/\Mcsm~is of order unity for the model with the slow rise, while the faster rising model has \Mej/\Mcsm\,$=2.6$. These are fairly similar to the ratios found by \citet{nic2014} for a number of other SLSNe I. However, these curves are unable to reproduce the transition to the shallower decline rate after 60\,d. By maximum light, the forward and reverse shocks have finished traversing the CSM and ejecta, respectively, and therefore no further energy can be deposited. The light curves therefore decline exponentially from peak as the stored energy diffuses out.

\begin{table*}
\begin{center}
\caption{Two-component interaction fit parameters for Figure \ref{fig:csm2}.\label{tab:csm2}} 
\begin{tabular}{ccccccccc}
\hline
 	&	Profile		& \Mej/\M$^a$	& \Mcsm/\M$^a$	&  \Mni/\M		&	$E_{\rm k}/10^{51}$erg	& $R_{\rm int}$/$10^{13}$cm$^b$ 	& $\log (\rho_0$/g\,cm$^{-3}$)$^{c}$	& $\chi^2_{\rm red}$\\
\hline
1			&	wind$^e$		&	25.0		&	15.0			&	--		&	4.0					&	140				    	&	$-$12.3		&	1.13	 \\
2			&	wind			&	40.0$^f$	&	15.0			&	--		&	2.0$^g$				&	350				    	&	$-$14.1		&	0.26	 \\
\hline
\end{tabular}
\end{center}
$^a$Assuming $\kappa=0.2$\,cm$^{2}$\,g$^{-1}$; 
$^b$Interaction radius = inner radius of CSM density profile; 
$^c$$\rho_0 = \rho_{\rm CSM}(r = R_{\rm int})$; 
$^d$$\rho_{\rm CSM}(r) = \rho_0$; 
$^e$$\rho_{\rm CSM}(r) = \rho_0 r^{-2}$; 
$^f$$M_{\rm ej,2}= M_{\rm ej,1}+M_{\rm CSM,1}$; 
$^g$$E_{\rm k,2} = E_{\rm k,1} - E_{\rm rad,1}$.

\end{table*}

The middle panel shows models using the wind density profile. In general, the wind scenario favours similar or slightly lower \Mej~and \Mcsm~compared to the shell models, with \Mej\,$\approx30$\,\M~giving a good fit along with \Mcsm\,$=15$-18\,\M~to cover the range of possible rise times. The kinetic energy in the wind model is $4\times10^{51}$\,erg. The wind models are more sensitive to the interaction radius, $R_{\rm int}$, compared to the shell models. For the inferred radii of $\approx 3 \times10^{14}$\,cm, a wind speed of 10\,\kms~would imply that the mass loss occurred over a period of 10-200\,d before explosion. However, these models suffer from the same problem at late times as do the shell models, in that they decline too quickly. Comparing the temperature evolution of each model (bottom panel), we find that both the wind and shell models give a reasonable match to observations, with the slow rising models predicting cooler temperatures that are closer to the data. While the quality of the light curve fits do not allow us to distinguish between wind and shell models, there exists significantly more literature describing the interaction of SN ejecta with a wind density profile. Therefore the remaining analysis will focus more on the wind model. This will also be useful when we come to model the radio light curve in section \ref{sec:radio}.

No explicit velocity comparison is possible, since the simple interaction model assumes a stationary CSM. However, as is often the case when modelling SLSNe, the exclusively high-velocity lines seen in the spectrum may be difficult to reconcile with having $\ga10$\,\M~of slower-moving material around the star. In an attempt to address this issue more quantitatively, we can use some of the analytic equations given by \citet{che2011} in conjunction with our fits. Those authors parameterised the wind as a function of radius as $\rho_w=\dot{M}/(4\pi v_w)r^{-2} \equiv   5.0 \times10^{16} D_* r^{-2}$, where $\rho_w$ is the density of the wind, $\dot{M}$ is the mass-loss rate, $v_w$ is the wind velocity, and
\begin{equation}
D_* \equiv \frac{\dot{M}}{10^{-2}\,{\rm M}_\odot\,{\rm yr}^{-1}} \left(\frac{v_{\rm wind}}{10\,{\rm km\,s}^{-1}}\right)^{-1}.
\label{eq:mdot}
\end{equation}
We found that the peak luminosity is quite sensitive to $D_*$; the value for the wind models in Figure \ref{fig:csm} is $D_* = 19.4$ (or equivalently, $\dot{M}/v_{10}\approx0.2$\,\M\,yr$^{-1}$, where $v_{10}$ is the wind velocity in units of 10\,\kms). In the \citet{che2011} framework, this can be used to calculate an effective diffusion radius:
\begin{equation}
R_d = 4.0 \times 10^{14}\, \kappa_{0.34}^{0.8}\, E_{\rm k,51}^{0.4}\, M_{\rm ej,10}^{-0.2}\, D_*^{0.6},
\end{equation} 
where the subscripts refer to normalisation of the parameters to 0.34\,cm$^2$\,g$^{-1}$, $10^{51}$\,erg and 10\,\M~as used by those authors. Our light curve models give estimates for all of these parameters (Table \ref{tab:csm}), leading to $R_d = 7.3 \times 10^{15}$\,cm, which is reassuringly close to the radii determined from blackbody fits (Figure \ref{fig:bol}).

\begin{figure}
\centering
\includegraphics[width=8.7cm,angle=0]{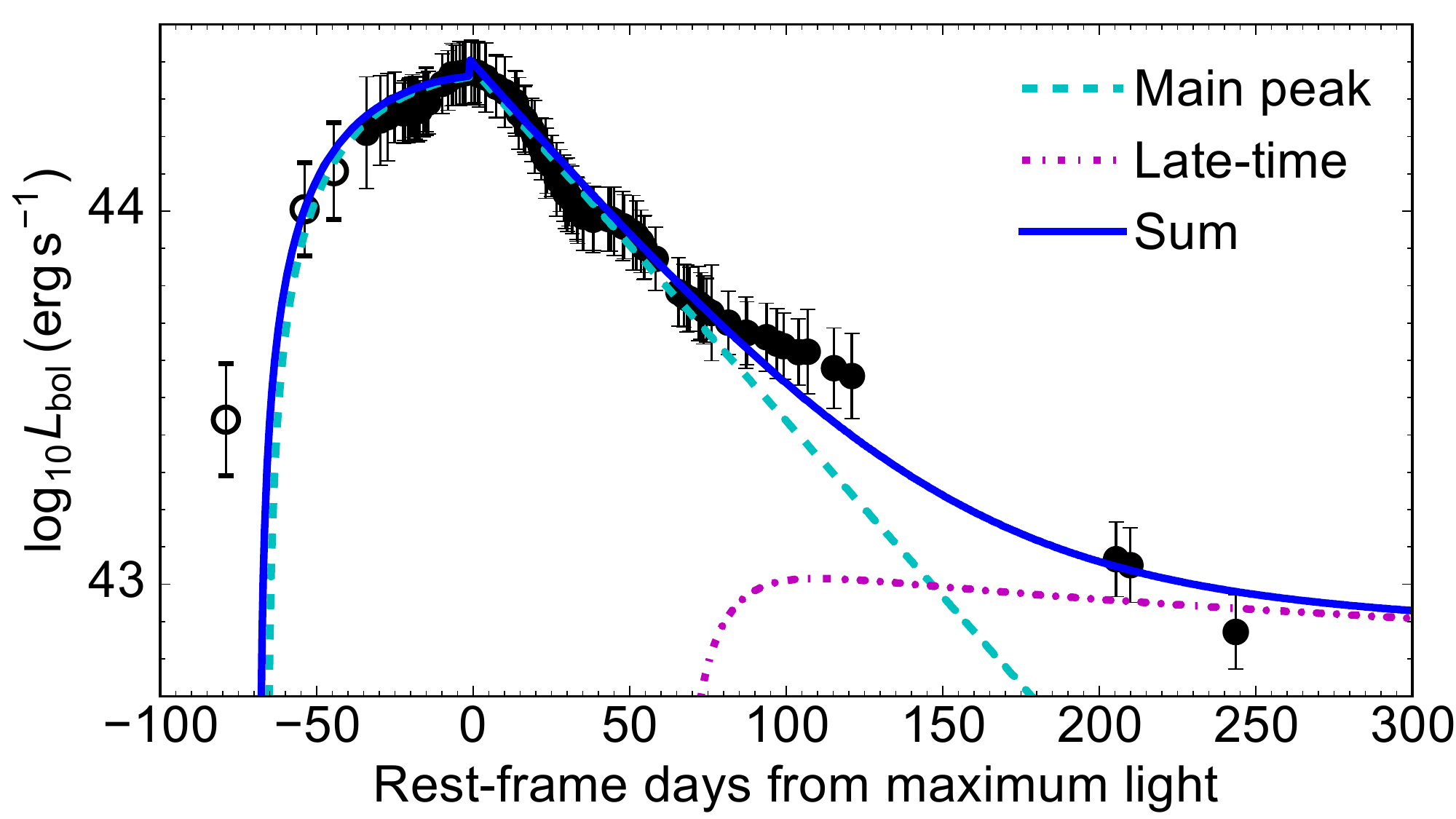}
\figcaption{Example of a model with two detached shells of CSM to power the peak and tail phases of the light curve. The parameterised mass-loss rate for the region shell is $\dot{M}/v_{10}=0.2$\,\M\,yr$^{-1}$, while the outer region has $\dot{M}/v_{10}=0.05$\,\M\,yr$^{-1}$. See Table \ref{tab:csm2} for a full description of the model parameters.\label{fig:csm2}}
\end{figure}

\citet{che2011} found that the SN behaviour depended on whether the terminal radius of the wind, $R_w$, was larger or smaller than $R_d$. They predicted that for $R_w > R_d$, a dense shell forms deep in the wind, leading to low velocities, and continued late-time interaction that gives a flattening in the light curve. On the other hand, if $R_w < R_d$, the outermost layers are accelerated by radiation pressure, and there is little power input after maximum light. The authors suggest that SLSNe I are examples of the latter situation, whereas SLSNe IIn correspond to the former. For SN 2015bn, our models give $R_w = 5.6$-6.4$\,\times10^{15}$\,cm, depending on the CSM mass. Thus we are in the regime $R_w \la R_d$, possibly resolving the lack of narrow lines in the spectrum. In this case we do not expect to see a flattening of the light curve, which is indeed borne out by our model fits (though not by the data). Therefore another energy source must power the tail. $R_w \la R_d$ was also found to be the case for SN 2010gx \citep{che2011}; however \citet{chom2011} applied the same analysis to PS1-10ky and PS1-10awh and found $R_w \ga R_d$. Within this approximate framework, the more robust constraint seems to be that $R_w \approx R_d$, which is required for efficient production of radiation \citep{chom2011}. Finally, the \citet{che2011} model allows us to calculate a shock velocity of $\approx7000$\,\kms~assuming an 80\,d rise time based on the observed data.

In order to recover the slow decline observed after +50\,d, we re-ran the fits with \Mni~as an additional free parameter, and found that adding 2-4\,\M~of \Ni~to each model could reproduce the late-time behaviour. This is shown in Figure \ref{fig:csm}. For the wind model, a lower $\rho_{\rm CSM}$ then gave a match the light curve shape around peak, with a corresponding increase in the interaction radius to keep the peak brightness constant (this is equivalent to keeping the pre-explosion mass-loss rate and velocity constant). The interaction models with \Ni-decay give good fits to the light curve, although the values of \Mni~are extremely large -- normal core-collapse SNe have \Mni\,$\approx0.1$\,\M, and even LGRB-SNe synthesize \Mni\,$\la0.5$\,\M~\citep[e.g.~see the recent review by][]{cano2016}.

The alternative scenario sees the shallower tail phase of the light curve powered by further interaction with mass-loss at a larger radius. This is appealing, since we do not need to invoke both a large CSM mass and a large \Ni~mass. We construct such a model by first fitting an interaction model to the data at $t<60$\,d, and then subtracting this from the whole light curve. We then fit the residuals at $t>60$\,d as an interaction with additional CSM. Most of the model parameters for this second interaction are fixed by the fit parameters of the first interaction. In the following discussion, we use the subscripts 1 and 2 to refer to the fits around peak and to the late-time residuals, respectively. The fixed parameters are $M_{\rm ej,2}= M_{\rm ej,1}+M_{\rm CSM,1}$; $E_{\rm k,2} = E_{\rm k,1} - E_{\rm rad,1}$ (where $E_{\rm rad}$ is the total energy lost through radiation, $\approx2\times10^{51}$\,erg); $R_{\rm int,2} = R_{\rm CSM,1} + \sqrt{(10E_{\rm k,2})/(3M_{\rm ej,2})} \Delta t$ (where $R_{\rm CSM}$ is the outer radius of the CSM, and $\Delta t$ is the time between shock breakout from the inner CSM and the second interaction -- this is approximately equal to the time from maximum light until the beginning of the tail phase, i.e.~$\approx60$\,d). Therefore the only free parameters are $M_{\rm CSM,2}$ and $\rho_{0,2}$. For a given density, the CSM mass only affects the duration of the second interaction but not the luminosity, thus this parameter is only weakly constrained by the data. However, we find that we require $M_{\rm CSM,2}\ga 10$\,\M.

The best-fitting model is shown in Figure \ref{fig:csm2}. While the late-time fit is not as good as the \Ni-powered model in Figure \ref{fig:csm}, we caution that we are now pushing the simplified analytic CSM model to its limits, and this fit is intended only to give an order-of-magnitude estimate for the mass-loss rate required to give the shallower tail phase. In this case we find $\dot{M}/v_{10} \approx 0.05$\,\M\,yr$^{-1}$. (see Table \ref{tab:csm2} for a full description of the model parameters). As this model is only approximate, it is possible that a continuous CSM could also fit the data, for example a dense inner shell, attached to an outer wind component. Regardless, the luminosity at late times seems to require a mass-loss rate $\dot{M}/v_{10} \ga 0.01$\,\M\,yr$^{-1}$. However, this model raises questions about the spectroscopic evolution. While it is conceivable that the single-shell model could avoid showing narrow lines in certain circumstances, it seems likely that the outer CSM, \emph{which remains unshocked around light curve maximum}, would imprint low-velocity features on the observed spectrum. This is essentially the same conclusion reached by \citet{che2011}, as discussed above. One possible way to avoid seeing narrow lines at early times could be if this outer CSM were a face-on disk or torus.

\subsection{Radioactive decay models}

In section \ref{sec:bol}, it was observed that the late-time decline rates of SN 2015bn and a number of other SLSNe I look very similar to the radioactive decay of \Co, the daughter nucleus of \Ni. We also saw above that introducing a few solar masses of \Ni~to the interaction models gave a good fit to the tail luminosity. However, powering this phase by \Co~alone would imply that $\approx15$\,\M~of \Ni~were synthesised in the explosion, using the scaling relations of \citet{ham2003}. This would require the pair-instability explosion of a stellar core of over 110\,\M, according to the models of \citet{kas2011}. Such a scenario was ruled out for PTF12dam by \citet{nic2013}, based on the rise time and blue spectra. In addition to the spectroscopic and photometric similarity between these two events, the undulations in the light curve of SN 2015bn seem to be incompatible with a PISN, as such a huge ejected mass would be expected to wash out any undulating structure in the light curve due to the very long diffusion time \citep[e.g.][]{fra2013}. Thus it is difficult to conceive of a plausible model in which the tail phase of SN 2015bn is predominantly powered by \Co-decay. However, the conspiracy among an increasing number of SLSNe I to match this decline rate is certainly intriguing, and we therefore examine radioactively powered models in some detail here.

\begin{figure}
\centering
\includegraphics[width=8.7cm,angle=0]{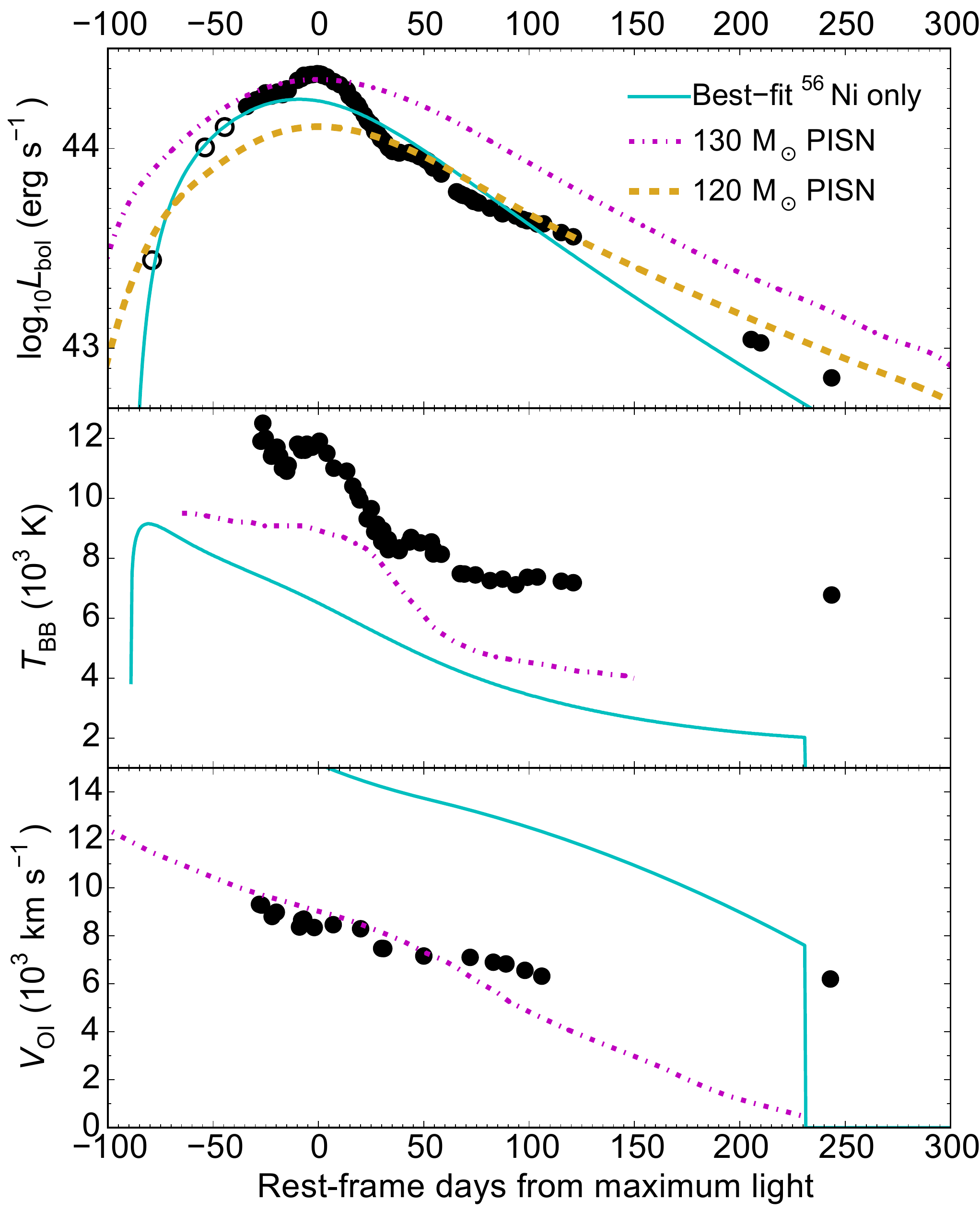}
\figcaption{\Ni-decay-powered models for SN 2015bn. Top: Modelling the bolometric luminosity. We fit a diffusion model with \Mej\,$=26.8$\,\M, \Mni\,$=25.8$\,\M, \E\,$=3\times10^{52}$\,erg. Also shown are hydrogen-free PISN simulations from \citet{kas2011} Middle: Comparisons of model temperatures with the blackbody fits. Bottom: Comparisons of model velocities with those measured from the O\,\textsc{i}\,$\lambda$7774 line.\label{fig:ni}}
\end{figure}

We fitted the light curve of SN 2015bn with a \Ni-powered model, finding that reasonable fits were only possible with \Mni\,$\approx$\,\Mej. The best-fit model shown in Figure \ref{fig:ni} has \Mej\,$\approx27.0$\,\M, \Mni\,$\approx26$\,\M. Note that only one fit is shown, as the best-fitting models including/excluding the first data point are almost identical. In order to decrease the diffusion time to fit the initial decline from peak, a large explosion energy of \E\,$=3\times10^{52}$\,erg was required. This in turn reduced the $\gamma$-ray trapping efficiency at late times, leading to a final decline rate that is faster than \Co-decay. The model is unconvincing both in terms of the unrealistic \Ni-fraction in the ejecta and a poor overall fit to the light curve. Moreover, this model predicts a low photospheric temperature, and a very high velocity. The high velocity is due to the large value of \E. In general, this fit is inferior to the magnetar and interaction models.

\begin{figure}
\centering
\includegraphics[width=8.7cm,angle=0]{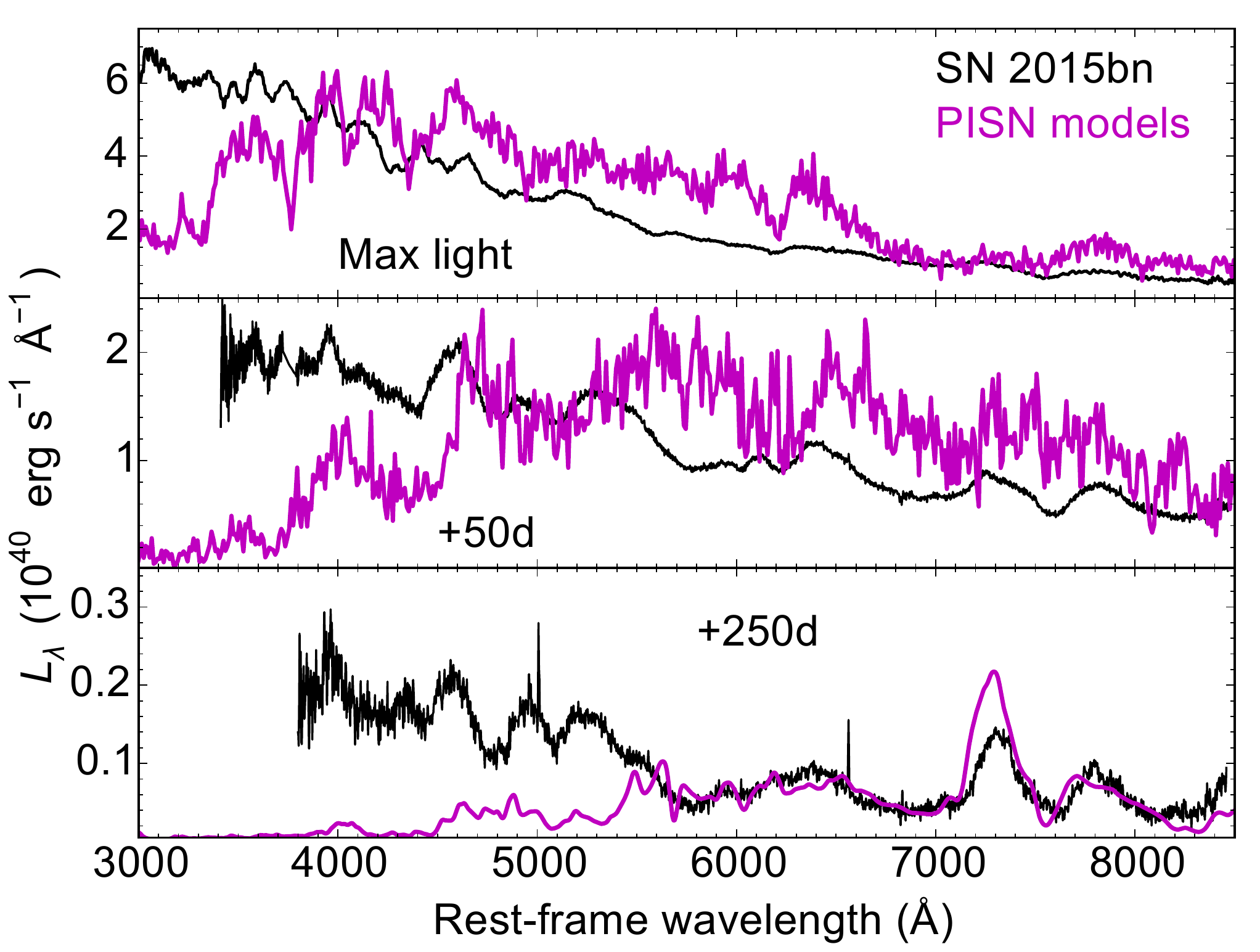}
\figcaption{Comparison of selected spectra with PISN models from \citet[top, middle]{kas2011} and \citet[bottom]{jer2016}. The \citet{jer2016} model was calculated at 400\,d after explosion; we scale the luminosity between 6000-7000\,\AA~to match the observed flux in our spectrum of SN 2015bn at 325\,d. The PISN suffers from extreme iron line-blanketing at blue wavelengths, and overall shows much redder colours than SN 2015bn. However, the nebular model (bottom) gives a good match at $\lambda\ga6000$\,\AA.\label{fig:neb}}
\end{figure}

We also compare to hydrodynamic simulations of PISNe calculated by \citet{kas2011}. The peak luminosity is well-matched by the explosion of a 130\,\M~stripped-envelope PISN, synthesising 40\,\M~of \Ni. It also matches the decline rate at times $\ga50$-100\,d, but the faster decline observed in the data from 0-50\,d is difficult to reconcile with this model. The rise time of this model is also too long relative to the data, even including the earliest point. The absolute luminosity of SN 2015bn at this phase is in better agreement with the 120\,\M~model, which has 24\,\M~of \Ni, similar to our own diffusion fit (the decay energy in PISN models remains fully trapped at late times due to the large ejected mass). However, this lower-mass model severely underestimates the peak luminosity and does not reproduce the behaviour for times earlier than +50\,d. One possibility that could perhaps match the light curve shape would be to add strong CSM interaction around peak to the 120\,\M~PISN model. However, this would require invoking an extreme CSM mass as well as the extreme ejecta and \Ni~masses, and a detailed investigation of such a model is beyond the scope of this paper. The temperature in the more massive (hotter) 130\,\M~model is cooler than SN 2015bn by about 3000\,K at all times. The velocity in this model gives a fairly good match until about +50\,d, but the rapid decline as the photosphere recedes looks quite different from the slow decline exhibited by SN 2015bn.

In Figure \ref{fig:neb}, we compare the +243\,d spectrum of SN 2015bn to the model spectrum of a 130\,\M~hydrogen-poor PISN, at a similar phase, from \citet{jer2016}. This model differs from the nebular PISN model shown by \citet{gal2009}, because the \citeauthor{jer2016} model is computed using realistic ejecta models from \citet{heg2002}, whereas the \citeauthor{gal2009} models used tuned parameters to fit the spectrum of SN 2007bi, resulting in relative abundances that did not match PISN explosion models \citep{nic2013}. The new models also include line blocking, which is extensive in these high-density ejecta and damps UV/blue emission. The model shown gives a good match to SN 2015bn between $\sim6000$-8000\,\AA~(oxygen and calcium). We note that strong oxygen and calcium lines in the nebular phase are not unique to PISNe, but are generically seen in core-collapse SNe \citep[e.g.][]{fil1997}. The matches to the line widths and overall ratios are good. This match at red wavelengths is much better than was seen in similar comparisons between the PISN models and SN 2007bi and PTF12dam \citep{jer2016}. However, there is a huge discrepancy between the model and data below $\approx5500$\,\AA. In the PISN model, line-blanketing by the large iron mass suppresses the flux in the blue, whereas the late spectrum of SN 2015bn retains a pronounced blue continuum that is brighter than the model by an order of magnitude. This discrepancy has previously been pointed out for other slow-declining SLSNe by \citet{des2012,nic2013,jer2016}. We note that line formation at bluer wavelengths may be more complex compared to in the red, but it is unclear what model deficiencies could account for the scale of this difference between model and data. Despite the similarity in the light curve slope and spectrum above 6000\,\AA, there remain many important differences between the observational data and models powered exclusively by \Ni.

\subsection{The nature of the undulations}\label{sec:bumps}

Having discussed the possible power sources for the bulk of the light curve, we now turn to interpretations of the shoulder and knee. In both cases, the temperature evolution suggests that an additional heat source is active over these times, as all of the power sources we have considered (\Ni-decay, magnetar, or shock heating from interaction) predict a monotonic decline from maximum. The discussion in this section will focus on the two models that gave the more convincing fits in the previous section (magnetar spin-down and CSM interaction), beginning with interaction.

Similar light curve undulations to those observed in SN 2015bn were also seen in the light curve of SN 2009ip, which is thought to be either a Type IIn SN or a SN imposter \citep{pas2013}. Either way, the luminosity at maximum light was powered by interaction with pre-expelled material \citep[e.g.][]{mau2013,fra2013,smi2013,mar2014,gra2014}. However, \citet{fra2013} note that an interaction between more than a few solar masses of ejecta and CSM (which would certainly be required to match the high luminosity in SN 2015bn) would struggle to match the undulations, as these would tend to be washed out by the long diffusion time. Therefore the most probable way to explain these using CSM interaction is to introduce multiple collisions with additional shells or clumps of material.

All of the aforementioned studies of SN 2009ip found that at least some of the features in the light curve could be explained by successive collisions with mass expelled in previous \emph{observed} outbursts of the progenitor. If we believe that the historical CSS detection of SN 2015bn at $\approx 185$\,d before explosion was an outburst of the progenitor, and that interaction with this material powers the knee or shoulder, the blackbody radii at these epochs (Figure \ref{fig:bol}) would imply that the pre-expelled material had a velocity of $\sim3000$\,\kms. This would be compatible with pulsational pair-instability mass ejections \citep{woo2007}.

However, the blackbody radius in SN 2015bn immediately begins to decrease at the onset of the knee phase. This may be difficult to explain in an interaction model, as the collision with a massive shell should initially be optically thick \citep[e.g.][]{mor2012}. In SNe powered by interaction, the radius of the emitting surface generally shows an increase as the shock propagates through the CSM \citep[e.g.][]{fas2000,mar2014}. This may not be too problematic, as only a relatively small amount of CSM is needed to power the undulations. For example taking the knee feature, we find that SN 2015bn is over-luminous by $2\times10^{43}$\,\ergs~for 15\,d, compared to a smooth decline. Assuming a velocity $v=7000$\,\kms~from the spectrum at this phase, we can use the common scaling relation \citep{qui2007,smi2007} $L\approx \frac{1}{2} M_{\rm CSM} v^2 / t_{\rm rise}$, giving $M_{\rm CSM,knee} \sim 0.05$\,\M. Such a low CSM mass is probably not compatible with pulsational-PISN shells; the final pulses before explosion in all of the \citet{woo2007} models eject at least $\ga1$\,\M. The low mass, relative to the ejecta/CSM masses in any of our model fits, might help to explain why we see little change in the continuum (with the temperature approximately constant at this phase), but this modest CSM mass may be expected to result in narrow emission lines, as the optical depth will be low compared to the case of massive CSM. However, as for other SLSNe I, no narrow lines are observed in the spectrum of SN 2015bn.

Next we consider the magnetar model. Before investigating any magnetar-specific means of producing the undulations in the light curve, it should be pointed out that the low-mass CSM interaction described above could equally apply to a light curve that was primarily powered by magnetar spin-down. Nevertheless, the magnetar scenario does have other means of producing fluctuations in the light curve. One such mechanism is the magnetar-powered shock breakout described by \citet{kas2015}. In this model, the central overpressure from the magnetar wind drives a second shock through pre-exploded ejecta, which breaks out at large radius and hence can give a bright optical display. However, their equation 26 shows that this breakout should happen within 20\,d of explosion for any sensible combination of ejecta mass, explosion energy and magnetar parameters. Moreover, the breakout would not be noticeable in the light curve if it occurred near maximum brightness. Therefore this model is only applicable to early-time bumps, and not to the shoulder or knee in SN 2015bn.

\citet{met2014} found that for a magnetar-powered SN, the hard radiation field should drive ionization fronts outwards from the base of the ejecta. Being the most abundant element in Type Ic SN ejecta, oxygen was considered to be the dominant source of electrons at early times. Their model predicted that when the O\,\textsc{ii} layer reached the ejecta surface, the opacity to UV photons would decrease, leading to a UV-breakout. This occurred on a timescale of tens of days after explosion, and this timescale increases with ejecta mass. Comparing with the properties of SN 2015bn during the shoulder phase, this model seems to consistently explain its timing, the fact that it is only apparent in the blue and UV bands, and the presence of O\,\textsc{ii} lines in the spectrum at this time. The fast decline in the colour temperature during this phase also indicates that it could be a form of breakout event.

A key prediction of this model is that the UV breakout should be followed by X-ray breakout tens to hundreds of days later. This occurs when the O\,\textsc{iv} ionization front reaches the surface. For the model computed by \citeauthor{met2014} with a 2\,ms spin period (needed to power the peak optical luminosity of SN 2015bn), X-ray breakout occurred 240\,d after explosion. However, that model assumed an ejecta mass of only 3\,\M, a factor of 3-5 lower than the masses inferred from our fits to SN 2015bn (Table \ref{tab:mag}). The breakout timescale is a strong function of ejecta mass, and thus X-ray breakout is not expected to occur over the timescale of our observations. However, we are carrying out an X-ray monitoring campaign for SN 2015bn to look for signatures of ionization breakout; the results will be published in a future study by R.~Margutti et al.

During the knee plateau from +30-50\,d, the photospheric radius starts to decrease, and equivalent widths of spectral lines from heavier species such as iron and silicon increase. This seems to indicate that the photosphere is beginning to undergo significant recession into the ejecta. The velocity evolution slows at around this time, suggesting that the velocity profile is relatively flat within a certain radius. \citet{kas2010} predicted that the magnetar wind would sweep up most of the ejecta into a dense shell with uniform velocity. Moreover, they show that the temperature jumps sharply at the edge of this shell. The knee, and the associated increase in temperature and flattening in velocity, could therefore be explained by the photosphere crossing into this hotter region as it recedes. If SLSNe I are generally powered by magnetars, one may wonder why this distinctive behaviour is not seen more often. The answer could be that, in lower mass events, the magnetar wind sweeps up essentially \emph{all} of the ejecta, such that there is no discontinuity. Alternatively, it could simply be that most SLSN light curves do not have sufficient temporal sampling for us to catch variations on this timescale.

Another possibility to be considered is that one or both of the undulations are powered by recombination, similar to the plateaus in Type IIP SNe. However, the high temperature and the spectra clearly rule out hydrogen recombination. 
One plausible candidate that could be consistent with the shoulder is oxygen recombination. Oxygen in SNe is expected to be mostly singly-ionized at temperatures between 12000-15000\,K, and neutral below this \citep{hat1999,qui2013}. SN 2015bn is seen to drop below $\sim12000$\,K at precisely this phase. The early spectroscopic evolution may also support this: the O\,\textsc{ii} lines in SN 2015bn are weak at $\sim$20\,d pre-peak compared with those in PTF12dam and other SLSNe at similar phases (see Figures \ref{fig:slow} and \ref{fig:fast}). Instead, the early spectrum closely matches PTF12dam at around maximum light, when these lines are close to disappearing. This could suggest that  O\,\textsc{ii} is being depleted by recombination. The photospheric recession during the knee phase may also point towards recombination. However, it is unclear in this scenario why the features would be most pronounced in the UV bands, or which ion could recombine to power the knee ($T_{\rm BB}\approx 8500$\,K).

%\begin{figure}
%\centering
%\includegraphics[width=8.7cm,angle=0]{lomb.pdf}
%\figcaption{Top: the bolometric light curve of SN 2015bn from $-$30 to +120 days, after subtracting polynomial fits (coloured lines) and a magnetar fit (black line). Middle: Lomb-Scargle periodogram \citep{lomb1976,sca1982} calculated for each of the polynomial-detrended light curves. Dashed lines give the 3$\sigma$ contours calculated by Monte-Carlo simulation. Bottom: same as above, but for the magnetar-detrended light curve. In all cases, we see significant power between 40-60\,d. The peak at $\sim100$ days reflects the timescale of the overall broad light curve decline, and disappears with better detrending.}
%\label{fig:ls}
%\end{figure}

Undulations or plateaus in the light curve evolution could also in principle
result from an abrupt change in the continuum optical/UV opacity, as might
arise due to the sudden emergence of an ionization front through the
ejecta \citep{met2014}.  If the ejecta remain
largely neutral at early times, the optical opacity will be due primarily to bound-bound
transitions and could be relatively modest ($\kappa \lesssim 0.05$ cm$^{2}$
g$^{-1}$, depending on the composition).  However, once the ejecta become
ionized and the number of free electrons increases, the resulting electron
scattering opacity would come to dominate, increasing $\kappa$ by a factor
of several.  This sudden rise in opacity could slightly delay the escape
of radiation from the magnetar nebula or CSM interaction shocks,
producing a plateau or lull in the light curve decline.  Exploring this
scenario further would require a detailed radiative transfer
calculation, which is beyond the scope of this paper.

\begin{figure}
\centering
\includegraphics[width=8.7cm,angle=0]{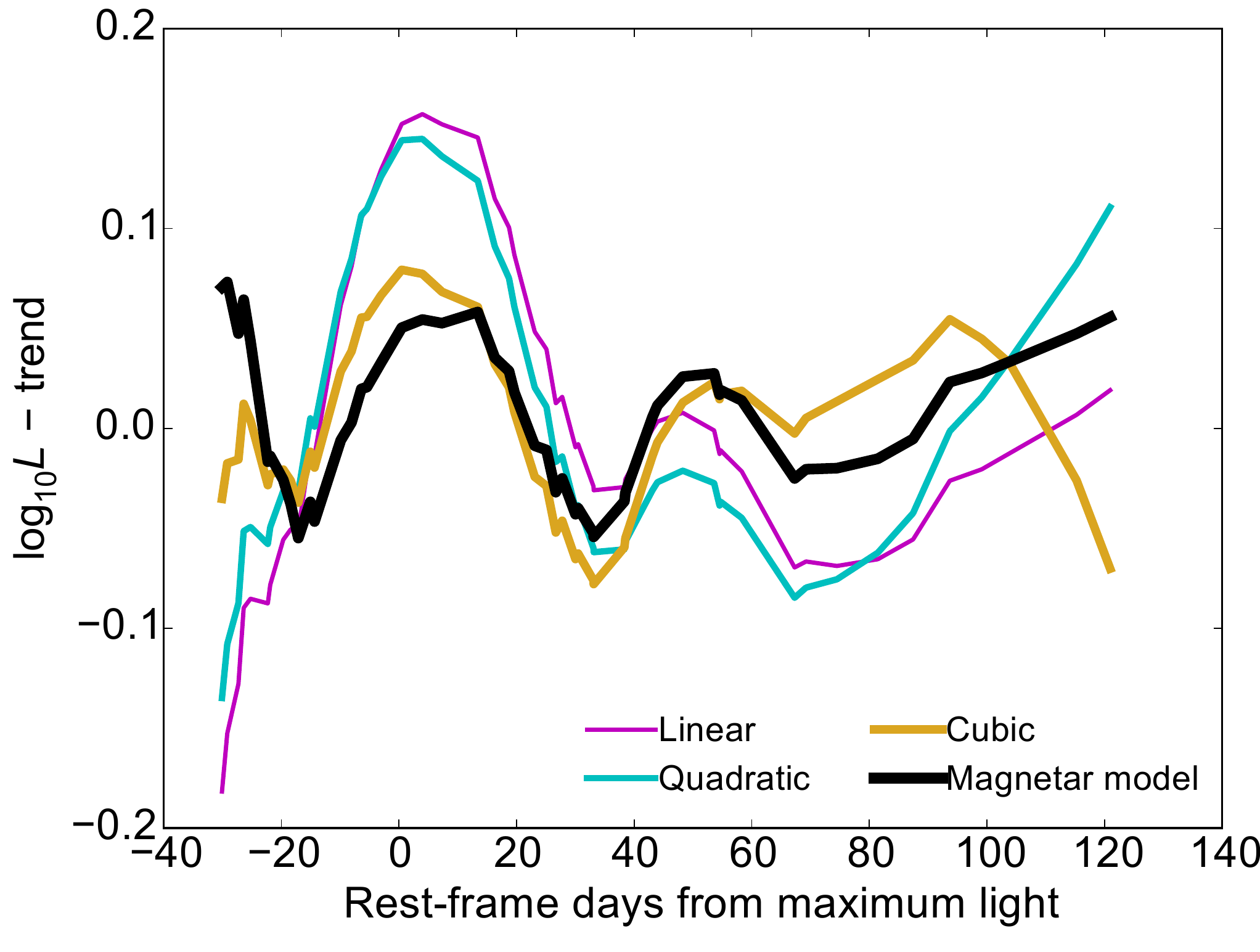}
\figcaption{The bolometric light curve of SN 2015bn from $-$30 to +120 days, after subtracting polynomial fits (coloured lines) and a magnetar fit (black line). The residuals (undulations) occur on timescales of 30-50\,d.\label{fig:ls}}
\end{figure}

To see the undulations more clearly, and to look for any hints of periodicity, we removed the gross structure of the light curve to examine the residuals. We did this by fitting and then subtracting low-order polynomials (we tried first-, second- and third-order). We also experimented with subtracting one of the magnetar-powered fits (solid line in Figure \ref{fig:mag}). Figure \ref{fig:ls} shows the bolometric light curve after subtraction of the various fits. The detrended data are seen to exhibit variation on a timescale of $\approx30$-50\,d. A dominant timescale of $\la50$\,d would be inconsistent with a very massive model, such as a PISN. However, it is fully consistent with the diffusion timescale in an ejecta of $\approx7$-15\,\M~\citep{arn1982,nic2015b}, which is similar to what we infer from our magnetar fits. Thus the undulations seem to be compatible with variations in the velocity/density structure in ejecta similar to that in our magnetar models.

If the undulations are really periodic, this could indicate that they are caused by interaction with a spiral density perturbation, perhaps caused by binary interaction \citep{fra2013}. Evidence of close binary interaction would certainly be an intriguing part of the picture for understanding SLSN progenitors. Unfortunately, the data do not span a sufficiently long time interval ($\Delta t \approx 150$\,d) to reliably pick out periodicity on a timescale on this order ($t_0 \approx 50$\,d\,$\approx \Delta t/3$). For example, \citet{mart2014} applied a periodogram analysis to SN 2009ip, but ignored peaks in the power spectrum with periods $\ga \Delta t/3$. Although the undulations in the light curve of SN 2015bn are an important clue to the nature of this explosion, sampling over much longer timescales will be needed to robustly test for periodicity or the presence of a dominant timescale.

\section{Radio non-detections}\label{sec:radio}

We observed SN 2015bn with the Karl G.~Jansky Very Large Array \citep[VLA;][]{per2011} on 2015 Dec 11 \citep{ale2016}. This date is 238\,d after optical maximum, and approximately 320-335\,d after explosion if we assume a similar light curve morphology to LSQ14bdq. This epoch is useful for discriminating between a number of competing models, especially as interaction models are expected to be optically thick around 5-20\,GHz before optical maximum owing to synchrotron self-absorption \citep{che1998}, and particularly free-free absorption if the surrounding medium is very dense \citep{che1982}.

Observations were carried out while the VLA was in its most compact configuration (D configuration, where all antennae are within $\approx 0.6$\,km of the array centre), and obtained in two frequency bands: C-band (mean frequency of 5.5\,GHz) and K-band (22\,GHz). Each band has a total bandwidth of 2\,GHz (split up into two sub-bands), and we observed for 30\,min on-source in each. No source was detected at the position of SN 2015bn in either band. The C band observation was degraded at low frequencies by a bright radio source $\approx 6$' from the SN coordinates, so we concentrate on the high-frequency sub-band (mean frequency 7.4\,GHz). The K band data were unaffected by this contaminating source. After self-calibration, we derived the following 3$\sigma$ upper limits on the radio emission from SN 2015bn: $F_{\rm 7.4GHz}<75\,\mu$Jy; $F_{\rm 22GHz}<40\,\mu$Jy.

\subsection{Comparison to gamma-ray bursts}

We first tested whether these limits can constrain the presence of an off-axis $\gamma$-ray burst associated with SN 2015bn. \citet{lun2014} found that SLSNe I and LGRBs tend to occur in similar environments, though recently \citet{ang2016} have proposed that the only similarity is low metallicity. In any case, both types of explosions display similar spectra lacking in hydrogen and helium \citep[e.g.][]{pas2010}, and both may be powered by central engines. Recently, \citet{gre2015} presented a luminous ($M\approx-20$) SN associated with an ultra-long GRB, but it was not clear from the relatively featureless spectrum whether this was a `classical' SLSN I. The presence of a radio afterglow from a confirmed SLSN would firmly establish the SLSN-GRB connection; alternatively, a lack of radio emission could instead indicate that either no stable jet forms, or it does not break out of the massive progenitor, and thus does not accelerate the outer ejecta to relativistic velocity.

\begin{figure}
\centering
\includegraphics[width=8.7cm,angle=0]{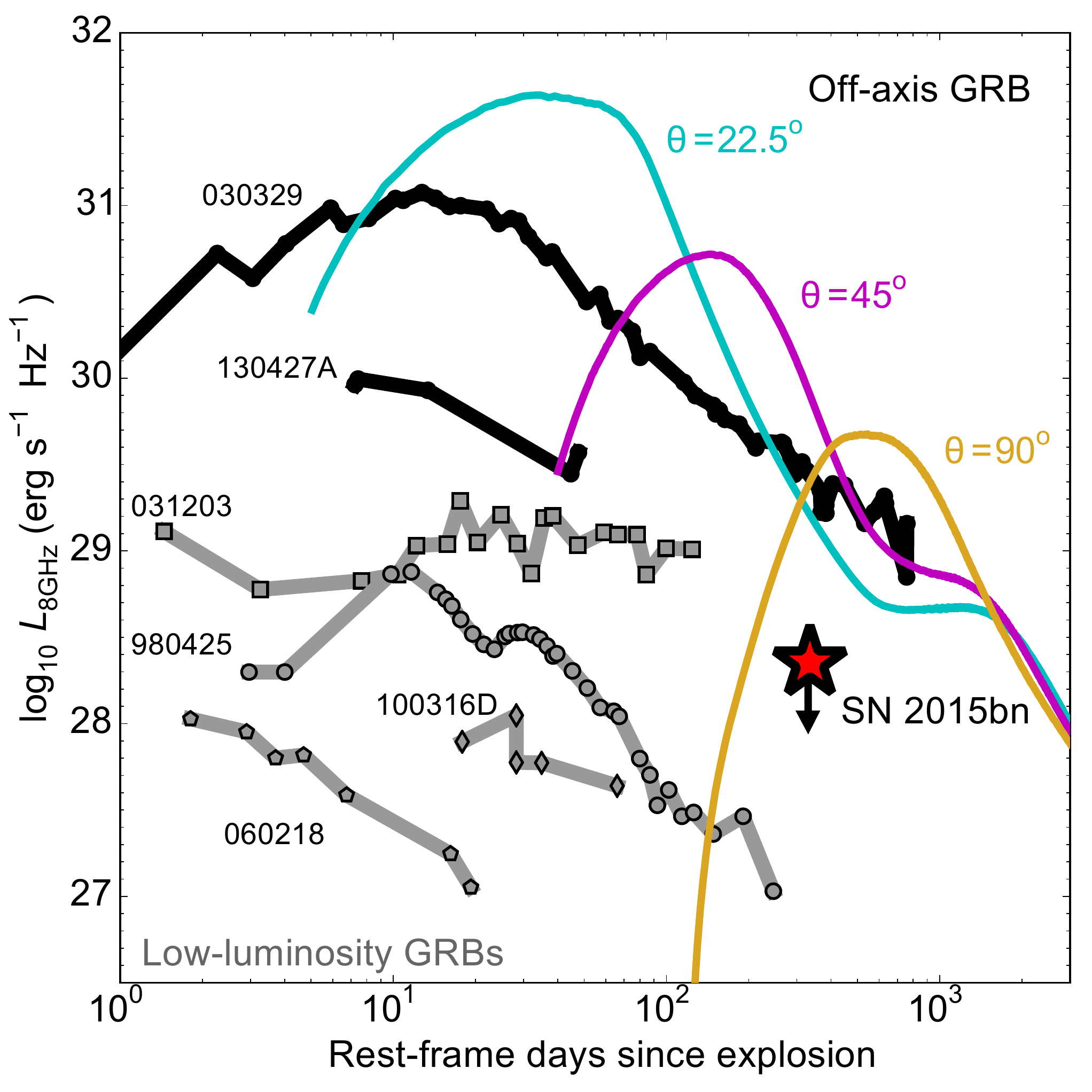}
\figcaption{Radio non-detection of SN 2015bn compared to the predictions of off-axis LGRB models with $E_{\rm k,jets}=2 \times 10^{51}$\,erg and a particle density of 1\,cm$^{-3}$. The limit on the radio emission from SN 2015bn robustly rules out a healthy jet. A low-luminosity GRB cannot be excluded. Comparison sample: \citet{mar2013}, and references therein.\label{fig:grb}}
\end{figure}

We compare models for LGRB-SNe at a variety of viewing angles in Figure \ref{fig:grb}. The simulations are described by \citet{van2010}, and assume typical LGRB parameters: total kinetic energy in the two jets of $E_{\rm k,jets}=2 \times 10^{51}$\,erg; a homogeneous particle density of $n=1$\,cm$^{-3}$ in the surrounding medium; equal fractions of the total energy density in magnetic fields and relativistic electrons ($\epsilon_B = \epsilon_e = 0.1$); jet half-opening angle $\theta_{\rm jet} = 0.2$\,rad; isotropic-equivalent energy $E_{\rm iso} = 10^{53}$\,erg; and a power-law slope of $p=2.5$ for the accelerated electron distribution. For explosions viewed further from the jet axis, the radio emission is weaker, and peaks later. The models are calculated at a frequency of 8.5\,GHz, which gives a good match to the observer-frame 7.4\,GHz limit for SN 2015bn.

Our deep radio limit for SN 2015bn rules out even the most off-axis model, making SN 2015bn the first SLSN for which we can robustly exclude a luminous LGRB. However, we also compare to a sample of observed `low-luminosity' (\textit{ll}) GRBs, for which the emission does not seem to be strongly collimated as in LGRBs. \textit{ll}-GRBs may form a separate population from high-luminosity LGRBs, and dominate the volumetric GRB rate locally \citep{gue2007,lia2007}. As demonstrated by the figure, we cannot rule out a \textit{ll}-GRB associated with SN 2015bn, and can only exclude the luminous and highly-collimated jets of a `healthy' LGRB. Similarly deep radio observations of the next nearby SLSN I before or around optical maximum should be able to constrain the presence of a \textit{ll}-GRB.

\subsection{Comparison to models for the optical luminosity}

Radio observations can also be useful in breaking the degeneracies between the competing models for the optical light curve. While the magnetar and interaction models give similarly good fits to the optical data, their predicted radio signatures are very different. In general, radio synchrotron emission is generated by electrons accelerated to relativistic velocities at the forward shock \citep{che1982,wei2002,che2006}.

If SLSNe I are otherwise fairly normal Type Ic SNe that are reheated by a magnetar engine, we expect the radio properties to resemble those of typical Type Ic SNe. For those objects, the CSM density is relatively low, with radio light curves implying mass-loss rates of around $10^{-7}$-$10^{-5}$\,\M\,yr$^{-1}$ \citep[if the wind velocity is $10^3$\,\kms; e.g.][]{ber2002,sod2007,dro2015}. The interaction is dominated by the outermost, fastest ejecta ($v\ga0.15c$), which carries only $\sim10^{-5}$ of the total kinetic energy \citep{mat1999}, while the dominant source of absorption is synchrotron self-absorption \citep{che1998}. Most normal SNe Ic (i.e.~no LGRB) that have been detected in the radio peaked at $L_{\rm 8GHz} \la10^{28}$\,erg\,s$^{-1}$\,Hz$^{-1}$, on a timescale of tens of days \citep{ber2003,sod2010,mar2014b}, though there are some exceptions \citep[e.g.][]{cor2014,kam2016}. Generally, the predicted luminosity at $\sim300$\,d is well below our limit for SN 2015bn. Radio emission from the magnetar itself is not expected, as Galactic magnetars are not detected as radio sources \citep[e.g.][]{gae2001}, and any signal that was emitted would be obscured by free-free absorption in the ionised inner regions of the ejecta \citep{chom2011}. In conclusion, our radio observation is not constraining for the magnetar model.

For the alternative model in which the optical luminosity is powered by interaction with CSM, the masses and densities involved are much higher; our best-fit wind model has a mass-loss rate higher than that inferred for normal SNe Ic by several orders of magnitude (if normalized to the same wind velocity). In this regime, all of the ejecta interacts with the massive CSM, and therefore the kinetic energy involved is the total energy of the ejecta, $\approx {\rm few} \times 10^{51}$\,erg (section \ref{sec:mod}). 
For the very dense mass-loss in our CSM fits, we expect that the radio emission will be obscured by free-free absorption until well after maximum light \citep[for example radio emission in Type II SNe -- which have mass-loss rates intermediate between Type Ic SNe and our models here -- peaks at $\ga100$\,d after explosion;][]{wei2002}.

For each of the models shown in Figure \ref{fig:csm}, the forward shock reaches the outer edge of the CSM at the time of optical maximum; the declining light curve is then powered by diffusion of the stored shock energy. This is similar to the statement that $R_w \approx R_d$ (section \ref{sec:csm}). \citet{nic2014} fitted a number of SLSN light curves with this interaction model, and also found that good fits required shock breakout at around maximum light. In this scenario, we might expect to see radio emission around optical maximum, when the shock is near the outer surface of the CSM. Once the shock expands past the CSM, there is no further particle acceleration. Adiabatic expansion rapidly reduces the energy density in relativistic particles and magnetic fields, and the radio luminosity falls off rapidly \citep[$L_\nu \propto t^{-6}$;][]{chom2011} -- therefore we expect no radio emission if there is no more mass to interact with outside of $R_w$.

However, the late-time light curve of SN 2015bn could indicate continued interaction with CSM further out from the progenitor, for example a tail of lower density material causing the change in decline rate after around 70\,d. In Figure \ref{fig:csm2}, we showed a model that was powered by continued interaction with an extended outer region of CSM that had $\dot{M}/v_{10} = 0.05 $\,\M\,yr$^{-1}$, where $v_{10}$ is the wind velocity normalized to 10\,\kms. Several other SLSNe also show a more gradual decline at late times (Figure \ref{fig:bc}). Radio observations at this phase can help to distinguish whether the slow luminosity decline is indeed powered by continued interaction, rather than \Ni-decay or a magnetar. If interaction is the culprit, as would be confirmed by bright radio emission, we can then use the radio luminosity to probe the mass-loss history of the progenitor.

\begin{figure}
\centering
\includegraphics[width=8.7cm,angle=0]{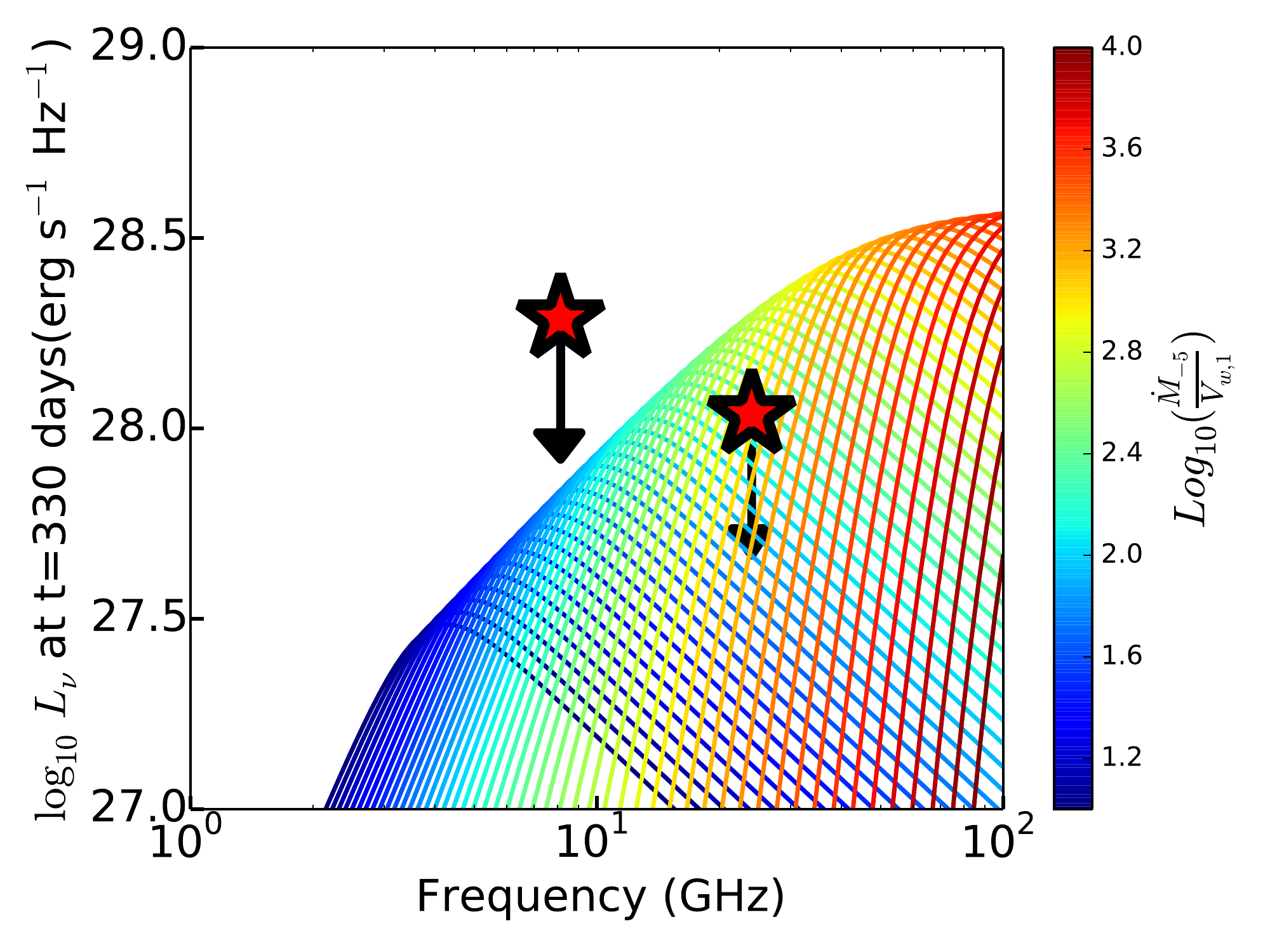}
\figcaption{Radio spectra at the approximate time of our observation of SN 2015bn for interaction models with \E\,$=4 \times 10^{51}$\,erg and $v_{\rm SN} = 8000$\,\kms, and a variety of mass-loss rates, parameterised as $\dot{M}/10^{-5}$\,\M\,yr$^{-1}$ ($v_w/10$\,\kms)$^{-1}$. The low-frequency limit is not constraining, but the high-frequency limit can exclude a range of mass-loss rates.\label{fig:radio_spec}}
\end{figure}

\begin{figure}
\centering
\includegraphics[width=8.7cm,angle=0]{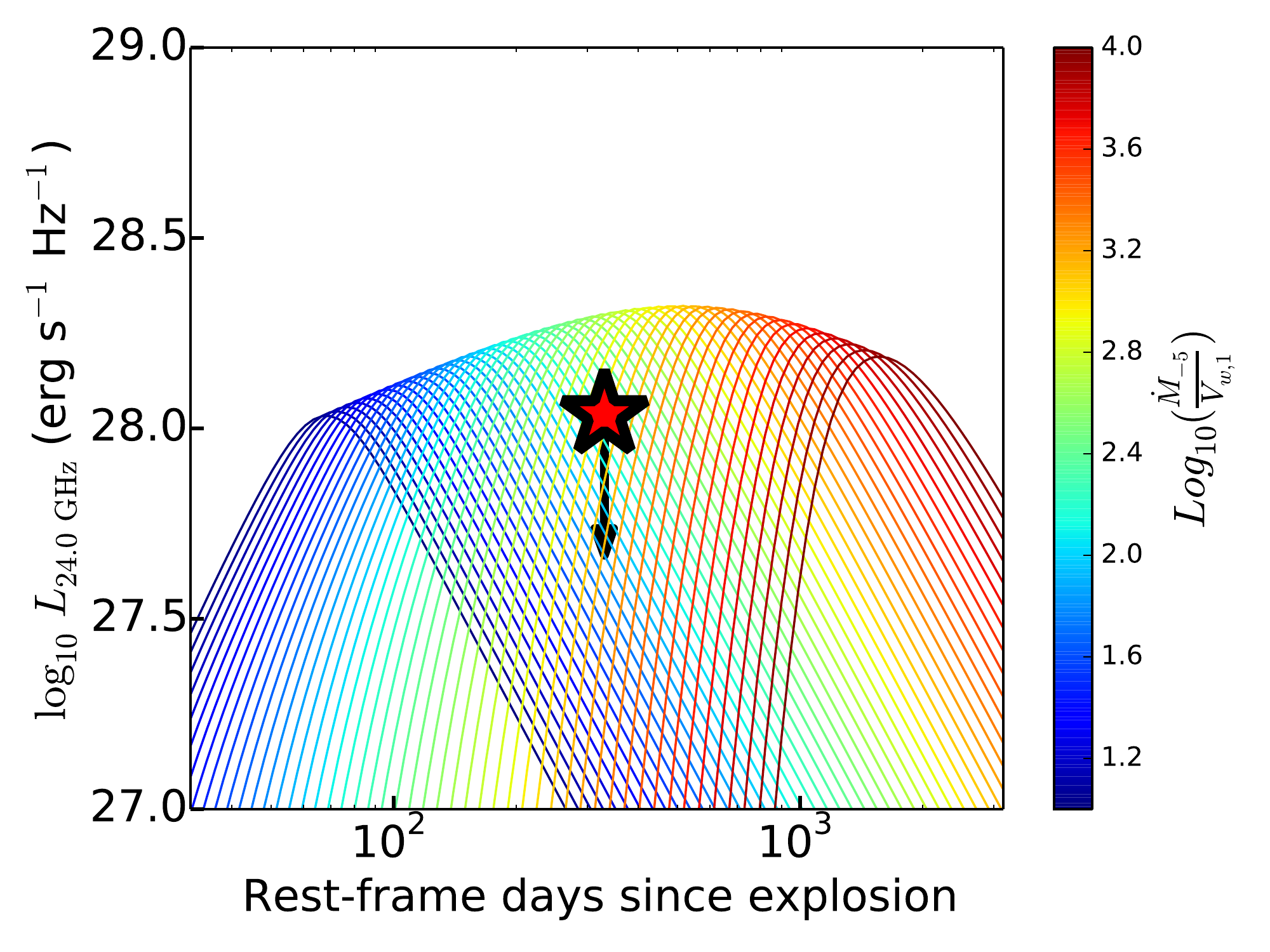}
\figcaption{The radio light curves of interaction models at 24\,GHz. Our limit for SN 2015bn excludes models with mass-loss rates between $10^{-2.7}<\dot{M}<10^{-2.0}$\,\M\,yr$^{-1}$. Models with $\dot{M}\sim10^{-1}$\,\M\,yr$^{-1}$ (as implied by the optical light curve) peak later than our observation, at around 3 years after explosion.\label{fig:radio_lc}}
\end{figure}

We tested this scenario by comparing our deep radio limit to the predictions of interaction-powered models. Based on our estimate from Figure \ref{fig:csm2} (see also Table \ref{tab:csm2}), we assume that the late-time luminosity is powered by continued interaction with a wind density profile located outside of $R_{\rm int}=10^{15}$\,cm. We can safely ignore the properties of any inner, denser regions of CSM, as the radio observations were carried out 200\,d after the transition to the shallower decline rate, indicating the shock should long have left the dense region and will interact only with the outer material. Our model, which includes free-free absorption and synchrotron self-absorption \citep[for further details on the model, see][]{kam2014,kam2016}, conservatively assumes a shock velocity of 5000\,\kms, which is at the lower end of the range we find in our fits to the optical light curve.

The radio spectral energy distributions of these models are shown in Figure \ref{fig:radio_spec}. Comparing to our limits for SN 2015bn, we can see that it is our high-frequency observation that proves to be constraining. Therefore we compare the 22\,GHz observation to the radio light curves in Figure \ref{fig:radio_lc}. These models are calculated at 24\,GHz to give a good match to the rest-frame frequency of the data. As demonstrated by the figure, the observational limit excludes models with mass-loss rates of $10^{-2.7}\la\dot{M}/v_{10}\la10^{-2}$\,\M\,yr$^{-1}$. Using a faster shock velocity does not greatly impact upon these results, but would generate brighter radio light curves that would rule out a slightly wider range in mass-loss rates. The largest mass-loss rate that we can exclude, $\dot{M}/v_{10} = 0.01$\,\M\,yr$^{-1}$, is close to our estimate for that needed to power the late-time optical light curve through continued ejecta-CSM interaction. Models with $\dot{M}/v_{10}\approx 0.1$\,\M\,yr$^{-1}$ reach radio maximum at  $10^3$ days after explosion. Therefore radio observations at $\nu\ga20$\,GHz, carried out over the coming months and years, corresponding to progressively denser CSM, should be able to confirm or rule out the continued-interaction model we used to fit the optical light curve. This is perhaps the first easily testable prediction that can potentially distinguish between interaction- and magnetar-powered models.

Inspection of Figure \ref{fig:radio_spec} shows that for high mass-loss rates ($\dot{M}/v_{10} \sim 0.1 $\,\M\,yr$^{-1}$), the synchrotron flux is expected to peak in the millimeter range, suggesting that future observations of SLSNe should use both radio and millimeter observations to constrain the SED. The Atacama Large Millimeter/sub-millimeter Array (ALMA) should be able to reach the required sensitivity for the next SLSN at $z \approx 0.1$.

\section{Host galaxy}\label{sec:host}

The host of SN 2015bn is visible as a faint, compact blue source in SDSS. This is not surprising, as all SLSNe I to date have been found in low-metallicity dwarf galaxies \citep{nei2011,chen2013,lun2014,chen2014,lel2015}, with the exception of ASASSN-15lh \citep{dong2016}, which is also distinct because of its luminosity. The SDSS magnitudes from Data Release 12 \citep{alam2015} are $u=23.11\pm0.39$, $g=22.30\pm0.10$, $r=22.06\pm0.13$, $i=22.06\pm0.19$, and $z=21.63\pm0.44$. The source is marginally detected by \textit{GALEX} in the NUV band (but not in FUV) at $m_{\rm NUV}=23.62\pm0.59$. Using the SDSS colour transformations of \citet{jor2006}, we find an absolute $B$-band magnitude $M_B\approx -16.0$. This is similar to the faintest SLSN hosts identified by \citet{lun2014} and \citet{lel2015}. SN 2015bn outshines its host by 2-5 magnitudes over the course of the 2016 observing season. Late, deep SN spectra show weak, narrow emission lines of \Ha, \Hb, [O\,\textsc{ii}]\,$\lambda$3727 and [O\,\textsc{iii}] $\lambda\lambda$4959,5007 from the host galaxy, from which we derived the redshift of $z=0.1136$. We measured the line fluxes in the two latest spectra at +106 and +243\,d. The mean of these fluxes gives a Balmer decrement \Ha/\Hb\,$=2.41\pm0.65$, where the large error is due to the difficulty in measuring the very weak \Hb. Comparing this to the theoretical ratio for Case B recombination, \Ha/\Hb\,$=2.86$ \citep[assuming an electron temperature of 10000\,K and density of 100\,cm$^{-3}$;][]{ost1989}, we find that the internal dust extinction in the host, while admittedly quite uncertain, is consistent with our assumed value of $E(B-V)=0$ (see section \ref{sec:rf}).

We measured the offset of SN 2015bn from the host centroid as follows. First we aligned the deep NTT $g$-band image of the SN with the SDSS pre-explosion stack and mapped to a common pixel scale using 10 field stars and the \textsc{iraf} tasks \textsc{geomap} and \textsc{geotran}. The SN is offset by $1.7\pm0.9$ EFOSC2 pixels in the North-East direction, where the error is calculated from the uncertainty in the image alignment. This corresponds to $0\farcs4 \pm 0\farcs2$. The angular scale at the distance of SN 2015bn is 2.1\,kpc$/\arcsec$, resulting in a physical offset of $\la1$\,kpc from the center of the host galaxy. This small offset is similar to the median offset of the SLSN sample from \citet{lun2015}. The half-light radius in the SDSS image is $0\farcs7$, giving a normalized offset for SN 2015bn of $r/r_{50}\approx0.6$, again similar to the values in \citeauthor{lun2015}, as well as to the mean for LGRBs, $\langle r/r_{50}\rangle=0.67$ \citep{bla2016}.

It is common practice in the literature to estimate the metallicity using the strong-line $R_{23}$ method \citep{pag1979,mcg1991,kob1999}, particularly as weaker, temperature-sensitive lines are generally difficult to detect from the faint hosts of SLSNe \citep{chen2014}. This diagnostic uses ratios of [O\,\textsc{ii}]\,$\lambda$3727, [O\,\textsc{iii}]\,$\lambda\lambda$4959,5007 and \Hb. The disadvantage of using $R_{23}$ is that the calibration is double-valued, with a high- and low-metallicity branch. Detection of weak lines can help to break this degeneracy; however no such lines have been detected for the SN 2015bn host at the current depth of our observations. We find that the lower branch gives $12+\log({\rm O/H})=8.05$, while the higher branch gives 8.60. There is an uncertainty of about 0.2\,dex associated with these values due to a noise spike contaminating the [O\,\textsc{ii}]\,$\lambda$3727 line. Nevertheless, these values are much in line with other SLSN I hosts \citep{lun2014,chen2014,lel2015}. We note that when it has been possible to distinguish between the two branches of the $R_{23}$ relation, the lower value is generally favoured for SLSN hosts.

\begin{figure}
\centering
\includegraphics[width=8.7cm,angle=0]{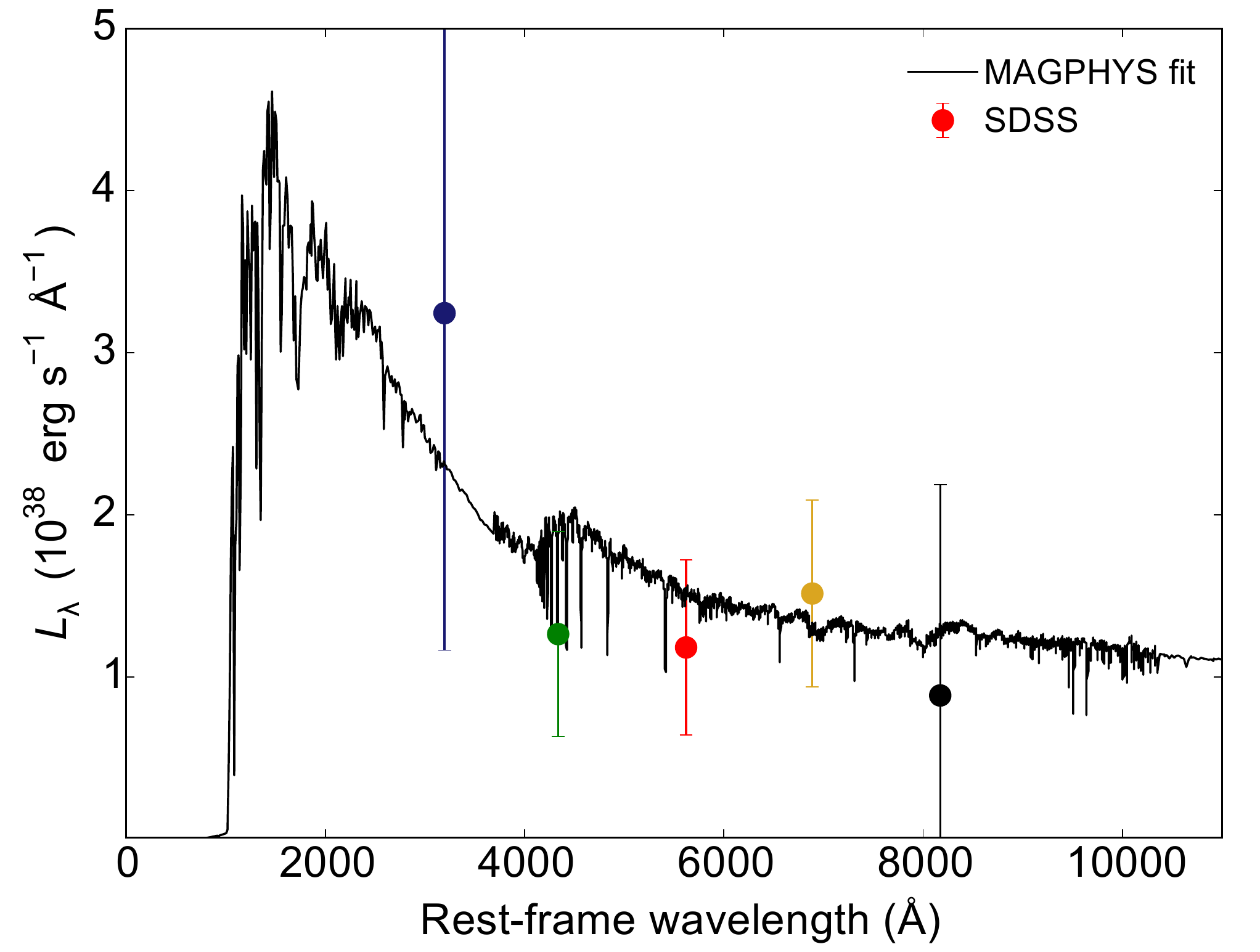}
\figcaption{Best-fitting \textsc{magphys} \citep{dacun2008} model for the host of SN 2015bn. Coloured points are the SDSS \textit{ugriz} (petrosian) magnitudes from Data Release 12 \citep{alam2015}.\label{fig:host}}
\end{figure}

We fitted the SDSS magnitudes\footnote{We neglect the GALEX point, as it is such a marginal detection} of the host galaxy using the stellar population code \textsc{magphys} \citep{dacun2008}. This code employs a library of stellar evolution and population models from \citet{bru2003}, treating stellar and dust contributions to the total luminosity separately. The formal best-fitting model is shown in Figure \ref{fig:host}. \textsc{magphys} also calculates likelihood distributions for key model parameters. The median stellar mass is found to be $4.2\times10^8$\,\M, with a $1\sigma$ range of $1.8\times10^8$-$1.1\times10^9$\,\M. This is similar to many of the SLSN I host galaxies studied by \citet{lun2014} and \citet{lel2015}. The $r$-band light-weighted age of the stellar population is $\sim 1.7^{+2.3}_{-1.2}$\,Gyr, which is older than the majority of SLSN hosts, but consistent with the older galaxies in the \citet{lun2014} sample. However, we note that this is only weakly constrained, due to the large uncertainties on the SDSS photometry. A more robust constraint will only be possible after SN 2015bn has faded. Finally, the model star-formation rate (SFR) of $0.55\pm0.18$\,\M\,yr$^{-1}$ is typical of these galaxies. We find a somewhat lower SFR if we use the measured \Ha~flux and the scaling relation of \citet{ken1998}: SFR\,$=7.9\times 10^{42} (L_{{\rm H}\alpha}$/erg\,s$^{-1}$) = 0.036\,\M\,yr$^{-1}$. In conclusion, the host of SN 2015bn fits the distinctive profile of the usual galaxies that play host to SLSNe I.

\section{Conclusions}\label{sec:conc}

We have presented a detailed study of SN 2015bn, one of the closest and brightest SLSNe I yet discovered. Using a wide range of facilities, we collected a densely sampled light curve in the UV, optical and NIR, along with spectroscopy covering all phases of its evolution. The resulting dataset provides a new benchmark for observations of SLSNe I, motivating detailed study. A deep radio observation several hundred days after explosion gives a physically restrictive limit on both the presence of relativistic jets and extended dense mass-loss.

SN 2015bn is a slowly-declining SLSN, and in many ways a typical one. The late-time decline rate is very similar to PTF12dam and SN 2007bi, as are the persistent blue colours. The SED matches a hot blackbody in the optical, while the UV light is strongly affected by absorption lines at maximum light. The spectroscopic evolution also shows remarkable similarity to the other slow decliners. We follow the evolution of all of the strongest lines in the spectrum until $>100$\,d from maximum light, finding an extremely gradual evolution except for a period between +7-30\,d, when a steep temperature decline triggers a transition from a spectrum dominated by O\,\textsc{ii} and Fe\,\textsc{iii} to one dominated by Fe\,\textsc{ii}, Ca\,\textsc{ii}, Si\,\textsc{ii} and Mg\,\textsc{i}]. Line velocities show a very gradual decline over time. The host galaxy is clearly a faint, blue dwarf -- similar to the hosts of virtually all SLSNe I.

Yet at the same time, SN 2015bn reveals some surprising differences that may offer new insight into the nature of these explosions. The light curve shows a number of distinctive undulations, that seem to indicate a complex density structure -- for example detached CSM shells at large radii, or a magnetar wind and ionisation fronts. We found that these fluctuations were much more pronounced at UV wavelengths.

In the spectrum, we tentatively see evidence that there could be residual hydrogen in some SLSNe I, with some early absorption features possibly matching \Ha~and \Hb. We also see a line in the NIR that could be consistent with He\,\textsc{i}, which has been seen in only one previous SLSN I \citep{ins2013}. However, other possibilities exist to explain the potential H and He features.
We make a systematic comparison of the very late spectra of SLSNe I (fast and slow) with normal and broad-lined SNe Ic, the latter being associated with LGRBs. We find many similarities that suggest a link between these various classes, as well as some differences in the sodium, silicon and calcium lines that may provide clues to the variations in physical conditions between events.

Applying physical models to the light curve, we found that both magnetar and CSM-interaction power sources could reproduce the data. Our best-fit magnetar model suggests that the first detection of SN 2015bn is during the `bump' phase, and gives a good fit to the temperature and velocity evolution. For the interaction model, either a transition to a lower-density outer layer of CSM, or several solar masses of \Ni~in the ejecta, seem to be required to match the tail phase. Purely \Ni-powered models, including PISN models, give a good match only at late times, and the extremely large ejecta mass associated with the pair-instability means that they struggle to reproduce the short-timescale behaviour around maximum light. 

One of the greatest mysteries surrounding SLSNe I is the nature of their apparent connection to LGRBs. This connection is both observational, in their similar spectra (Figure \ref{fig:ic}) and host galaxies  (\citealt{lun2014}; but see \citealt{ang2016}), and theoretical, in the central engine models \citep[e.g.][]{met2015}. Our radio limit, for the first time, explicitly rules out a healthy off-axis LGRB for an observed SLSN. This suggests that, if both classes are powered by engines, the SLSN engine may fail to drive a jet through the stellar envelope. However, we cannot exclude a low-luminosity GRB, so the possibility of a weak jet remains.

On the other hand, if the optical light curve is powered by interaction, model fits can predict the time at which the CSM becomes optically thin to radio emission. The mass-loss rate implied by the late-time bolometric light curve is $\dot{M}/v_{10}\ga10^{-2}$\,\M\,yr$^{-1}$, while our radio limits rule out a wind with $10^{-2.7}\dot{M}/v_{10}\la10^{-2}$\,\M\,yr$^{-1}$. Continued radio observations of SN 2015bn over the next few years will present a unique chance to improve constraints on interaction-powered models for SLSNe and probe the mass-loss histories of their progenitors.

The depth, cadence and wavelength coverage of the data presented here make SN 2015bn the most thoroughly observed SLSN I to date. This dataset should provide a valuable resource for modelling efforts and detailed comparisons with SLSNe discovered in the future. The undulating light curve sub-structure, along with the recent discovery of the fast early bumps, shows the importance of observing SLSNe with a high cadence, despite their light curves being on the whole broader than normal SNe Ic.

We are continuing to observe SN 2015bn as it slowly fades away. Deep spectroscopy in the nebular phase and continued radio follow-up will offer even tighter constraints on the explosion mechanism. This work shows the importance of following SLSNe over a wide range in wavelength, in order to break the degeneracy between magnetar and interaction models and constrain possible progenitors.

\acknowledgments

We thank Nidia Morrell for observations at Magellan. S.J.S.~acknowledges funding from the European Research Council under the European Union's Seventh Framework Programme (FP7/2007-2013)/ERC Grant agreement n$^{\rm o}$ [291222] and STFC grants ST/I001123/1 and ST/L000709/1.
This work is based (in part) on observations collected at the European Organisation for Astronomical Research in the Southern Hemisphere, Chile as part of PESSTO, (the Public ESO Spectroscopic Survey for Transient Objects Survey) ESO program 188.D-3003, 191.D-0935.
The Pan-STARRS1 Surveys (PS1) have been made possible through contributions of the Institute for Astronomy, the University of Hawaii, the Pan-STARRS Project Office, the Max-Planck Society and its participating institutes, the Max Planck Institute for Astronomy, Heidelberg and the Max Planck Institute for Extraterrestrial Physics, Garching, The Johns Hopkins University, Durham University, the University of Edinburgh, Queen's University Belfast, the Harvard-Smithsonian Center for Astrophysics, the Las Cumbres Observatory Global Telescope Network Incorporated, the National Central University of Taiwan, the Space Telescope Science Institute, the National Aeronautics and Space Administration under Grant No.~NNX08AR22G issued through the Planetary Science Division of the NASA Science Mission Directorate, the National Science Foundation under Grant No.~AST-1238877, the University of Maryland, and Eotvos Lorand University (ELTE). Operation of the Pan-STARRS1 telescope is supported by the National Aeronautics and Space Administration under Grant No.~NNX12AR65G and Grant No.~NNX14AM74G issued through the NEO Observation Program.
Based on observations made with the Nordic Optical Telescope, operated by the Nordic Optical Telescope Scientific Association at the Observatorio del Roque de los Muchachos, La Palma, Spain, of the Instituto de Astrofisica de Canarias. A.G.-Y.~is supported by the EU/FP7 via ERC grant No.~307260, the Quantum Universe I-Core programme by the Israeli Committee for Planning and Budgeting and the ISF; by Minerva and ISF grants; by the Weizmann-UK `making connections' programme; and by the Kimmel and YeS awards.
B.D.M. is supported by NSF grant AST-1410950 and the Alfred P. Sloan Foundation.
Support for L.G.~is provided by the Ministry of Economy, Development, and Tourism's Millennium Science Initiative through grant IC120009 awarded to The Millennium Institute of Astrophysics (MAS), and CONICYT through FONDECYT grant 3140566.
This work was partly supported by the European Union FP7 programme through ERC grant number 320360.
K.~M.~acknowledges support from the STFC through an Ernest Rutherford Fellowship.
A.M.~acknowledges funding from CNRS.
Development of ASAS-SN has been supported by NSF grant AST-0908816 and CCAPP at the Ohio State University. ASAS-SN is supported by NSF grant AST-1515927, the Center for Cosmology and AstroParticle Physics (CCAPP) at OSU, the Mt.~Cuba Astronomical Foundation, George Skestos, and the Robert Martin Ayers Sciences Fund.
B.S.~is supported by NASA through Hubble Fellowship grant HF-51348.001 awarded by the Space Telescope Science Institute, which is operated by the Association of Universities for Research in Astronomy, Inc., for NASA, under contract NAS 5-26555. C.S.K.~is supported by NSF grants AST-1515876 and AST-1515927.
T.W.-S.H.~is supported by the DOE Computational Science Graduate Fellowship, grant number DE-FG02-97ER25308.
V.A.V.~is supported by a NSF Graduate Research Fellowship.
P.S.C.~is grateful for support provided by the NSF through the Graduate Research Fellowship Program, grant DGE1144152. 
P.B.~is supported by the National Science Foundation Graduate Research Fellowship Program under Grant No. DGE1144152.
D.A.H., C.M.~and G.H.~are supported by NSF grant 1313484.

\bibliographystyle{apj}

\bibliography{/Users/matt/Documents/Papers/mybib}

\clearpage

\appendix

\setcounter{figure}{0}
\setcounter{table}{0}

\section{Photometric data}

\begin{table*}
\centering
\caption{Ground-based photometry in SDSS-like $ugriz$ filters (AB magnitudes)} \label{tab:ugriz}
\scriptsize
\begin{tabular}{cccccccc}
\hline
MJD	& Phase$^a$  & $u$	&	$g$	&	$r$	&	$i$	&	$z$	&	Telescope \\
\hline	
57048.7	&	-47.9	&	--	&	--	&	--	&	17.62 (0.07)	&	--	&	PS1	\\
57049.7	&	-47.0	&	--	&	--	&	--	&	17.59 (0.07)	&	--	&	PS1	\\
57053.5	&	-43.6	&	--	&	--	&	17.33 (0.13)	&	--	&	--	&	PS1	\\
57068.47	&	-30.1	&	--	&	16.94 (0.01)	&	--	&	--	&	--	&	PS1	\\
57069.45	&	-29.2	&	--	&	16.88 (0.01)	&	--	&	--	&	--	&	PS1	\\
57071.6	&	-27.3	&	--	&	16.89 (0.05)	&	--	&	17.12 (0.09)	&	--	&	LCOGT-1m	\\
57072.59	&	-26.4	&	--	&	16.80 (0.04)	&	16.98 (0.06)	&	17.18 (0.07)	&	17.44 (0.13)	&	LCOGT-1m	\\
57073.88	&	-25.3	&	--	&	16.81 (0.09)	&	17.02 (0.06)	&	17.16 (0.08)	&	--	&	LCOGT-1m	\\
57077.17	&	-22.3	&	17.08 (0.05)	&	16.87 (0.04)	&	16.95 (0.06)	&	17.10 (0.03)	&	17.21 (0.04)	&	LT+IO:O	\\
57077.45	&	-22.0	&	--	&	16.88 (0.01)	&	--	&	--	&	--	&	PS1	\\
57077.62	&	-21.9	&	--	&	16.83 (0.03)	&	16.92 (0.06)	&	17.13 (0.06)	&	--	&	LCOGT-1m	\\
57078.47	&	-21.1	&	--	&	16.88 (0.01)	&	--	&	--	&	--	&	PS1	\\
57080.08	&	-19.7	&	16.98 (0.04)	&	16.82 (0.03)	&	16.94 (0.05)	&	17.05 (0.02)	&	17.23 (0.02)	&	LT+IO:O	\\
57081.5	&	-18.4	&	--	&	16.78 (0.08)	&	17.00 (0.06)	&	17.05 (0.04)	&	--	&	LCOGT-1m	\\
57082.94	&	-17.1	&	17.11 (0.07)	&	16.79 (0.04)	&	16.92 (0.05)	&	17.01 (0.02)	&	17.18 (0.03)	&	LT+IO:O	\\
57085.22	&	-15.1	&	--	&	16.63 (0.09)	&	16.95 (0.08)	&	16.98 (0.06)	&	--	&	LCOGT-1m	\\
57086.01	&	-14.4	&	17.01 (0.04)	&	16.77 (0.08)	&	16.90 (0.05)	&	17.01 (0.07)	&	17.13 (0.04)	&	LT+IO:O	\\
57090.97	&	-9.9	&	--	&	16.68 (0.03)	&	16.82 (0.13)	&	16.95 (0.11)	&	--	&	LCOGT-1m	\\
57093.02	&	-8.1	&	16.78 (0.04)	&	16.63 (0.03)	&	16.78 (0.05)	&	16.92 (0.02)	&	17.07 (0.03)	&	LT+IO:O	\\
57094.87	&	-6.4	&	--	&	16.64 (0.04)	&	16.75 (0.07)	&	16.90 (0.09)	&	--	&	LCOGT-1m	\\
57095.47	&	-5.9	&	--	&	16.70 (0.02)	&	--	&	--	&	--	&	PS1	\\
57095.93	&	-5.5	&	16.79 (0.03)	&	16.61 (0.04)	&	16.77 (0.05)	&	16.89 (0.02)	&	17.05 (0.02)	&	LT+IO:O	\\
57098.61	&	-3.0	&	--	&	16.62 (0.06)	&	16.80 (0.05)	&	16.89 (0.04)	&	--	&	LCOGT-1m	\\
57102.52	&	0.5	&	--	&	16.61 (0.04)	&	16.72 (0.06)	&	16.87 (0.06)	&	--	&	LCOGT-1m	\\
57103.4	&	1.3	&	--	&	16.67 (0.01)	&	--	&	--	&	--	&	PS1	\\
57104.43	&	2.2	&	--	&	16.73 (0.01)	&	--	&	--	&	--	&	PS1	\\
57106.44	&	4.0	&	--	&	16.67 (0.14)	&	16.84 (0.13)	&	16.92 (0.13)	&	--	&	LCOGT-1m	\\
57110.16	&	7.3	&	--	&	16.75 (0.05)	&	16.85 (0.06)	&	16.88 (0.08)	&	--	&	LCOGT-1m	\\
57116.93	&	13.4	&	16.98 (0.10)	&	16.83 (0.04)	&	16.91 (0.07)	&	16.95 (0.07)	&	17.09 (0.03)	&	LT+IO:O	\\
57120.15	&	16.3	&	--	&	16.96 (0.10)	&	16.99 (0.06)	&	17.03 (0.08)	&	--	&	LCOGT-1m	\\
57122.85	&	18.7	&	17.29 (0.07)	&	17.00 (0.06)	&	17.01 (0.05)	&	17.05 (0.02)	&	17.14 (0.03)	&	LT+IO:O	\\
57123.87	&	19.6	&	--	&	17.07 (0.15)	&	17.12 (0.14)	&	17.20 (0.12)	&	--	&	LCOGT-1m	\\
57127.72	&	23.1	&	--	&	17.18 (0.19)	&	17.21 (0.14)	&	17.18 (0.14)	&	--	&	LCOGT-1m	\\
57129.94	&	25.1	&	17.62 (0.02)	&	17.14 (0.02)	&	17.16 (0.05)	&	17.19 (0.02)	&	17.20 (0.02)	&	LT+IO:O	\\
57131.71	&	26.7	&	--	&	17.33 (0.15)	&	17.30 (0.14)	&	17.39 (0.10)	&	--	&	LCOGT-1m	\\
57132.93	&	27.8	&	17.76 (0.03)	&	17.24 (0.03)	&	17.20 (0.05)	&	17.22 (0.02)	&	17.21 (0.02)	&	LT+IO:O	\\
57135.44	&	30.0	&	--	&	17.42 (0.14)	&	17.28 (0.08)	&	17.33 (0.06)	&	--	&	LCOGT-1m	\\
57135.94	&	30.5	&	17.92 (0.03)	&	17.32 (0.02)	&	17.26 (0.05)	&	17.27 (0.02)	&	17.32 (0.02)	&	LT+IO:O	\\
57137.27	&	31.7	&	--	&	17.18 (0.01)	&	--	&	--	&	--	&	PS1	\\
57138.86	&	33.1	&	18.09 (0.04)	&	17.35 (0.03)	&	17.29 (0.05)	&	17.31 (0.02)	&	17.35 (0.02)	&	LT+IO:O	\\
57138.88	&	33.1	&	--	&	17.41 (0.06)	&	17.33 (0.07)	&	17.35 (0.07)	&	--	&	LCOGT-1m	\\
57144.71	&	38.4	&	--	&	17.60 (0.08)	&	17.50 (0.09)	&	17.40 (0.11)	&	--	&	LCOGT-1m	\\
57144.89	&	38.5	&	18.09 (0.08)	&	17.55 (0.06)	&	17.41 (0.06)	&	17.34 (0.03)	&	17.37 (0.05)	&	LT+IO:O	\\
57148.31	&	41.6	&	--	&	--	&	17.29 (0.02)	&	--	&	--	&	PS1	\\
57150.03	&	43.1	&	--	&	17.55 (0.03)	&	17.43 (0.05)	&	17.32 (0.03)	&	--	&	LCOGT-1m	\\
57150.96	&	44.0	&	18.21 (0.03)	&	17.51 (0.03)	&	17.39 (0.05)	&	17.35 (0.02)	&	17.37 (0.03)	&	LT+IO:O	\\
57155.76	&	48.3	&	--	&	17.60 (0.03)	&	17.44 (0.06)	&	17.43 (0.07)	&	--	&	LCOGT-1m	\\
57161.7	&	53.6	&	--	&	17.66 (0.03)	&	17.66 (0.08)	&	17.42 (0.10)	&	--	&	LCOGT-1m	\\
57162.73	&	54.5	&	--	&	17.84 (0.03)	&	17.66 (0.06)	&	--	&	--	&	LCOGT-1m	\\
57162.9	&	54.7	&	18.47 (0.05)	&	17.78 (0.03)	&	17.56 (0.05)	&	17.51 (0.02)	&	17.49 (0.02)	&	LT+IO:O	\\
57166.92	&	58.3	&	18.55 (0.04)	&	17.86 (0.04)	&	17.68 (0.06)	&	17.58 (0.03)	&	17.52 (0.03)	&	LT+IO:O	\\
57173.38	&	64.1	&	--	&	--	&	17.75 (0.06)	&	17.69 (0.06)	&	--	&	LCOGT-1m	\\
57176.95	&	67.3	&	18.95 (0.05)	&	18.18 (0.03)	&	17.90 (0.06)	&	17.75 (0.03)	&	17.72 (0.03)	&	LT+IO:O	\\
57179.11	&	69.2	&	--	&	18.18 (0.05)	&	17.88 (0.06)	&	17.85 (0.05)	&	--	&	LCOGT-1m	\\
57184.96	&	74.5	&	--	&	18.28 (0.02)	&	18.06 (0.05)	&	17.97 (0.06)	&	--	&	LCOGT-1m	\\
57192.71	&	81.5	&	--	&	18.41 (0.03)	&	18.12 (0.06)	&	17.98 (0.04)	&	--	&	LCOGT-1m	\\
57199.38	&	87.4	&	--	&	18.39 (0.04)	&	18.23 (0.06)	&	18.07 (0.06)	&	--	&	LCOGT-1m	\\
57206.34	&	93.7	&	--	&	18.33 (0.09)	&	18.17 (0.09)	&	17.98 (0.11)	&	--	&	LCOGT-1m	\\
57212.39	&	99.1	&	--	&	18.42 (0.04)	&	18.31 (0.06)	&	18.22 (0.04)	&	--	&	LCOGT-1m	\\
57217.73	&	103.9	&	--	&	18.53 (0.04)	&	18.37 (0.06)	&	18.25 (0.09)	&	--	&	LCOGT-1m	\\
57230.35	&	115.3	&	--	&	18.68 (0.05)	&	18.42 (0.07)	&	18.26 (0.06)	&	--	&	LCOGT-1m	\\
57236.71	&	121.0	&	--	&	18.71 (0.10)	&	18.53 (0.09)	&	18.39 (0.10)	&	--	&	LCOGT-1m	\\
57373.33$^{b}$	&	243.7	&	21.24 (0.07)	&	20.67 (0.14)	&	20.50 (0.13)	&	20.18 (0.10)	&	19.96 (0.13)	&	NTT+EFOSC2	\\
\hline
\end{tabular}

$^a$ In rest-frame days from MJD\,=\,56102; $^{b}$ Magnitude after removal of host galaxy flux

\end{table*}

\begin{table*}
\centering
\caption{Ground-based photometry from SLT in $BVRI$ filters (Vega magnitudes)} \label{tab:BVRI}
\footnotesize
\begin{tabular}{cccccc}
\hline
MJD	& Phase$^a$ 	&	$B$	&	$V$	&	$R$	&	$I$	 \\
\hline	
57074.21	&	-25.0	&	16.99 (0.09)	&	16.82 (0.05)	&	16.91 (0.09)	&	16.71 (0.05)	 \\
57079.34	&	-20.3	&	16.99 (0.08)	&	16.85 (0.08)	&	16.86 (0.06)	&	16.75 (0.04)	 \\
57080.33	&	-19.5	&	16.94 (0.05)	&	16.75 (0.07)	&	16.80 (0.08)	&	16.72 (0.02)	 \\
57081.33	&	-18.6	&	16.93 (0.06)	&	16.81 (0.06)	&	16.83 (0.07)	&	16.77 (0.07)	 \\
57082.24	&	-17.7	&	17.05 (0.17)	&	16.76 (0.10)	&	16.81 (0.08)	&	16.67 (0.04)	 \\
57084.24	&	-15.9	&	17.07 (0.12)	&	16.81 (0.09)	&	16.81 (0.06)	&	16.60 (0.04)	 \\
57085.26	&	-15.0	&	17.00 (0.14)	&	16.77 (0.15)	&	16.75 (0.05)	&	16.68 (0.12)	 \\
57094.29	&	-6.9	&	16.81 (0.04)	&	16.62 (0.06)	&	16.71 (0.06)	&	16.55 (0.03)	 \\
57095.27	&	-6.0	&	16.85 (0.05)	&	--	&	16.71 (0.07)	&	--					 \\
57097.21	&	-4.3	&	16.76 (0.08)	&	16.61 (0.09)	&	16.64 (0.09)	&	16.57 (0.11)	 \\
57098.21	&	-3.4	&	16.80 (0.07)	&	16.56 (0.06)	&	16.65 (0.10)	&	16.50 (0.04)	 \\
57099.25	&	-2.5	&	16.81 (0.05)	&	16.61 (0.08)	&	16.64 (0.07)	&	16.53 (0.04)	 \\
57101.31	&	-0.6	&	16.80 (0.05)	&	16.58 (0.07)	&	16.69 (0.07)	&	16.50 (0.05)	 \\
57104.18	&	2.0	&	16.77 (0.08)	&	16.67 (0.09)	&	16.68 (0.08)	&	16.51 (0.05)	 \\
57110.21	&	7.4	&	16.87 (0.07)	&	16.66 (0.08)	&	16.71 (0.08)	&	16.51 (0.08)	 \\
57113.29	&	10.1	&	16.85 (0.10)	&	16.72 (0.16)	&	16.72 (0.12)	&	16.52 (0.06)	 \\
57117.25	&	13.7	&	16.96 (0.12)	&	16.65 (0.08)	&	16.85 (0.09)	&	--			 \\
57118.21	&	14.6	&	17.14 (0.16)	&	16.81 (0.16)	&	--	&	16.49 (0.07)			 \\
57120.21	&	16.4	&	17.02 (0.18)	&	16.78 (0.07)	&	16.84 (0.09)	&	16.62 (0.05)	 \\
57126.18	&	21.7	&	17.28 (0.07)	&	16.88 (0.04)	&	16.93 (0.07)	&	16.69 (0.07)	 \\
57128.16	&	23.5	&	17.33 (0.06)	&	17.03 (0.07)	&	16.97 (0.07)	&	16.82 (0.04)	 \\
57134.24	&	29.0	&	17.57 (0.07)	&	17.21 (0.08)	&	17.10 (0.08)	&	16.89 (0.13)	 \\
57137.21	&	31.6	&	17.66 (0.07)	&	17.24 (0.09)	&	17.15 (0.07)	&	16.95 (0.09)	 \\
57139.21	&	33.4	&	17.86 (0.07)	&	17.34 (0.07)	&	17.16 (0.08)	&	17.00 (0.06)	 \\
57141.17	&	35.2	&	17.88 (0.13)	&	17.48 (0.09)	&	17.27 (0.07)	&	17.05 (0.09)	 \\
57152.16	&	45.0	&	17.76 (0.07)	&	17.41 (0.08)	&	17.24 (0.08)	&	17.05 (0.09)	 \\
57155.15	&	47.7	&	17.89 (0.06)	&	17.43 (0.08)	&	17.22 (0.07)	&	17.03 (0.12)	 \\
57160.05	&	52.1	&	17.92 (0.07)	&	17.59 (0.09)	&	17.38 (0.06)	&	--			 \\
57161.01	&	53.0	&	18.03 (0.06)	&	--	&	17.44 (0.09)	&	17.15 (0.07)			 \\
57175.02	&	65.6	&	18.60 (0.22)	&	17.96 (0.09)	&	17.53 (0.05)	&	17.33 (0.05)	 \\
57178.11	&	68.3	&	--	&	18.01 (0.09)	&	17.79 (0.08)	&	17.44 (0.07)			 \\
57182.12	&	71.9	&	--	&	17.99 (0.15)	&	17.72 (0.05)	&	17.54 (0.15)			 \\
57183.02	&	72.8	&	18.55 (0.08)	&	17.96 (0.08)	&	17.83 (0.08)	&	17.56 (0.10)	 \\
57184.03	&	73.7	&	18.65 (0.08)	&	18.01 (0.10)	&	17.88 (0.07)	&	17.53 (0.07)	 \\
57185.02	&	74.6	&	18.56 (0.07)	&	18.06 (0.05)	&	17.78 (0.07)	&	17.61 (0.09)	 \\
57199.06	&	87.2	&	18.77 (0.16)	&	18.39 (0.15)	&	17.96 (0.12)	&	--			 \\
57210.04	&	97.0	&	18.76 (0.20)	&	18.48 (0.11)	&	18.10 (0.11)	&	17.85 (0.16)	 \\
57221.0	&	106.9	&	18.66 (0.20)	&	18.38 (0.20)	&	18.38 (0.09)	&	--		 \\
\hline
\end{tabular}

$^a$ In rest-frame days from MJD\,=\,56102

\end{table*}

\begin{table*}
\centering
\caption{\textit{Swift} UVOT photometry (Vega magnitudes)} \label{tab:swift}
\footnotesize
\begin{tabular}{cccccccc}
\hline
MJD	& Phase$^a$  & $uvw2$	&	$uvm2$	&	$uvw1$	&	$u$	&	$b$	&	$v$ \\
\hline	
57072.39	&	-26.6	&	17.38 (0.05)	&	16.85 (0.04)	&	16.57 (0.05)	&	15.93 (0.04)	&	16.91 (0.04)	&	16.86 (0.06)	\\
57074.17	&	-25.0	&	17.34 (0.05)	&	16.92 (0.05)	&	16.54 (0.04)	&	15.94 (0.03)	&	--	&	--	\\
57076.17	&	-23.2	&	17.40 (0.07)	&	16.91 (0.12)	&	16.56 (0.08)	&	16.00 (0.04)	&	--	&	--	\\
57078.32	&	-21.3	&	17.37 (0.07)	&	17.09 (0.08)	&	16.49 (0.05)	&	15.97 (0.04)	&	--	&	--	\\
57080.52	&	-19.3	&	17.43 (0.07)	&	17.05 (0.07)	&	16.50 (0.05)	&	15.94 (0.04)	&	--	&	--	\\
57084.34	&	-15.9	&	17.36 (0.07)	&	16.87 (0.07)	&	16.52 (0.05)	&	15.95 (0.04)	&	--	&	--	\\
57089.98	&	-10.8	&	17.37 (0.07)	&	16.73 (0.06)	&	16.36 (0.06)	&	15.96 (0.06)	&	16.73 (0.06)	&	16.53 (0.09)	\\
57090.48	&	-10.3	&	--	&	--	&	--	&	15.81 (0.03)	&	--	&	--	\\
57093.76	&	-7.4	&	17.18 (0.06)	&	16.72 (0.05)	&	16.38 (0.06)	&	15.73 (0.05)	&	16.64 (0.05)	&	16.55 (0.08)	\\
57098.99	&	-2.7	&	17.23 (0.07)	&	16.78 (0.09)	&	16.36 (0.07)	&	15.74 (0.05)	&	16.58 (0.05)	&	16.61 (0.09)	\\
57100.98	&	-0.9	&	--	&	--	&	--	&	15.69 (0.06)	&	--	&	--	\\
57106.87	&	4.4	&	17.38 (0.06)	&	16.78 (0.07)	&	16.40 (0.06)	&	15.78 (0.04)	&	16.63 (0.04)	&	16.59 (0.07)	\\
57111.50	&	8.5	&	17.34 (0.05)	&	17.08 (0.10)	&	16.53 (0.05)	&	15.85 (0.04)	&	16.72 (0.04)	&	16.51 (0.09)	\\
57110.70	&	7.8	&	--	&	--	&	--	&	15.89 (0.04)	&	--	&	--	\\
57118.42	&	14.7	&	17.67 (0.06)	&	17.33 (0.08)	&	16.79 (0.06)	&	16.09 (0.04)	&	16.86 (0.05)	&	16.80 (0.07)	\\
57122.08	&	18.0	&	17.92 (0.10)	&	17.41 (0.13)	&	16.93 (0.09)	&	16.26 (0.06)	&	16.99 (0.07)	&	16.89 (0.11)	\\
57122.79	&	18.7	&	--	&	--	&	--	&	16.28 (0.03)	&	--	&	--	\\
57166.35	&	57.8	&	--	&	--	&	--	&	17.65 (0.04)	&	--	&	--	\\
57185.21	&	74.7	&	--	&	--	&	--	&	18.14 (0.04)	&	--	&	--	\\
57206.83	&	94.1	&	--	&	--	&	--	&	18.33 (0.05)	&	--	&	--	\\
57330.72	&	205.4	&	--	&	--	&	--	&	19.92 (0.24)	&	--	&	--	\\
57335.79	&	209.9	&	--	&	--	&	--	&	19.73 (0.21)	&	--	&	--	\\
57120.82	&	16.9	&	17.88 (0.07)	&	17.41 (0.09)	&	16.97 (0.07)	&	16.18 (0.05)	&	17.02 (0.05)	&	16.74 (0.07)	\\
57128.63	&	23.9	&	18.27 (0.08)	&	17.96 (0.07)	&	17.33 (0.07)	&	16.62 (0.06)	&	17.19 (0.06)	&	16.93 (0.08)	\\
57131.53	&	26.5	&	--	&	18.18 (0.10)	&	--	&	16.74 (0.05)	&	--	&	--	\\
57136.31	&	30.8	&	18.62 (0.13)	&	18.58 (0.35)	&	17.79 (0.12)	&	16.89 (0.08)	&	17.55 (0.08)	&	17.06 (0.11)	\\
57144.47	&	38.1	&	19.13 (0.11)	&	18.66 (0.09)	&	18.14 (0.09)	&	17.09 (0.07)	&	17.66 (0.07)	&	17.26 (0.10)	\\
57151.19	&	44.2	&	19.16 (0.10)	&	19.02 (0.13)	&	18.14 (0.09)	&	17.24 (0.06)	&	17.74 (0.06)	&	17.47 (0.09)	\\
57160.63	&	52.6	&	19.24 (0.13)	&	18.95 (0.11)	&	18.20 (0.11)	&	17.39 (0.08)	&	17.92 (0.08)	&	17.45 (0.12)	\\
57172.41	&	63.2	&	20.10 (0.24)	&	19.49 (0.16)	&	18.56 (0.15)	&	18.11 (0.15)	&	18.59 (0.15)	&	17.78 (0.17)	\\
57231.37	&	116.2	&	--			&	--			&	--			&	18.40 (0.32)	&	--	&	--	\\
57232.37	&	117.1	&	20.06 (0.23)	&	20.32 (0.25)	&	19.27 (0.24)	&	--	&	18.90 (0.25)	&	18.33 (0.35)	\\
57330.72$^{b}$	&	205.4	&	--			&	--			&	--			&	20.02 (0.24)	&	--	&	--	\\
57335.79$^{b}$	&	209.9	&	--			&	--			&	--			&	19.84 (0.21)	&	--	&	--	\\
\hline
\end{tabular}

$^a$ In rest-frame days from MJD\,=\,56102; $^{b}$ Magnitude after removal of host galaxy flux

\end{table*}

\begin{table*}
\centering
\caption{Ground-based NIR photometry (Vega magnitudes)} \label{tab:JHK}
\footnotesize
\begin{tabular}{cccccc}
\hline
MJD	& Phase$^a$  & $J$	&	$H$	&	$K$	&	Telescope \\
\hline	
57092.17	&	-8.83	&	16.70 (0.12)	&	16.64 (0.10)	&	16.43 (0.19)	&	NTT+SOFI \\
57102.17	&	0.15	&	16.60 (0.09)	&	16.65 (0.12)	&	--	&	NTT+SOFI \\
57110.18	&	7.35	&	16.61 (0.07)	&	16.61 (0.27)	&	16.39 (0.30)	&	NTT+SOFI \\
57122.13	&	18.08	&	16.73 (0.12)	&	16.71 (0.26)	&	16.51 (0.24)	&	NTT+SOFI \\
57140.10	&	34.21	&	16.83 (0.09)	&	16.85 (0.20)	&	16.56 (0.20)	&	NTT+SOFI \\
57171.90	&	62.77	&	17.27 (0.12)	&	17.18 (0.13)	&	16.95 (0.14)	&	NOT+NOTCam \\
57205.70	&	93.12	&	17.50 (0.09)	&	17.48 (0.12)	&	17.19 (0.16)	&	NOT+NOTCam \\
57379.20$^{b}$	&	248.92	&	19.21 (0.08)	&	--	&	18.67 (0.06)	&	NOT+NOTCam \\
\hline
\end{tabular}

$^a$ In rest-frame days from MJD\,=\,56102; $^{b}$ Magnitude after removal of host galaxy flux

\end{table*}

\begin{table}
\centering
\caption{Publicly available CRTS photometry} \label{tab:crts}
\footnotesize
\begin{tabular}{ccc}
\hline
MJD	& Phase$^a$  &	$R^b$ \\
\hline
57014.00    & -79.0    &    18.87 (0.28)         \\  
57064.00    & -34.1    &    16.95 (0.18)           \\
57091.44    & -9.5     &   16.66 (0.04)             \\     
57097.44    & -4.1      &  16.63 (0.02)               \\   
57102.44    & 0.4     &   16.65 (0.02)                 \\ 
57109.44    & 6.7      &  16.65 (0.04)                  \\
57123.44    & 19.3      &  16.88 (0.02)                  \\
57129.44    & 24.6      &  17.04 (0.02)                  \\
57135.44    & 30.0      &  17.15 (0.05)                  \\
57155.44    & 48.0     &   17.33 (0.12)                  \\
57162.44    & 54.3      &  17.44 (0.11)                  \\
\hline

\end{tabular}

$^a$ In rest-frame days from MJD\,=\,56102 \\
$^b$ Average magnitude of $\approx4$ detections on each night

\end{table}

\begin{table}
\centering
\caption{ASAS-SN photometry from stacks of neighbouring epochs} \label{tab:asas}
\footnotesize
\begin{tabular}{ccc}
\hline
Mean MJD	& Phase$^a$  &	$V$ \\
\hline
57014.07	&	-78.96	&	$>$18.74 \\
57033.49	&	-61.52	&	$>$18.06 \\
57042.03	&	-53.85	&	17.50 (0.12) \\
57051.47	&	-45.38	&	17.19 (0.09) \\
57068.97	&	-29.66	&	16.81 (0.06) \\
57078.96	&	-20.69	&	16.94 (0.07) \\
57103.76	&	1.58		&	16.51 (0.07) \\
57120.45	&	16.57	&	16.67 (0.12) \\
57134.86	&	29.51	&	17.29 (0.13) \\
57158.80	&	51.01	&	17.41 (0.14) \\
57186.77	&	76.12	&	17.95 (0.28) \\
57378.08	&	247.92	&	$>$18.76 \\

\hline

\end{tabular}

$^a$ In rest-frame days from MJD\,=\,56102 \\

\end{table}

\begin{table}
\centering
\caption{SDSS sequence stars used to calibrate photometry} \label{tab:seq}
\footnotesize
\begin{tabular}{cccccccc}
\hline
Star	& RA  &	Dec	& $u$	&	$g$	&	$r$	&	$i$	&	$z$ \\
\hline
1	&	173.4232	&	0.7152961	&	17.28	&	16.23	&	15.88	&	15.76	&	15.73 \\
2	&	173.4245	&	0.7042278	&	16.48	&	15.10	&	14.86	&	14.38	&	14.30 \\
3	&	173.3985	&	0.7062970	&	16.31	&	15.01	&	14.27	&	14.23	&	14.06 \\
4	&	173.3756	&	0.7023557	&	17.03	&	15.74	&	15.27	&	15.12 	&	15.10 \\
5	&	173.3951	&	0.7397069	&	17.42	&	16.36	&	15.95	&	15.81	&	15.76 \\
6	&	173.4251	&	0.6638585	&	16.40	&	15.06	&	14.50	&	14.24	&	14.10 \\
7	&	173.4648	&	0.6588082	&	17.24	&	15.45	&	14.76	&	14.53	&	14.42 \\
8	&	173.4230	&	0.6537502	&	17.43	&	16.40	&	16.03	&	15.90	&	15.87 \\
9	&	173.4628	&	0.8027824	&	16.76	&	15.69	&	15.35	&	15.23	&	15.20 \\
10	&	173.3373	&	0.7143689	&	17.48	&	16.38	&	16.01	&	15.88	&	15.84 \\

\hline

\end{tabular}

NOTE: for the $UBVRI$ photometry, we transformed these magnitudes following \citep{jor2006}. \\
For $JHK$ photometry, we used all available 2MASS point sources in the field \citep{Skrutskie2006a}.

\end{table}

\begin{figure}[h]
\centering
\includegraphics[width=8.7cm,angle=0]{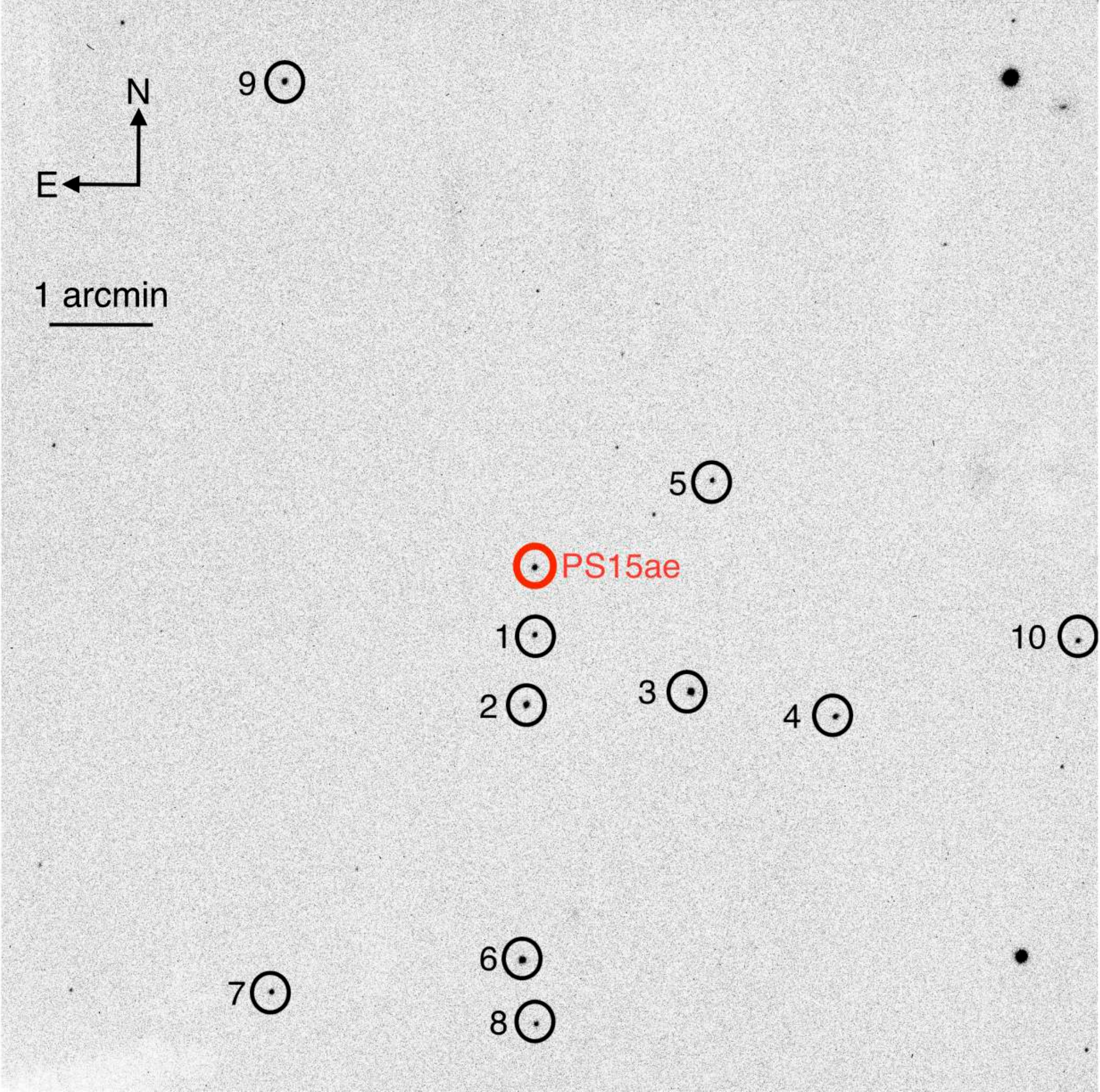}
\figcaption{$r$-band image of SN 2015bn from LT, showing the positions of the SDSS field stars used to calibrate the magnitudes.\label{fig:seq}}
\end{figure}

\begin{table}
\centering
\caption{Spectra of SN 2015bn\label{tab:spec}}
\begin{tabular}{cccccccc}
\hline
Date   & MJD    & Phase$^a$  &  Instrument  & 	Grism or Grating	&  Exposure time (s)	& Airmass    & Average resolution (\AA)\\
\hline			    		                    
2015-02-17 & 57071.3 & $-$28      & EFOSC2       &	Gr13 &    900	&	1.16      &  18  \\
2015-02-18 & 57072.3 & $-$27       & IMACS         &  G300-17.5	&    900       &      1.16	&	6     \\
2015-02-24 & 57077.6 & $-$22      & WiFeS         & R3000,B3000	&    1200     &     1.26	&	2  \\
2015-02-24 & 57078.3 & $-$21       & SOFI        & BG	&       4800	&	1.19   & 23   \\
2015-02-25 & 57079.3 & $-$20     & EFOSC2     & Gr11,Gr16	&    1800	&	1.21   & 13  \\
2015-03-01 & 57083.2 & $-$17      & SPRAT        & Wasatch600	&     900		&	1.67      &  18    \\
2015-03-11 & 57092.2 & $-$9       & SOFI   	  & BG	&     3240	&	1.16       &   23   \\
2015-03-11 & 57092.4 & $-$9    & MMT Blue Channel 	&    300GPM  &  300	&	1.35  &  6.5   \\
2015-03-11 & 57093.1 & $-$8    & EFOSC2       & Gr11,Gr16	& 	1800	&	1.42   &  13   \\
2015-03-12 & 57093.5 & $-$7    & WiFeS       &	 R3000,B3000	&  	3600  &	1.28	&	2     \\
%2015-03-16 & &     & FLOYDS       &    &     \\
2015-03-18 & 57100.3 & $-$2    & EFOSC2       & Gr11,Gr16	&  1800	&	1.31  &  13   \\
2015-03-24 & 57105.8 &  +3   & SNIFS       	& red+blue	&  1200		&	1.97  &   2  \\
2015-03-27 & 57109.3 & +7    & EFOSC2       	& Gr11,Gr16	&  1800	&	1.43  &  13   \\
2015-04-11 & 57124.2 & +20    & EFOSC2       & Gr11,Gr16	&  2100	&	1.51  &  13   \\
2015-04-22 & 57135.2 & +30    & IMACS       	&  G300-17.5	&  1200  &   1.66	&	6  \\
2015-04-23 & 57135.4 &  +30   & FLOYDS       & red+blue	& 3600	&	1.29   &  1.5   \\
2015-04-24 & 57136.0 & +31    & IMACS       	&  G300-17.5	&	1200 	&	1.25   &   6  \\
2015-04-28 & 57140.2 &  +34   & SOFI      	 & BG	&  3240	&	1.37  &  23   \\
2015-05-07 & 57149.9 & +43    & SPRAT        & Wasatch600	&  1800  &  1.14 	&	18  \\
2015-05-09 & 57151.9 &  +45   & SPRAT       & Wasatch600	& 1800  &   1.16	&	18  \\
2015-05-15 & 57157.2 &  +50   & IMACS       &  G300-17.5  	& 1200	&	1.61  &   6  \\
2015-05-23 & 57166.8 &  +58   & SNIFS       &   red+blue	&	1200		&	1.28    &    2 \\
%2015-05-31 & &     & FLOYDS       &    &     \\
2015-06-09 & 57182.7 &  +72   & SNIFS       &   red+blue	&  2000 	&	1.21  & 2    \\
2015-06-21 & 57194.7 &  +83   & SNIFS       &  red+blue 	&   1800		&	1.36  &  2   \\
2015-06-28 & 57201.7 &  +89   & SNIFS       &   red+blue	&  1800 	& 	1.65	&	2    \\
2015-07-08 & 57203.8 &  +98   & SNIFS       &   red+blue	&  1600  	&  	1.77	&	 2  \\
2015-07-16 & 57219.9 &  +106   & LDSS3       & VPH-all	& 1800	&	1.72   &  12   \\
2015-12-16 & 57372.4 &  +243   & IMACS       &  G300-17.5	&  900  & 	1.37	&  6  \\

\hline
\end{tabular}

$^a$ Phase in rest-frame days relative to epoch of maximum light.

\end{table}

\end{document}